\documentclass[twocolumn,twocolappendix]{aastex631}

\newcommand\kms{km\,s$^{-1}$}
\shortauthors{Chandra et al.}
\graphicspath{{./}{fig/}}
\newcommand\minesweeper{\texttt{MINESweeper}}
\newcommand\vtravel{$v_\mathrm{travel}$}
\newcommand{\GG}[1]{}
\newcommand\lunit{kpc\,km\,s$^{-1}$}

\usepackage{CJKutf8}

\defcitealias{Petersen2021}{PP21}
\defcitealias{Garavito-Camargo2019}{GC19}
\defcitealias{Garavito-Camargo2021}{GC21}
\defcitealias{Yaaqib2024}{YPP24}

\shorttitle{The Reflex Response to the LMC}

\begin{document}

\title{All-Sky Kinematics of the Distant Halo: The Reflex Response to the LMC}

\author[0000-0002-0572-8012]{Vedant~Chandra}
\affiliation{Center for Astrophysics $\mid$ Harvard \& Smithsonian, 60 Garden St, Cambridge, MA 02138, USA}

\author[0000-0003-3997-5705]{Rohan~P.~Naidu}
\altaffiliation{NASA Hubble Fellow}
\affiliation{MIT Kavli Institute for Astrophysics and Space Research, 77 Massachusetts Ave., Cambridge, MA 02139, USA}

\author[0000-0002-1590-8551]{Charlie~Conroy}
\affiliation{Center for Astrophysics $\mid$ Harvard \& Smithsonian, 60 Garden St, Cambridge, MA 02138, USA}

\author[0000-0001-7107-1744]{Nicolas~Garavito-Camargo} 
\affiliation{Center for Computational Astrophysics, Flatiron Institute, 162 5th Ave, New York, NY 10010, USA}

\author[0000-0003-3922-7336]{Chervin~Laporte}
\affiliation{LIRA, Observatoire de Paris, Universit\'e PSL, Sorbonne Universit\'e, Universit\'e Paris Cit\'e, CY Cergy Paris Universit\'e, CNRS, 92190 Meudon, France}

\affiliation{Institut de Ciencies del Cosmos (ICCUB), Universitat de Barcelona (IEEC-UB), Martí i Franquès 1, E-08028 Barcelona, Spain}

\affiliation{Kavli IPMU (WPI), UTIAS, The University of Tokyo, Kashiwa, Chiba 277-8583, Japan}

%%%%%%%%%%%

\author[0000-0002-7846-9787]{Ana~Bonaca}
\affiliation{The Observatories of the Carnegie Institution for Science, 813 Santa Barbara Street, Pasadena, CA 91101, USA}

\author[0000-0002-1617-8917]{Phillip~A.~Cargile}
\affiliation{Center for Astrophysics $\mid$ Harvard \& Smithsonian, 60 Garden St, Cambridge, MA 02138, USA}

\author[0000-0002-6993-0826]{Emily~Cunningham}
\altaffiliation{NASA Hubble Fellow}
\affiliation{Department of Astronomy, Columbia University, 550 West 120th Street, New York, NY, 10027, USA}
%\affiliation{Center for Computational Astrophysics, Flatiron Institute, 162 5th Ave, New York, NY 10010, USA}

\author[0000-0002-6800-5778]{Jiwon~Jesse~Han}
\affiliation{Center for Astrophysics $\mid$ Harvard \& Smithsonian, 60 Garden St, Cambridge, MA 02138, USA}

\author[0000-0002-9280-7594]{Benjamin~D.~Johnson}
\affiliation{Center for Astrophysics $\mid$ Harvard \& Smithsonian, 60 Garden St, Cambridge, MA 02138, USA}

\author[0000-0003-4996-9069]{Hans-Walter~Rix}
\affiliation{Max-Planck-Institut f{\"u}r Astronomie, K{\"o}nigstuhl 17, D-69117 Heidelberg, Germany}

\author[0000-0001-5082-9536]{Yuan-Sen~Ting \begin{CJK*}{UTF8}{gbsn}(丁源森)\end{CJK*}}
\affiliation{Research School of Astronomy \& Astrophysics, Australian National University, Cotter Road, Weston, ACT 2611, Australia}
\affiliation{School of Computing, Australian National University, Acton ACT 2601, Australia}
\affiliation{Department of Astronomy, The Ohio State University, Columbus, OH 43210, USA}

\author[0000-0002-0721-6715]{Rebecca~Woody}
\affiliation{Center for Astrophysics $\mid$ Harvard \& Smithsonian, 60 Garden St, Cambridge, MA 02138, USA}

\author[0000-0002-5177-727X]{Dennis~Zaritsky}
\affiliation{Steward Observatory and Department of Astronomy, University of Arizona, Tucson, AZ 85721, USA}

\correspondingauthor{Vedant Chandra}
\email{vedant.chandra@cfa.harvard.edu}

\begin{abstract}

\noindent 
The infall of the Large Magellanic Cloud (LMC) is predicted to displace the inner Milky Way (MW), imprinting an apparent `reflex motion' on the observed velocities of distant halo stars. 
We construct the largest all-sky spectroscopic dataset of luminous red giant stars from $50-160$~kpc, including a new survey of the southern celestial hemisphere.
We fit the full 6D kinematics of our data to measure the amplitude and direction of the inner MW's motion towards the outer halo.
The observed velocity grows with distance such that, relative to halo stars at $100$~kpc, the inner MW is lurching at $\approx 40$~\kms{} towards a recent location along the LMC's past orbit. 
Our measurements align with N-body simulations of the halo's response to a $1.8 \times 10^{11}\,M_\odot$ LMC on first infall, suggesting that the LMC is at least 15\% as massive as the MW.  
Our findings highlight the dramatic disequilibrium of the MW outskirts, and will enable more accurate measurements of the total mass of our Galaxy.

\end{abstract}

\keywords{Milky Way stellar halo (1060), Stellar kinematics (1608), Large Magellanic Cloud (903), Milky Way dynamics (1051)}

\section{Introduction} \label{sec:intro}

The LMC and SMC (hereafter the `Clouds') are by far the Milky Way's (MW's) most prominent satellite galaxies, and were likely the first external galaxies discovered \citep[e.g.,][]{Dennefeld2020}.
There is growing evidence that the Clouds form a binary pair --- with the LMC being  $\approx 10\times$ more massive than the SMC --- and are on their first orbital  infall towards the Milky Way \citep[][but see \citealt{Vasiliev2024} about the plausibility of a second infall scenario]{Kallivayalil2006, Besla2007, Besla2010, Kallivayalil2013}.
Multiple independent measurements estimate the total mass of the LMC to be between $1.0-2.5 \times 10^{11}\,M_\odot$ \citep{Penarrubia2016, Erkal2019a, Erkal2020, Vasiliev2021a, Shipp2021, CorreaMagnus2022, Koposov2023, Vasiliev2023, Fushimi2023, Watkins2024}, $10-25\%$ the mass of the Milky Way itself \citep[e.g.,][]{Bland-Hawthorn2016, Wang2020, Shen2022}. 

Given its large mass, the LMC is expected to significantly perturb our own Galaxy (see \citealt{Vasiliev2023} for a thorough review).
On a global scale, the LMC exerts a gravitational acceleration on various MW components as it approaches, and is predicted to shift the center-of-mass (COM) of the Galaxy significantly. 
While the disk and inner halo can quickly react to the changing COM by moving toward the LMC, stars in the outer halo are left behind due to their large distances and long orbital timescales ($\gtrsim 1$~Gyr). 
This results in an apparent `reflex motion' imprinted in the velocities of distant halo stars, which should become more prominent at larger distances \citep{Weinberg1995, Gomez2015, Erkal2019, Sheng2024, Kravtsov2024}.
The dominant observational effect is that distant halo stars should have apparent `upward' (positive $z$) velocities in the Galactocentric frame \citep{Laporte2018, Erkal2019, Garavito-Camargo2019, Garavito-Camargo2021, Petersen2020, Cunningham2020a, Cao2023}. 
\cite{Erkal2020} demonstrate that this effect differentially distorts the MW halo, and must be accounted for in dynamical analyses; neglecting the influence of the LMC can inflate mass estimates of the MW by up to $50\%$ (see also \citealt{CorreaMagnus2022}, \citealt{Chamberlain2023}, and \citealt{Kravtsov2024}).
On a more local scale, the LMC affects dark matter (DM) in the Milky Way's halo, imprinting a dynamical friction `wake' in the DM density along its past orbit \citep{Chandrasekhar1943, Garavito-Camargo2019, Tamfal2021, Foote2023}. 
These effects are predicted to leave a prominent signature in the density and kinematics of MW halo stars, with the perturbations being most prominent in the outer halo beyond the LMC at $\gtrsim 50$~kpc. 

In recent years, excellent progress has been made towards observationally detecting the influence of the LMC on MW halo stars. 
\cite{Belokurov2019a} used RR Lyrae and BHB stars to study a prominent overdensity in the southern sky, the Pisces Plume. 
This overdensity is elongated along the past orbit of the Clouds, and is systematically offset towards negative Galactocentric velocities, as predicted by their simulations of the LMC's wake. 
\cite{Conroy2021} constructed an all-sky sample of luminous red giant stars, and observed a wake-like overdensity in the southern hemisphere, along with a more diffuse overdensity in the northern hemisphere. 
Both of these features are predicted by the simulations of \citet[][hereafter \citetalias{Garavito-Camargo2019}]{Garavito-Camargo2019}, although the density contrast in the data is twice as strong as the simulations. 
\cite{Amarante2024} recently characterized the Pisces Plume with a large sample of BHB stars, although they do not detect a corresponding diffuse overdensity in the north. 

Attempts to measure the LMC's effects on the `bulk' MW halo can be significantly confounded by the presence of undetected coherent substructure in the outer halo.
The most prominent of these structures in the outer halo is the Sagittarius stream (Sgr), although 6D kinematics provide a powerful way to excise  Sgr stars on the basis of their angular momenta \citep{Johnson2020a, Penarrubia2021}.
The LMC and SMC are themselves predicted to contribute stripped stars to the Galactic halo, forming a stellar counterpart to the well-known gaseous Magellanic Stream \citep[MS; e.g.,][]{Besla2013, Petersen2022}. 
\cite{Zaritsky2020} discovered coherent debris in the H3 Survey \citep{Conroy2019b} at $\approx 50$~kpc, and argued that these might have a Magellanic origin.
\cite{Chandra2023b} subsequently discovered a population of stars that appear to be stripped from the Clouds (the Magellanic Stellar Stream, or MSS), spanning the entire extent of the trailing MS from 50-100~kpc. 
The MSS and other debris could partially constitute the Pisces Plume overdensity, but there remains a wake-like structure in the outer halo once the MSS is removed \citep{Chandra2023b, Amarante2024}.

Given the overlapping substructures present in the halo, full 6D kinematics are vital to disentangle coherent structures from the bulk MW halo. 
Recently, wide-field spectroscopic surveys --- combined with astrometry from the \textit{Gaia} space observatory --- have enabled the kinematic detection of the LMC's global influence on the MW halo.
\cite{Erkal2021} compiled Galactocentric radial velocities of halo stars beyond 30~kpc, and found that stars in the northern Galactic hemisphere are systematically redshifted relative to those in the south, as expected for the LMC-induced reflex motion. 
The amplitude of their measured velocity dipole was broadly consistent with a $1.5 \times 10^{11}\,M_\odot$ LMC. 
Stellar streams are also emerging as powerful tools to map the gravitational field of the Galaxy, including the LMC's influence \citep[e.g.,][]{Erkal2019a, Koposov2023, Lilleengen2023, Brooks2024}. 
\cite{Vasiliev2021a} comprehensively modelled the formation of the Sgr stream in the presence of the LMC, finding that the motion of the inner MW towards the LMC is a crucial ingredient to reproduce the observed properties of Sgr.

\citet[][hereafter \citetalias{Petersen2021}]{Petersen2021} fit a physical model of the reflex velocity to the 6D phase-space positions of RGB and BHB stars beyond 40~kpc, and measured a global velocity vector $v_\mathrm{travel} = 32 \pm 4$~\kms{} with an apex (maximum velocity) direction that points towards a location along the past orbit of the LMC. 
While the amplitude of their measured signal is consistent with simulations of a $\gtrsim 10^{11}\,M_\odot$ LMC, the measured apex direction is somewhat different from most simulations of the MW-LMC interaction \citep{Vasiliev2023}. 
Whereas \citetalias{Petersen2021} measure an apex direction in the opposite southern Galactic quadrant as the LMC today ($\ell_\mathrm{apex} > 0^\circ$), many simulations predict that the velocity dipole should point towards a more recent point on the LMC's orbit ($\ell_\mathrm{apex} < 0^\circ$) \citep[e.g.,][]{Garavito-Camargo2019, Vasiliev2021a}. 
More recent work by \citet[][hereafter \citetalias{Yaaqib2024}]{Yaaqib2024} finds an increasing \vtravel{} velocity towards larger distances using a sample spanning $20 < r_\mathrm{gal}/\mathrm{kpc} < 60$, with an apex direction that varies by tens of degrees as a function of distance. 

One major limitation of existing outer halo datasets is that they lack coverage at southern declinations, since all spectroscopic surveys probing the distant halo (e.g, the SEGUE survey, \citealt{Yanny2009}, used by \citetalias{Petersen2021} and \citetalias{Yaaqib2024}) have been conducted from the northern hemisphere. 
The H3 Spectroscopic Survey has significantly increased the number of distant halo stars with full 6D information, but this too is limited to declinations $\delta \gtrsim -20^\circ$ \citep{Conroy2019b}. 
Kinematic samples of the outer halo therefore have a large hole in the Galactic south, in precisely the quadrant where the LMC resides and exerts its maximal dynamical influence. 
It is therefore desirable to construct an all-sky dataset of outer halo stars, to obtain a global perspective on the Galaxy's kinematic response to the LMC. 

Over the past two years, we have been conducting a tailored spectroscopic survey of the most distant giants in the southern hemisphere, using the Magellan Echellette Spectrograph (MagE; \citealt{Marshall2008}) on the 6.5m Magellan Baade telescope. 
Our survey has currently observed $\approx 290$ giants with heliocentric distances $d_\mathrm{helio} > 50$~kpc, including $\approx 100$ giants beyond $d_\mathrm{helio} > 100$~kpc. 
In combination with past northern surveys like SEGUE and H3, this is the premier all-sky dataset to kinematically map out the furthest outskirts of our Galaxy. 
Previous results from our survey include identifying distant remnants of our Galaxy's last major merger \citep{Chandra2023a} and the discovery of the Magellanic Stellar Stream \citep{Chandra2023b}.

In this work, we combine our MagE survey data with existing data from the H3 and SEGUE surveys to construct the largest all-sky dataset of red giants with full 6D kinematics beyond 40~kpc, with all spectra homogeneously analyzed using the same pipeline (see $\S$\ref{sec:data}). 
We illustrate the all-sky kinematics of the outer halo in $\S$\ref{sec:velstruct}, demonstrating the dramatic dynamical disequilibrium of the MW outskirts.
Subsequently in $\S$\ref{sec:vreflex_model} we construct a framework to measure the LMC-induced reflex motion with a distance-dependent continuity model for the 6D kinematics of our sample.
We present our fitted model in $\S$\ref{sec:vreflex_results} and compare our results to N-body simulations and past measurements. 
We simultaneously measure the reflex-corrected bulk kinematic properties of the outer halo in $\S$\ref{sec:bulkvel}, and examine effects of the local dynamical wake of the LMC in $\S$\ref{sec:wake}. 
The broad implications of our findings are discussed in $\S$\ref{sec:discussion}, and our results are summarized in $\S$\ref{sec:conclusion}.

\section{Data}\label{sec:data}

This section describes the construction of an all-sky dataset of RGB stars with full 6D phase-space information beyond 40~kpc.
Throughout this work, we adopt a right-handed Galactocentric frame with a solar position $\mathbf{x}_\odot = (-8.12, 0.00, 0.02)$~kpc, and solar velocity $\mathbf{v}_\odot = (12.9, 245.6, 7.8)$~$\mathrm{km\,s^{-1}}$ \citep{Reid2004,Drimmel2018,GravityCollaboration2018}. 
Coordinate transformation are carried out using \texttt{astropy} \citep{AstropyCollaboration2013, AstropyCollaboration2018, AstropyCollaboration2022} and \texttt{gala} \citep{gala,adrian_price_whelan_2020_4159870}.

\subsection{MagE Survey}

Since 2022, we have been executing a tailored spectroscopic survey of luminous red giants in the outer halo. 
The selection procedure for our parent sample of red giant stars is detailed in \cite{Chandra2023a}, and our spectroscopic survey is further described in \cite{Chandra2023b}.
Briefly, we utilize \textit{Gaia} DR3 astrometry and unWISE photometry to filter out nearby dwarf stars and isolate giants off the Galactic plane ($|b| > 20^\circ$) with isochrone-predicted distances $\gtrsim 80$\,kpc \citep{Mainzer2014,Schlafly2019,GaiaCollaboration2021,Lindegren2021,GaiaCollaboration2022}. 

We obtained spectroscopic observations of a subset of the above selected giants with the Magellan Echellete Spectrograph (MagE; \citealt{Marshall2008}) on the 6.5m Magellan Baade Telescope at Las Campanas Observatory (PIs: Chandra \& Naidu). 
The stars are typically between $17 \lesssim G \lesssim 18.6$, and we utilize magnitude-dependent exposure times ranging from $10-40$~minutes per star to achieve signal-to-noise ratio (SNR)~$\approx 10~\mathrm{pixel}^{-1}$ at $5100$~\AA. 
Spectra are reduced with a fully-automated pipeline\footnote{\url{https://github.com/vedantchandra/merlin/releases/tag/v0.2}} built around the \texttt{PypeIt} utility \citep{pypeit:joss_pub}. 

Stellar parameters are estimated with the Bayesian \minesweeper{} code \citep{Cargile2020}. 
We fit a region of the MagE spectra (from 4800--5500~\AA) that contains the \ion{Mg}{1} triplet, where our linelists are best-calibrated. 
We also include archival optical--infrared broadband photometry in the likelihood, along with the \textit{Gaia} parallax \citep{Fukugita1996, Gunn1998, Skrutskie2006,  Mainzer2014, Chambers2016, GaiaCollaboration2021, GaiaCollaboration2022}. 
\minesweeper{} compares these observables to synthetic Kurucz model spectra, and additionally constrains the models to lie on \texttt{MIST} isochrones \citep{Kurucz1970, Kurucz1981, Choi2016}.
The posterior distribution of stellar parameters is sampled with \texttt{dynesty} \citep{Speagle2020}, producing measurements of the radial velocity $v_\mathrm{r}$, effective temperature $T_\mathrm{eff}$, surface gravity $\log{g}$, metallicity [Fe/H], [$\alpha$/Fe] abundance, and heliocentric distance. 
Repeat observations of bright radial velocity standard stars --- HIP4148 and HIP22787 --- over multiple nights indicate that our systematic RV accuracy floor is $\approx 1$\,\kms{}. 
For more details on the MagE outer halo survey, we refer to \cite{Chandra2023a,Chandra2023b}. 

\begin{figure}
    \centering
    \includegraphics[width=\columnwidth]{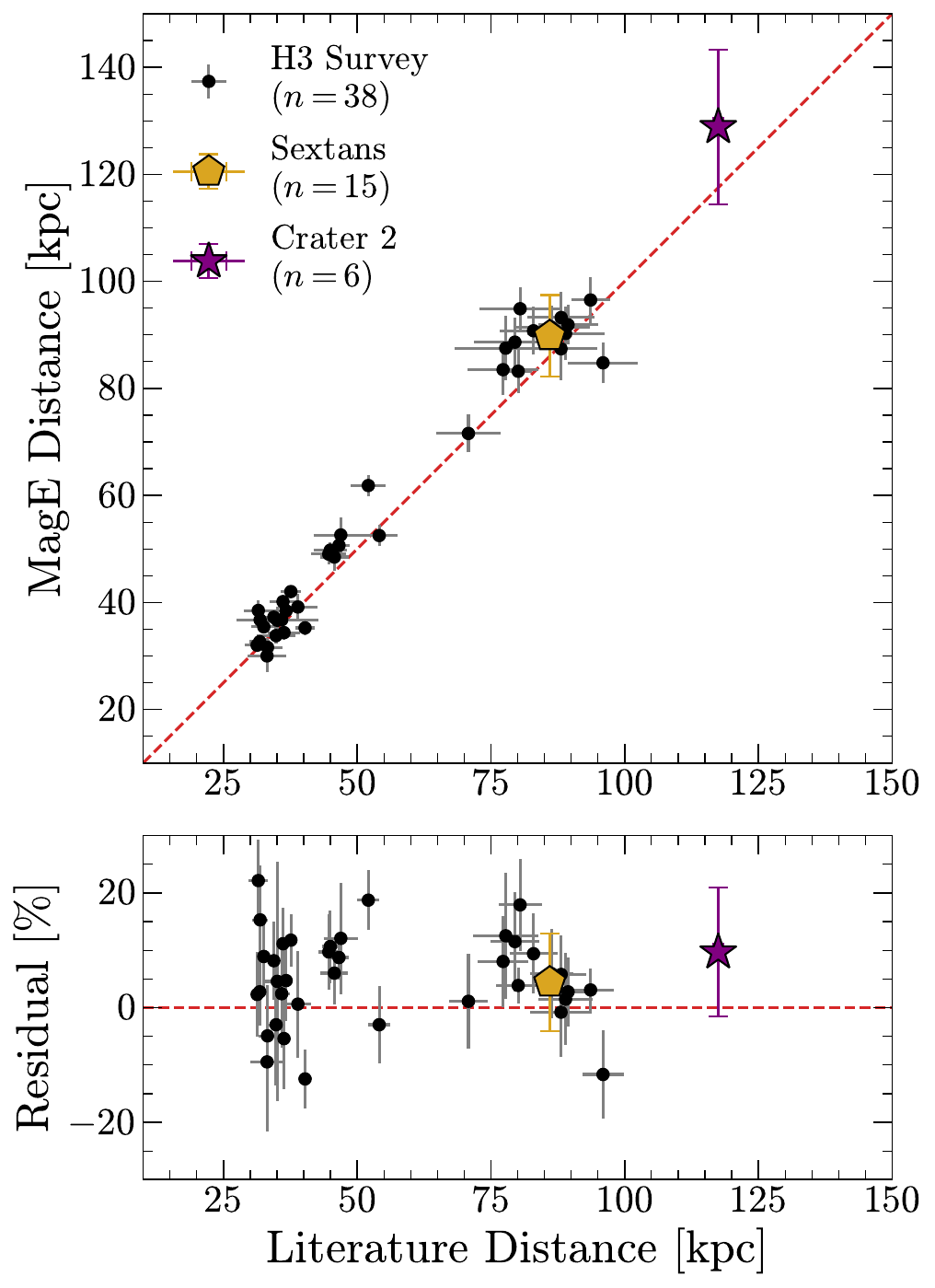}
    \caption{Validating the spectroscopic distances of stars in our MagE survey beyond 30~kpc.
    Black points show giant stars from the H3 Survey, which were fitted with a similar analysis pipeline as our MagE survey.
    The two colored points show mean distances of observed stars in the Crater~2 and Sextans dwarf galaxies, with their literature distances sourced from \cite{Torrealba2016b} and \cite{Munoz2018b} respectively.
    The lower panel shows the fractional residual, the difference between the MagE and literature distance divided by the literature distance.}
    \label{fig:dval}
\end{figure}

Figure~\ref{fig:dval} shows a cross-validation of the spectroscopic distance scale of our MagE survey. 
Apart from observing stars in common with the H3 Survey \citep{Conroy2019b}, we observed giants from our parent giant catalog belonging to the Sextans ($86 \pm 1$~kpc, \citealt{Irwin1990, Munoz2018b}) and Crater~2 ($117 \pm 1$~kpc, \citealt{Torrealba2016b}) dwarf galaxies. 
Figure~\ref{fig:dval} illustrates that we obtain reliable distance estimates out to $\gtrsim 100$~kpc.
The typical statistical distance uncertainties are smaller in the MagE dataset than H3 --- for stars in common --- since we aimed for higher SNR in the MagE survey, leading to more precise metallicities and consequently more precise isochrone distances.
Using the scatter in the relation shown in Figure~\ref{fig:dval} as a guide, we adopt an uncertainty floor of 10\% on our distance measurements.
This is consistent with the standard deviation of the measured distances in Crater~2 and Sextans, which are both $\approx 10\%$.

There is a $5\%$ systematic offset between the H3 and MagE spectroscopic distance scales (see bottom panel of Figure~\ref{fig:dval}), most likely caused by the MagE metallicity scale being $\approx 0.1$ dex lower than H3. 
Although we have not yet determined the cause of this offset, it is smaller than the typical statistical uncertainties on these distances.
We have verified that artificially shrinking the MagE distances by $5\%$ to the H3 scale has a negligible effect on the results presented here.
Furthermore, there is a $\approx 3$~\kms{} mean offset between MagE and H3 RVs of the same stars. 
We have likewise verified that shifting the MagE RVs to the H3 scale has a negligible effect on our results.

As of May 2024, we have observed $400$~stars in this survey, of which $291$ are spectroscopically confirmed to lie at heliocentric distances beyond 50~kpc, and $106$ are beyond 100~kpc. 
This is already the largest dataset of stars with precise metallicities and 6D phase-space information beyond 100~kpc, and by far the largest dataset of outer halo stars beyond $\approx 50$~kpc observed from the southern hemisphere. 
Our sample selection has a very low contamination rate from dwarfs ($\approx 10\%$). 
All data from this survey will eventually be made publicly available, once our primary scientific projects have been completed. 

\subsection{H3 \& SEGUE Data}

The H3 Spectroscopic Survey \citep{Conroy2019b} has been conducting a spectroscopic survey of halo stars with the Hectochelle instrument \citep{Szentgyorgyi2011} on the 6.5m MMT telescope at the Whipple Observatory in Arizona. 
We select stars observed by H3 up to January~2024 that have reliable stellar parameter fits from \minesweeper{}, and are not associated with cold substructures (stars with on-sky positions and velocities consistent with membership in dwarf galaxies, star clusters, or known stellar streams). 
Before applying any further selection cuts, there are $\approx 300$ giants beyond $d_\mathrm{helio} > 50$~kpc from H3. 
Typical RV uncertainties for these data are of order $0.5-1$~\kms{} in absolute terms \citep{Conroy2019b, Cargile2020}. 

The Sloan Extension for Galactic Understanding and Exploration (SEGUE; \citealt{Yanny2009}) survey observed a quarter of a million stars with the low-resolution BOSS spectrograph as a part of the Sloan Digital Sky Survey (SDSS; \citealt{York2000, Abazajian2009}).
We independently fit these spectra with the \minesweeper{} routine, from which $\approx 300$ stars with reliable parameters and distances $d_\mathrm{helio} > 50$~kpc are used in this work. 
Based on a comparison to literature parameters, we determine that SEGUE stars with SNR $<10$~per pixel do not produce reliable parameters, so we excise them from the dataset.
Due to the low resolution of BOSS, the SEGUE stars have typical RV uncertainties $\approx 5$~\kms{}. 

The dataset used here is not an exhaustive compilation of distant halo stars; rather, we have prioritized building a pure sample of stars with homogeneous stellar parameters derived with the \minesweeper{} pipeline. 
In cases where the same star has been observed by multiple surveys, we follow an order of precedence to select a single datapoint: MagE $>$ H3 $>$ SEGUE. 
This order is chosen to maximize RV accuracy; although the H3 Survey operates at a higher resolution than our MagE survey, the distant giants studied here typically have much higher SNR in our MagE data. 
Future work could incorporate the LAMOST survey, which has not been analyzed with \minesweeper{} yet \citep{Cui2012, Li2023lamm, Zhang2023lamk}.

\subsection{Sample Selection}\label{sec:sample}

\begin{figure}
    \centering
    \includegraphics[width=\columnwidth]{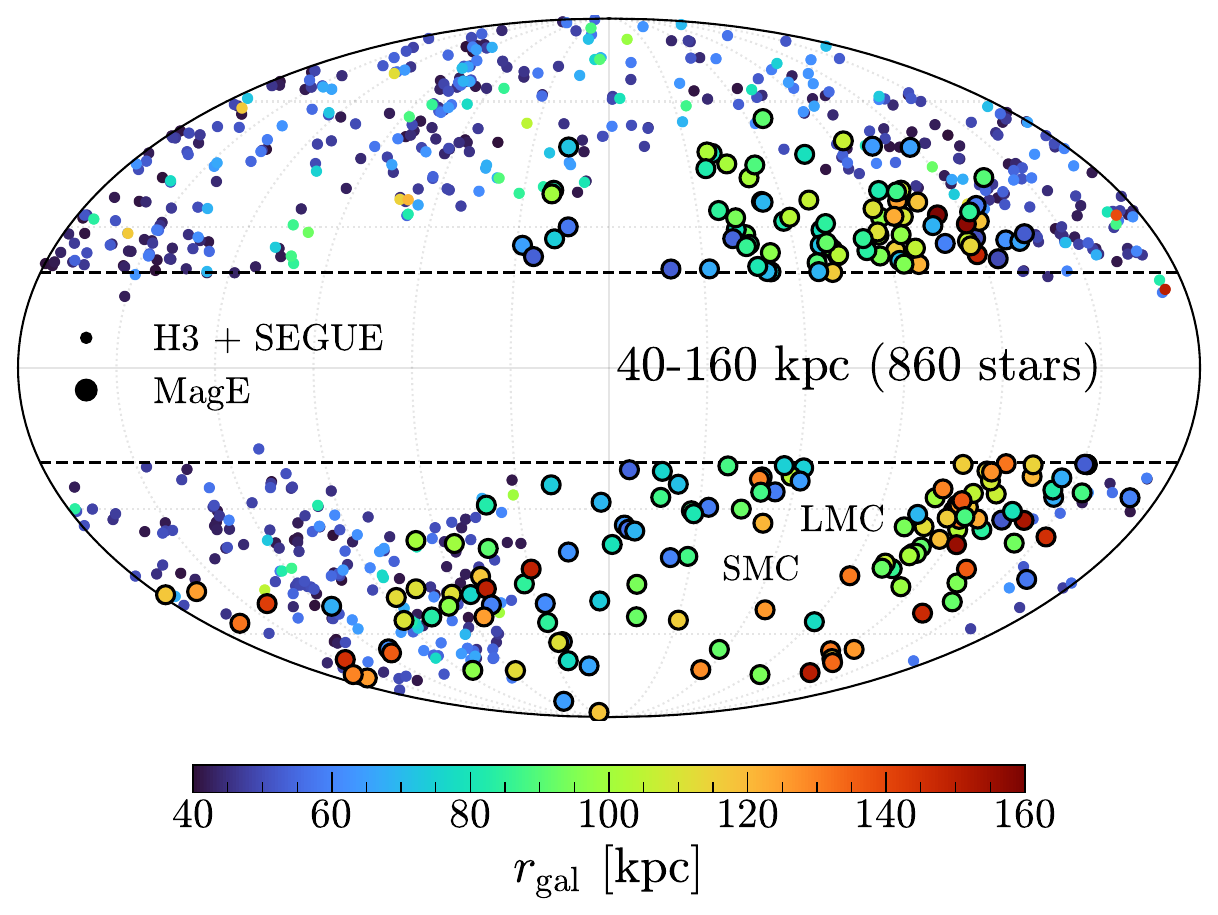}
    \includegraphics[width=\columnwidth]{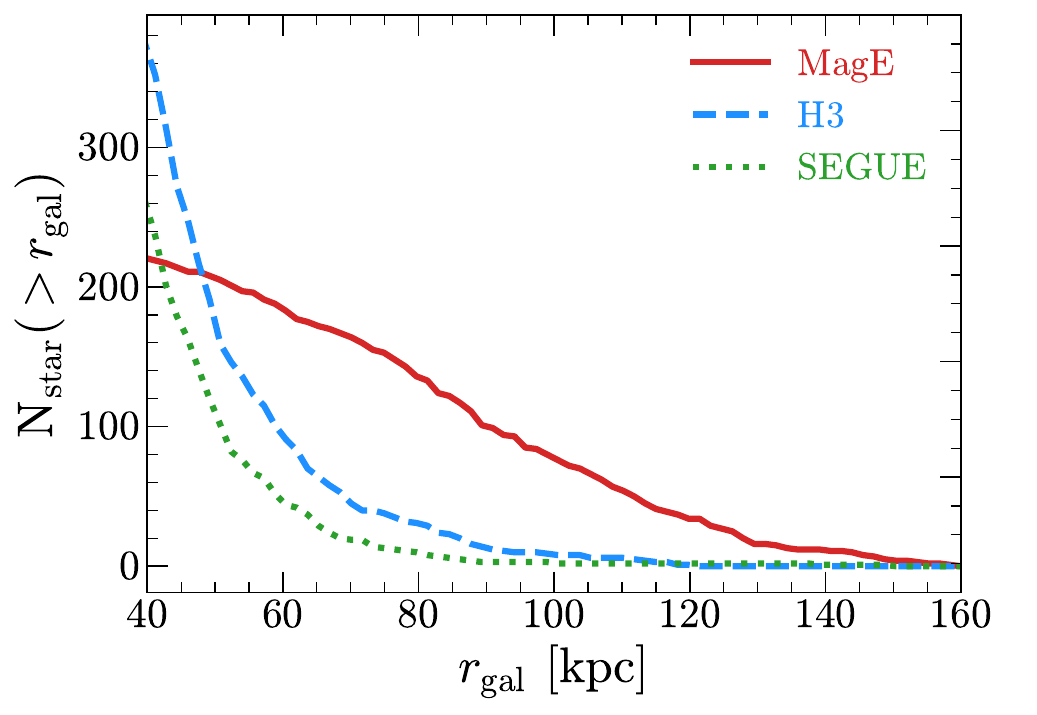}
    \caption{\textbf{Top:} Galactic map of our cleaned all-sky dataset of red giant stars beyond 40~kpc, colored by Galactocentric distance. 
    Stars with spectra from our MagE survey are outlined in black.
    \textbf{Bottom:} Cumulative number of stars in our sample beyond a given Galactocentric distance, split by spectroscopic survey. 
    Our MagE survey fills out the southern celestial hemisphere, and pushes to larger distances than previous surveys.}
    \label{fig:sample_map}
\end{figure}

Put together, our MagE+H3+SEGUE dataset contains $\approx 2600$ stars with $40 < r_\mathrm{gal}/\mathrm{kpc} < 160$. 
However, in order to reliably measure the halo's response to the LMC, it is vital to select a high-fidelity subset of these data with known substructures removed. 
To start with, we exclude $\approx 200$ stars with reported RV uncertainties $> 15$~\kms{} and fractional distance uncertainties $> 25\%$.

The most prominent stellar structure at these distances is the Sagittarius Stream \citep[Sgr; e.g.,][]{Majewski2003, Vasiliev2021a}. 
We kinematically exclude Sgr stars using the angular momentum cuts proposed by \citealt{Johnson2020a}, removing stars with $|b_\mathrm{Sgr}| < 25^\circ$ and $L_\mathrm{Y} < -2.5 - 0.3 \times L_\mathrm{Z}$, where the angular momenta are in units of \lunit. 
Within $25^\circ$ of the Sgr orbital plane (as defined by \citealt{Law2010a}), an additional excision is performed. 
Using the models of \cite{Vasiliev2021a} as a guide at each location of the stream, stars with distances within 10~kpc of the stream and radial velocities within 50~\kms{} of the stream (at the 3$\sigma$ level given reported uncertainties) are excised.

Stars with unphysical radial or tangential velocities $> 1000$\,\kms{} are removed, as are stars within $1^\circ$ of known dwarf galaxies and globular clusters \citep{McConnachie2012, Vasiliev2021b}. 
The 13 stars in our MagE dataset identified by \cite{Chandra2023b} as belonging to the Magellanic Stellar Stream are excised. 
Finally, although stars in our MagE survey are selected to lie far from the LMC and SMC (see Figure~\ref{fig:sample_map}), we search for any stars within $50^\circ$ of the LMC having proper motions and RVs consistent with belonging in an extended LMC halo. No stars confidently associated with the LMC are found.

Our final sample consists of $851$ field halo stars from SEGUE, H3, and MagE with $40 < r_\mathrm{gal}/\mathrm{kpc} < 160$. 
The top panel of Figure~\ref{fig:sample_map} shows their on-sky spatial distribution in the outer halo, and the bottom panel shows their cumulative distance distributions. 
The bottom panel of Figure~\ref{fig:sample_map} illustrates the high efficiency of our MagE survey in pushing to the furthest Galactic outskirts, far outnumbering the number of $r_\mathrm{gal} \gtrsim 60$~kpc stars compared to the H3 and SEGUE surveys. 
We adopt a systematic uncertainty floor of 1~\kms{} on RVs and 10\% on distances. 
Since most stars have reported catalog uncertainties smaller than this, the median uncertainties in our final catalog are 1~\kms{} for RVs and 10\% for distances.
The median absolute magnitude of stars in our sample is $M_\mathrm{G} = -1.3$, the median metallicity is [Fe/H]$= -1.3$, and the median alpha abundance is [$\alpha$/Fe]$=+0.2$.

\section{The Velocity Structure of the Outer Halo}\label{sec:velstruct}

\subsection{Radial Velocities}

\begin{figure}
    \centering
    \includegraphics[width=\columnwidth]{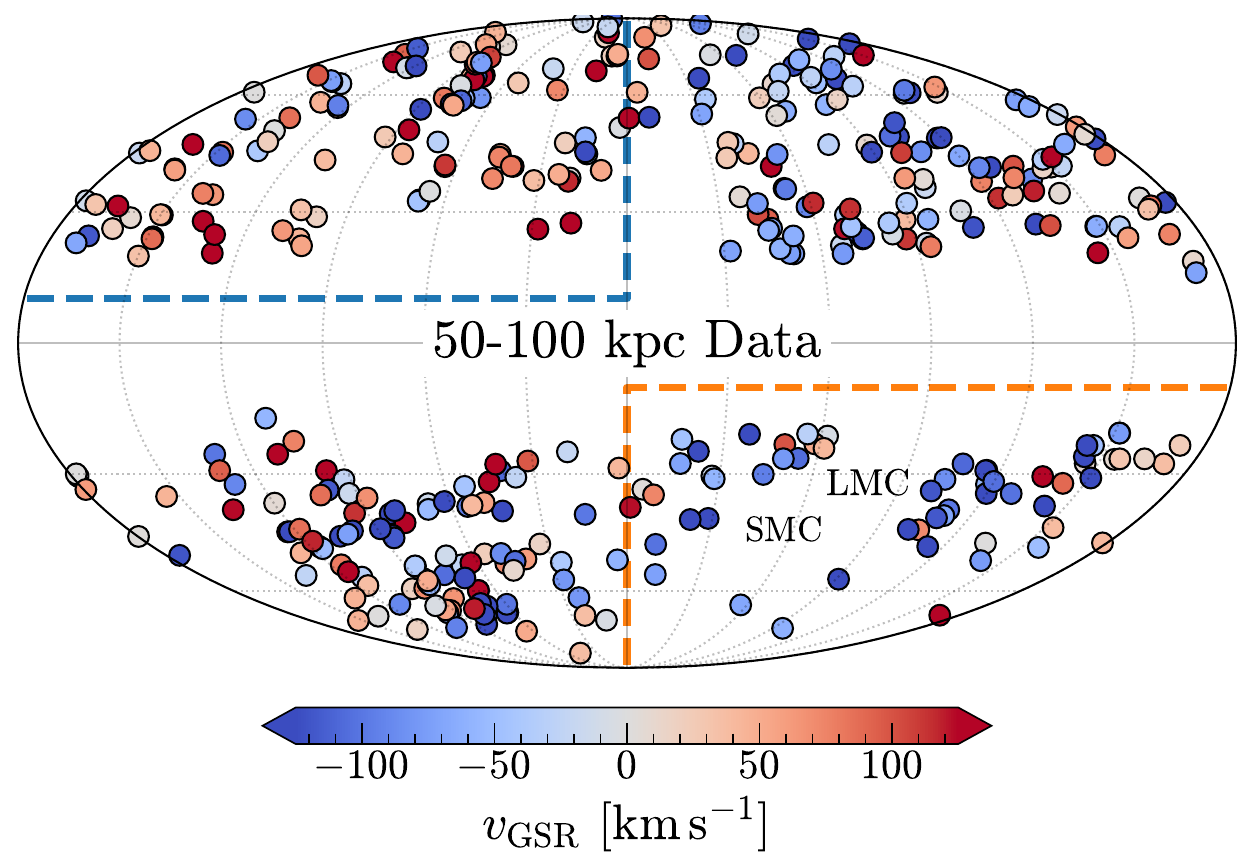}
    \includegraphics[width=\columnwidth]{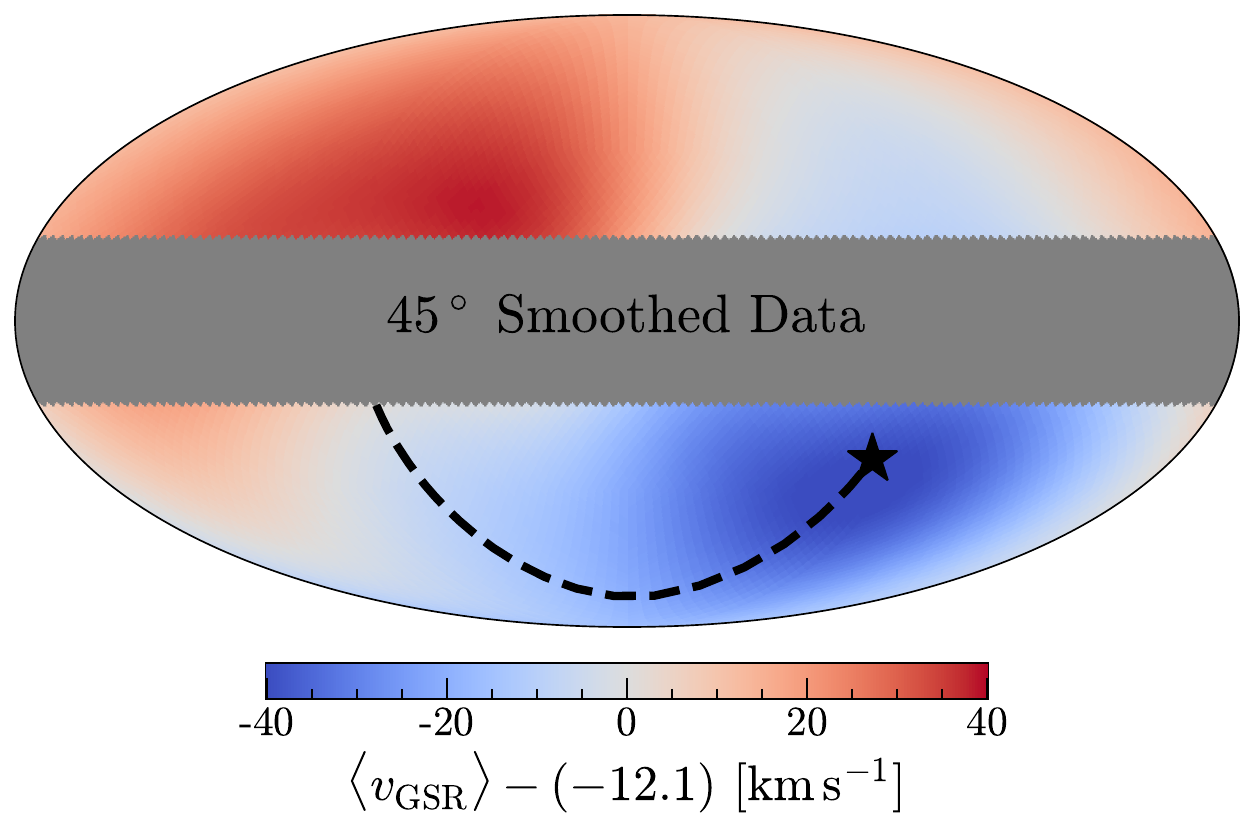}
    \includegraphics[width=\columnwidth]{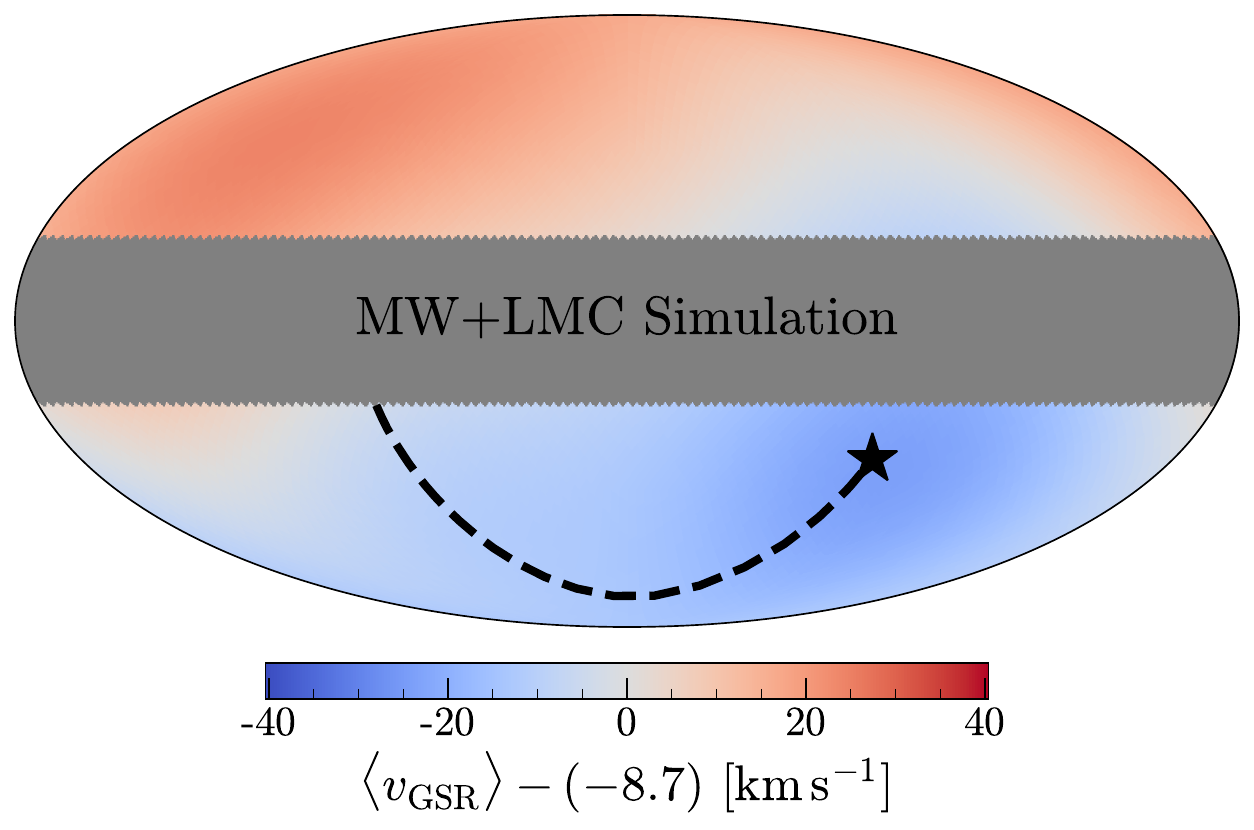}
    \caption{Comparing the radial velocity structure of our MW data to a numerical simulation of the halo's response to the LMC. 
    \textbf{Top:} Individual $v_\mathrm{GSR}$ measurements for stars in our dataset from $r_\mathrm{gal} = 50-100$~kpc.
    Dashed lines demarcate the quadrants used for the tomographic $v_\mathrm{GSR}$ measurements shown in Figure~\ref{fig:vgsr_rgal}.
    \textbf{Middle:} Average $v_\mathrm{GSR}$ map of our data, smoothed with a $45^\circ$ kernel.
    \textbf{Bottom:} Smoothed $v_\mathrm{GSR}$ computed using the MW+LMC3 simulation from \citetalias{Garavito-Camargo2019}, which assumed a $1.8 \times 10^{11}\,M_\odot$ halo mass for the LMC. 
    }
    \label{fig:sim_vgsr_map}
\end{figure}

We begin by considering the overall radial velocity structure of the halo. 
Figure~\ref{fig:sim_vgsr_map} shows the on-sky distribution of Galactocentric radial velocities $v_\mathrm{GSR}$ from our dataset, for stars with $r_\mathrm{gal} = 50-100$~kpc. 
We use this limited distance range for this figure instead of our full $40-160$~kpc sample to minimize the effect of the different selection functions of the surveys combined here. 
As Figure~\ref{fig:sample_map} illustrates, most stars beyond $100$~kpc are from our MagE survey and lie only in the southern hemisphere. 
The $r_\mathrm{gal} = 50-100$~kpc range ensures relatively even depth across the sky. 
In Appendix~\ref{sec:velmaps_distance}, we further investigate the distance-dependence of these smoothed velocity maps, finding consistent behavior across all distances. 

The top panel of Figure~\ref{fig:sim_vgsr_map} shows  $v_\mathrm{GSR}$ measurements for individual stars, and the middle panel shows the smoothed mean velocity map.
At each pixel on this map, the mean velocity of surrounding stars is computed, weighted by an angular Gaussian kernel with a radius of ~$45^\circ$. 
This effectively produces a smoothed velocity map in a continuous rather than binned manner. 
The maps are qualitatively unchanged but noisier if we use a smaller $15^\circ$ or $30^\circ$ kernel.
The sky-averaged mean velocity is subtracted out and indicated in the colorbar label.
The present-day location and past orbit of the LMC are overlaid, utilizing the orbital trajectory simulated by \citetalias{Garavito-Camargo2019} (model \texttt{LMC3} in their simulated suite, with a halo mass $1.8 \times 10^{11}\,M_\odot$). 

It is immediately apparent from Figure~\ref{fig:sim_vgsr_map} that the outer halo has a structured velocity distribution with variations across the sky.
The Galactocentric radial velocity exhibits a net negative $\approx -12$\,\kms{} signal. 
Furthermore, there is an unambiguous dipole signature in $v_\mathrm{GSR}$, with the southern halo systematically blueshifted, and the northern halo redshifted. 
Although this feature is most apparent in the smoothed map in the middle panel of Figure~\ref{fig:sim_vgsr_map}, it is visible even in the raw data in the top panel. 
This $v_\mathrm{GSR}$ dipole does not appear to be aligned purely in the north-south direction, but is rather tilted along an axis that approximately points towards the LMC. 
We emphasize that these are subtle mean trends in $v_\mathrm{GSR}$ brought out by averaging over large on-sky scales: the intrinsic radial velocity dispersion is 70$-$110~\kms{} depending on the sky location.
In Appendix~\ref{sec:velmaps_distance}, we demonstrate that this observation is robust to the distance range considered, and rule out this effect being an artifact due to the more distant coverage of the MagE data. 

From this $v_\mathrm{GSR}$ map alone, the reflex velocity imparted by the LMC is clearly visible.
Furthermore, the orientation of the signal in the data is consistent with numerical simulations of the interaction. 
The bottom panel of Figure~\ref{fig:sim_vgsr_map} shows the mean $v_\mathrm{GSR}$ map from the MW+LMC simulations of \citetalias{Garavito-Camargo2019}, specifically their \texttt{LMC3} model with a $1.8 \times 10^{11}\,M_\odot$ LMC virial mass. 
The data are quite similar to the results from this simulation --- particularly the tilted axis of the dipole --- although the observed dipole is slightly stronger than the simulated one. 
The simulation also exhibits a net negative sky-averaged $v_\mathrm{GSR} \approx -10$~\kms{}, hinting that this signal may be LMC-induced as well (as recently argued by \citetalias{Yaaqib2024}). 

\begin{figure}
    \centering
    \includegraphics[width=\columnwidth]{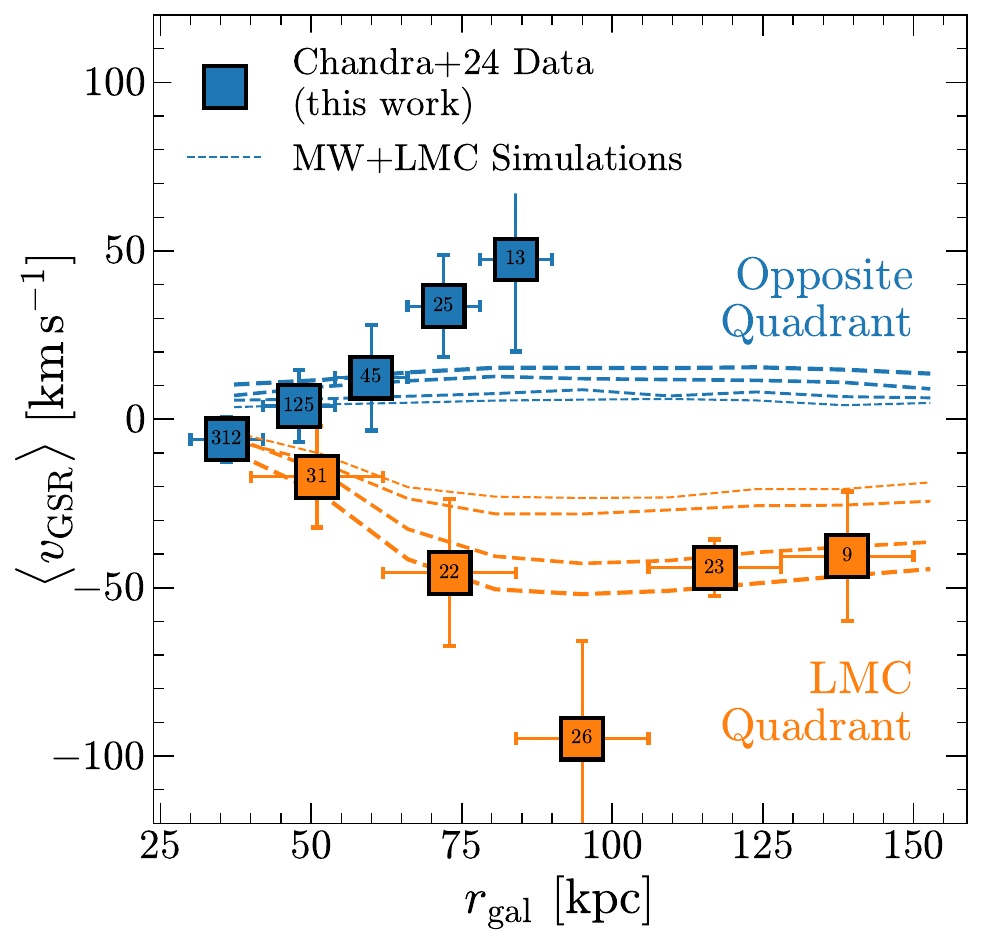}
    \caption{Mean $v_\mathrm{GSR}$ as a function of Galactocentric distance for the quadrants demarcated in the top panel of Figure~\ref{fig:sim_vgsr_map}: the quadrant containing the present-day location of the LMC, and the one diagonally opposing it. 
    1$\sigma$ error bars are obtained by bootstrapping, and the number of stars in each bin is indicated on the markers.
    Dashed lines show the same measurements made with the MW+LMC simulations of \citetalias{Garavito-Camargo2019}, with the line width being proportional to the LMC halo mass in each simulation: from thinnest to thickest corresponding to $[0.8, 1.0, 1.8, 2.5] \times 10^{11}\,M_\odot$. 
    }
    \label{fig:vgsr_rgal}
\end{figure}

Dashed lines in the top panel of Figure~\ref{fig:sim_vgsr_map} show the selection of two diagonally opposing quadrants that contain the regions of maximum mean radial velocity in both the data and simulations. 
The mean radial velocity in these quadrants is shown as a function of Galactocentric distance $r_\mathrm{gal}$ in Figure~\ref{fig:vgsr_rgal}: the quadrant containing the LMC in orange, and the diagonally opposing quadrant in blue. 
In each uncorrelated $r_\mathrm{gal}$ bin, the 3$\sigma$-clipped mean $v_\mathrm{GSR}$ is shown, with $1\sigma$ error bars computed via bootstrapping. 
Note that the measurement in the southern quadrant goes much further than that in the north, due to the unprecedented depth of our MagE survey (see Figure~\ref{fig:sample_map}).
An identical measurement is made on star particles in the MW+LMC simulation of \citetalias{Garavito-Camargo2019}, and from thinnest to thickest the dashed lines corresponds to LMC halo masses of $[0.8, 1.0, 1.8, 2.5] \times 10^{11}\,M_\odot$. 
These stars particles are `tagged on' to the dark matter particles simulated by \citetalias{Garavito-Camargo2019} (see their $\S$3.4), using the formalism of \cite{Laporte2013}.

Figure~\ref{fig:vgsr_rgal} emphasizes what is apparent in the maps of Figure~\ref{fig:sim_vgsr_map}: these quadrants have drastically different radial velocities over a wide range of distances.
Note that the bins uses to compute the average velocities in Figure~\ref{fig:vgsr_rgal} are entirely independent. 
The 80-100~kpc bin in the LMC quadrant exhibits markedly negative radial velocities and warrants future investigation.
This could be due to unresolved substructure at these distances, for example stars stripped from the LMC or SMC \citep[e.g.,][]{Belokurov2016, Chandra2023b}. 
In Appendix~\ref{sec:velmaps_distance}, we split the all-sky $v_\mathrm{GSR}$ maps by distance, showing that the overall behavior of these maps is consistent across distance bins.
Furthermore, we have verified that removing the 26 stars in the discrepant bin of Figure~\ref{fig:vgsr_rgal} does not significantly alter the results from our 6D model fit in $\S$~\ref{sec:vreflex_results}.
The observed behavior matches the MW+LMC simulations shown in Figure~\ref{fig:vgsr_rgal}: the amplitude of the negative radial velocity signal in the southern quadrant reaches $\approx -45$~\kms{} at $\approx 130$~kpc, most consistent with the simulations containing an LMC halo $\gtrsim 15\%$ the mass of the MW's.

\subsection{Tangential Velocities}

Typical tangential velocity uncertainties are more than a factor of ten larger than radial velocity uncertainties for the giants in our dataset. 
Furthermore, any random or systematic uncertainties in the distance estimates propagate to the calculated tangential velocities.   
Therefore, it is challenging to straightforwardly interpret tangential velocity maps as we have done for radial velocities in the preceding section. 
However, the \textit{mean} tangential velocity can be a powerful tracer of the LMC's influence, particularly in the Galactic latitude ($v_\mathrm{t,b}^\ast$) direction \citep[e.g.,][]{Erkal2021, Sheng2024}.
This component of the tangential velocity should capture the apparent `upward' (Galactocentric $z > 0$) reflex motion of halo stars as the inner Galaxy lurches `downward'. 

\begin{figure}
    \centering
    \includegraphics[width=\columnwidth]{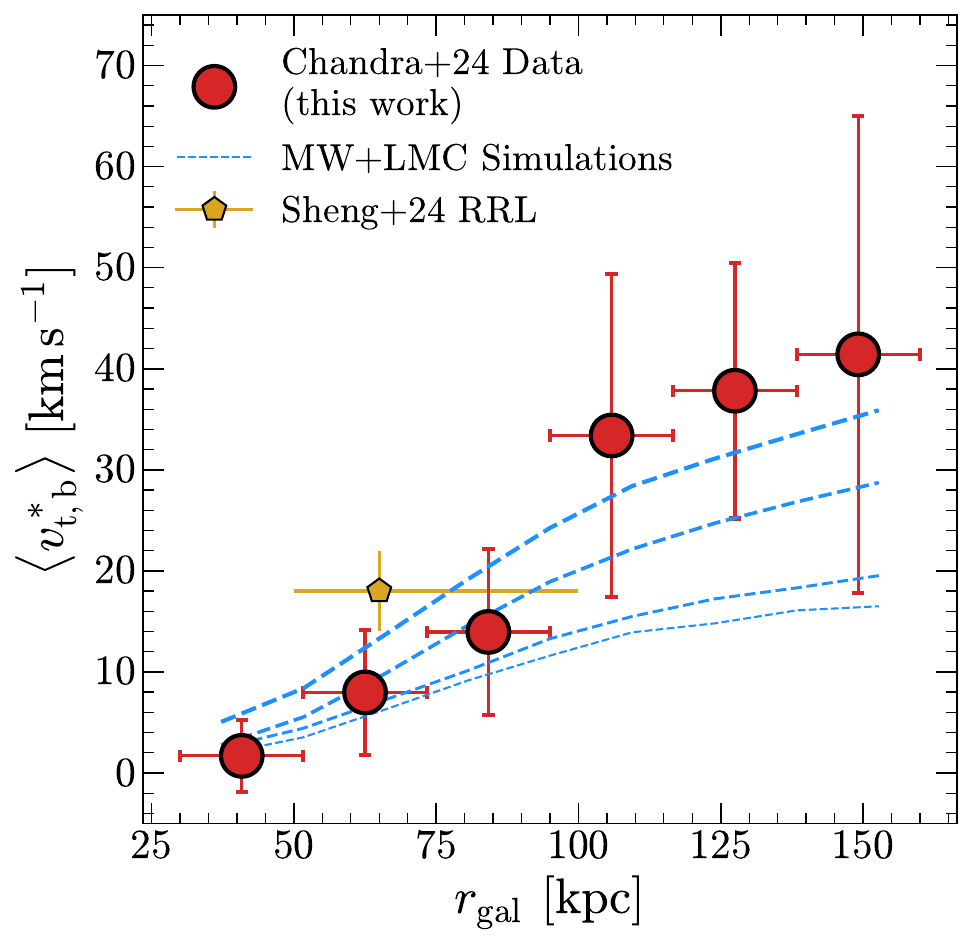}
    \caption{All-sky mean tangential velocity along the Galactic latitude coordinate as a function of Galactocentric distance, with 1$\sigma$ error bars from bootstrapping.
    The recent measurement from \cite{Sheng2024} using RR Lyrae stars is shown in gold.
    Dashed lines show the same measurements made with the MW+LMC simulations of \citetalias{Garavito-Camargo2019}, with the line width being proportional to the LMC halo mass in each simulation: from thinnest to thickest $[0.8, 1.0, 1.8, 2.5] \times 10^{11}\,M_\odot$.}
    \label{fig:vtb_rgal}
\end{figure}

Figure~\ref{fig:vtb_rgal} shows the all-sky mean tangential velocity along the Galactic latitude $\langle v_\mathrm{t,b}^\ast \rangle$ as a function of Galactocentric radius $r_\mathrm{gal}$. 
These tangential velocities have been corrected for the Solar reflex motion. 
There is a striking trend of increasing $\langle v_\mathrm{t,b}^\ast \rangle$ as a function of $r_\mathrm{gal}$, saturating at a maximum of $\langle v_\mathrm{t,b}^\ast \rangle \approx 50$~\kms{} beyond 100~kpc. 
As in Figure~\ref{fig:vgsr_rgal}, the dashed lines show simulation-based predictions from \citetalias{Garavito-Camargo2019}. 
The increasing $\langle v_\mathrm{t,b}^\ast \rangle$ over distance is a tell-tale signature of the LMC's reflex motion in proper motions, and this is the first measurement of the growing tangential velocity to date \citep{Vasiliev2023}.
\cite{Sheng2024} recently measured the mean tangential velocity of RR Lyrae stars from 50-100~kpc, and that measurement is shown for comparison in Figure~\ref{fig:vgsr_rgal}. 
The observed velocities shown in Figure~\ref{fig:vtb_rgal} are consistent with the most massive LMCs in the \citetalias{Garavito-Camargo2019} simulations, with halo masses $> 10^{11}\,M_\odot$. 

% The mean velocity trends presented 
% In subsequent sections, we use the full 6D phase-space information of our dataset to quantitatively measure the reflex motion of the MW disk relative to these outer halo stars. 

\section{Modelling the Global Velocity of the Outer 
Halo}\label{sec:vreflex_model}

The preceding section demonstrated spatial variations in the mean velocities of halo stars --- both on-sky and as a function of distance --- that can be attributed to the LMC-induced reflex motion. 
These trends already point to a relatively massive $\gtrsim 10^{11}\,M_\odot$ LMC (see Figures~\ref{fig:vgsr_rgal} and \ref{fig:vtb_rgal}).
In this section, we develop a physically-motivated model of the inner Galaxy's travel velocity \vtravel{} towards the outer halo. 
% Our model encapsulates the variation in the apparent reflex motion as a function of distance, tomographically 
As Figures~\ref{fig:vgsr_rgal} and \ref{fig:vtb_rgal} demonstrate, the apparent reflex motion of halo stars should sharply grow as a function of distance. 
Therefore, we develop a flexible continuity model that allows the travel velocity \vtravel{} and all associated parameters to linearly vary across the distance range covered by our sample.

\subsection{Global Velocity Model with Reflex Motion}

Our goal is to fit the full 6D kinematics of observed stars to measure the amplitude and direction of the apparent velocity \vtravel{} of the inner galaxy relative to its distant outskirts. 
\citetalias{Petersen2021} developed an elegant formalism that models \vtravel{} as a velocity vector in a rotated Galactocentric reference frame, and fit this model with a Gaussian likelihood for the observed parameters. 
We adapt the basic ingredients of the \citetalias{Petersen2021} model formalism here, developing a likelihood for the observed on-sky position, distance, radial velocity, and proper motion of each star in terms of the model parameters. 

Briefly, at a given $r_\mathrm{gal}$, the stellar halo is assumed to have a mean 3D velocity vector in Galactocentric coordinates
\begin{equation}\label{eqn:reflexmodel}
\langle \mathbf{v} \rangle = \mathbf{v_\mathrm{travel}} + \langle v_r \rangle + \langle v_\phi \rangle + \langle v_\theta \rangle
\end{equation}
The conversion from heliocentric to Galactocentric coordinates is described at the start of $\S$\ref{sec:data}.
The reflex motion is encoded in the $\mathbf{v_\mathrm{travel}}$ parameter, which is expressed by a scalar amplitude $v_\mathrm{travel}$ and two angular coordinates encoding the apex (maximum velocity) direction on the sky.
Note that the travel velocity $v_\mathrm{travel}$ represents the velocity of the inner MW toward the distant halo. 
The `reflex' velocity is the negative of the $v_\mathrm{travel}$ vector, i.e., the apparent motion of distant halo stars towards observers on Earth. 

The direction of \vtravel{} can be encoded by any on-sky angular quantity. 
Although \citetalias{Petersen2021} used the Galactic coordinates $\ell{}$ and $b$ directly, we find that this leads to inefficient convergence of our fitting routine near the pole at $b \approx -90^\circ$, resulting in posterior distributions `pinched' at the southern pole. 
We therefore use the rotated Magellanic Stream coordinate frame from \cite{Nidever2008}. 
The $B_\mathrm{MS} = 0^\circ$ latitude of this frame closely follows the LMC orbit shown in Figure~\ref{fig:sim_vgsr_map}, and the present-day LMC is located at $L_\mathrm{MS} = 0^\circ$. 
We have verified that this choice of frame does not influence the best-fit model parameters in any way, but enables more efficient sampling (away from any poles) and well-behaved posterior distributions in our probabilistic model-fitting routine.

Apart from the amplitude and direction of the reflex motion, the model in Equation~\ref{eqn:reflexmodel} includes mean velocities in each of the spherical Galactocentric components --- $\langle v_\mathrm{r} \rangle$, $\langle v_\mathrm{\phi}\rangle$, $\langle v_\mathrm{\theta} \rangle$.
This allows for any departures in the bulk motion of the halo from the simple \vtravel{} model, for example the halo's overall velocity prior to the arrival of the LMC.

Finally, we include intrinsic dispersion parameters for each Galactic velocity component --- $\sigma_\mathrm{v,r}$,  $\sigma_\mathrm{v,\ell}$,  $\sigma_\mathrm{v,b}$ --- that are added in quadrature to the observed uncertainties in the model likelihood. 
Apart from measuring the intrinsic dispersion of the halo stars, these parameters can also absorb any underestimation in the measurement uncertainties. 

We depart from the \citetalias{Petersen2021} model in that we explicitly expand all model parameters in terms of Galactocentric distance, making a \textit{tomographic} measurement of the reflex motion. 
We adopt a simple linear expansion of each model parameter, such that the number of parameters is increased by a factor of two. 
Each parameter can now have different values at $r_\mathrm{gal} = 40$ and at $r_\mathrm{gal} = 120$, with a linear trend between those two values.
This results in two independent parameter vectors $S_{40}$ and $S_{120}$, each containing 9 parameters: $v_\mathrm{reflex}$, $\ell_\mathrm{MS, apex}$, $b_\mathrm{MS, apex}$,  $\langle v_\mathrm{r} \rangle$, $\langle v_\phi \rangle$, $\langle v_\theta \rangle$, ${\sigma_\mathrm{v,r}}$, ${\sigma_\mathrm{v,\ell}}$, ${\sigma_\mathrm{v,b}}$.
We only fit stars out to $r_\mathrm{gal} = 120$ with this model, reducing the sample size to $821$ stars --- $373$ from H3, $261$ from SEGUE, and $187$ from MagE.
Restricting the fit to $r_\mathrm{gal} < 120$ ensures relatively even coverage across the sky at the upper distance range, since almost all stars beyond this distance are from our southern MagE survey. 
We have tested three other choices of the fitted distance range ($40-100$~kpc, $30-120$~kpc, $30-100$~kpc), and find that they do not alter the main conclusions of this work. 
In the $30-120$~kpc fit, the overall amplitude of the fitted $v_\mathrm{travel}$ is lower at all distances. 
We speculate that this is due to the linearity assumption breaking down at $r_\mathrm{gal} < 40$~kpc, since the $v_\mathrm{travel}$ signature is expected to flatten at these distances (\citetalias{Garavito-Camargo2019}, \citetalias{Yaaqib2024}). 
The $40-120$~kpc sample has the most homogeneous sky coverage across all distances, and is adopted as our fiducial sample for the rest of this work.

To prevent outliers from biasing our model fit, we clip velocity outliers from our sample. 
Specifically, we split the sample into 3 distance bins bounded by the following edges: $[40, 60, 90, 120]$~kpc. 
In each distance bin, we compute the sigma-clipped mean and standard deviation of each velocity component: one radial and two tangential. 
We remove stars that are $\geq 3.5\sigma$ away from the mean in either velocity component. 
This results in 34 stars being removed from the sample, the vast majority of which are within $r_\mathrm{gal} < 60$~kpc. 
These stars could belong to undetected substructures (or known structures like Sagittarius, missed by our angular momentum selection for Sgr members), or could simply have spurious stellar parameters or velocity measurements. 
Our fitted sample of stars from $40-120$~kpc therefore has 772 stars. 

For each star, the velocity model is calculated using the model parameters at that distance --- linearly evaluated using a given parameter set $\{S_{40},S_{120}\}$ --- and the likelihood is computed. 
This continuity model avoids the pitfalls of binning the data in distance. 
The substantial distance uncertainties are bound to place stars in incorrect bins, smearing out the signal by placing (intrinsically more numerous) nearby stars at larger distances, damping the inferred reflex velocity. 
The continuity model utilizes statistical information from the entire sample at once to uncover the distance-dependence of the model parameters, under the assumption that the true parameters vary linearly and continuously over Galactocentric distance. 

Conceptually, the continuity model reduces the variance at any given distance, while potentially increasing the bias --- for example, the parameters at nearby distances are by definition influenced by data at larger distances.
By applying our model to the simulations from \citetalias{Garavito-Camargo2019}, we have verified that the linear continuity model is sufficient to capture the overall trend of the \vtravel{} parameters, as well as the velocity dispersion parameters.
In Appendix~\ref{sec:datamatch} we fit a simulation that matches the size and spatial sampling of our dataset, affirming that our methodology is robust against uneven spatial sampling. 

To compute the likelihood, the mean velocity model (Equation~\ref{eqn:reflexmodel}) is projected into the `observed' heliocentric coordinates, the radial velocity and proper motions.
The conversion from velocity to proper motion uses the heliocentric distance to each star, and is governed by the proportionality constant $k \equiv 4.74$~\kms{}\,kpc$^{-1}$\,(mas\,yr$^{-1}$)$^{-1}$. 
The mean velocity model defines the Gaussian likelihood for our model fit:
\begin{equation}\label{eqn:lik}
    \mathcal{L}\left(\ell, b, d_\mathrm{helio}, v_r, \mu_\ell, \mu_b \mid S_\mathrm{40}, S_\mathrm{120} \right)
\end{equation}
The likelihood utilizes the full proper motion covariance matrix provided by \textit{Gaia}, and propagates our measured distance uncertainties into the tangential velocity uncertainties (see Equations~5-10 in \citetalias{Petersen2021}). 
The uncertainties of the on-sky positions are assumed to be negligible.
For more details on the model formalism and likelihood, we refer to \citetalias{Petersen2021} and \citetalias{Yaaqib2024}.

The \vtravel{} continuity model has 18 free parameters: 9 describing the model parameters at $r_\mathrm{gal} = 40$~kpc, and 9 at $r_\mathrm{gal} = 120$~kpc. 
We adopt uninformative priors on all parameters. 
For the $B_\mathrm{MS,apex}$ latitude, a uniform prior on $\cos{[B_\mathrm{MS,apex}]}$ is used, and for the intrinsic dispersion parameters a uniform prior on the logarithm of the dispersion is used. 
An initial maximum-likelihood fit is run using \texttt{scipy.optimize} \citep{Virtanen2020}.
The posterior distribution of model parameters is sampled with the nested sampling routine \texttt{dynesty} \citep{Speagle2020}. 
The full corner plot of posterior distributions is shown in Appendix~\ref{sec:corner}, and the best-fit parameters are summarized in Table~\ref{tab:fit_summary}. 
The best-fit parameters from the linear continuity model are also evaluated at a reference distance of $r_\mathrm{gal} = 100$~kpc, and listed in Table~\ref{tab:fit_summary_dref}.
\begin{deluxetable}{ccc}
\label{tab:fit_summary}
\tablecaption{Summary of the best-fit parameters from our \vtravel{} continuity model. Each parameter has a free value at $r_\mathrm{gal} = 40$~kpc and at $r_\mathrm{gal} = 120$~kpc, and is assumed to vary linearly between those values as a function of distance.}
\tablehead{\colhead{Parameter} & \colhead{$r_\mathrm{gal} = 40$~kpc} & \colhead{$r_\mathrm{gal} = 120$~kpc}}
\tablewidth{\columnwidth}
\startdata
$v_\mathrm{travel}$ [km$\,$s$^{-1}$] & $6_{-4}^{+5}$ & $47_{-12}^{+12}$ \\
$\ell_\mathrm{MS, apex}$ [deg] & $-77_{-22}^{+21}$ & $-10_{-15}^{+17}$ \\
$b_\mathrm{MS, apex}$ [deg] & $12_{-19}^{+20}$ & $3_{-14}^{+14}$ \\
$\langle v_\mathrm{r} \rangle$ [km$\,$s$^{-1}$] & $-8_{-5}^{+5}$ & $-25_{-10}^{+9}$ \\
$\langle v_\mathrm{\phi} \rangle$ [km$\,$s$^{-1}$] & $14_{-5}^{+5}$ & $-14_{-8}^{+8}$ \\
$\langle v_\mathrm{\theta} \rangle$ [km$\,$s$^{-1}$] & $-1_{-5}^{+5}$ & $-3_{-12}^{+12}$ \\
${\sigma_\mathrm{v,r}}$ [km$\,$s$^{-1}$] & $106_{-4}^{+4}$ & $69_{-5}^{+6}$ \\
${\sigma_\mathrm{v,\ell}}$ [km$\,$s$^{-1}$] & $78_{-4}^{+4}$ & $20_{-4}^{+5}$ \\
${\sigma_\mathrm{v,b}}$ [km$\,$s$^{-1}$] & $56_{-3}^{+3}$ & $33_{-5}^{+6}$ \\
\enddata
\tablecomments{The marginalized posterior distribution of each parameter is summarized using the 16th, 50th, and 84th quantiles, corresponding to $1\sigma$ uncertainties for a Gaussian distribution.}
\end{deluxetable} %

To ensure that our model implementation is consistent with previous measurements from \citetalias{Petersen2021}, we fit their dataset using our model in Appendix~\ref{sec:pp21fit}, reproducing their results.
As an additional check to verify our results, we fit all stars in our dataset with a distance-independent 9-parameter \vtravel{} model, directly analogous to that used by \citetalias{Petersen2021} and \citetalias{Yaaqib2024}. 
These results are presented in Appendix~\ref{sec:vreflex_all}, and are broadly consistent with the 18-parameter continuity model results presented in the rest of this section. 

\begin{deluxetable}{ccc}
\label{tab:fit_summary_dref}
\tablecaption{Summary of the best-fit parameters from our \vtravel{} continuity model (see Table~\ref{tab:fit_summary}), linearly evaluated at a reference distance of $100$~kpc. For convenience and ease of comparison, we also provide the apex direction transformed to Galactic coordinates, and the \vtravel{} velocity in Galactocentric coordinates (see Figure~\ref{fig:vtravel_galcen}).}
\tablehead{\colhead{\hspace{1cm}Parameter}\hspace{1cm} & \colhead{\hspace{1cm}$r_\mathrm{gal} = 100$~kpc}\hspace{1.5cm}}
\tablewidth{\columnwidth}
\startdata
$v_\mathrm{travel}$ [km$\,$s$^{-1}$] & $37_{-9}^{+9}$ \\
$\ell_\mathrm{MS, apex}$ [deg] & $-27_{-10}^{+11}$ \\
$b_\mathrm{MS, apex}$ [deg] & $6_{-9}^{+9}$ \\
$\langle v_\mathrm{r} \rangle$ [km$\,$s$^{-1}$] & $-21_{-7}^{+6}$ \\
$\langle v_\mathrm{\phi} \rangle$ [km$\,$s$^{-1}$] & $-7_{-6}^{+6}$ \\
$\langle v_\mathrm{\theta} \rangle$ [km$\,$s$^{-1}$] & $-2_{-9}^{+9}$ \\
${\sigma_\mathrm{v,r}}$ [km$\,$s$^{-1}$] & $77_{-4}^{+5}$ \\
${\sigma_\mathrm{v,\ell}}$ [km$\,$s$^{-1}$] & $28_{-4}^{+5}$ \\
${\sigma_\mathrm{v,b}}$ [km$\,$s$^{-1}$] & $38_{-4}^{+4}$ \\
\hline
$\ell_\mathrm{apex}$ [deg] & $-80_{-18}^{+18}$ \\
$b_\mathrm{apex}$ [deg] & $-59_{-9}^{+11}$ \\
\hline
$v_\mathrm{travel, x}$ [km$\,$s$^{-1}$] & $3_{-5}^{+5}$ \\
$v_\mathrm{travel, y}$ [km$\,$s$^{-1}$] & $-17_{-5}^{+6}$ \\
$v_\mathrm{travel, z}$ [km$\,$s$^{-1}$] & $-31_{-9}^{+9}$ \\
\enddata
\tablecomments{The marginalized posterior distribution of each parameter is summarized using the 16th, 50th, and 84th quantiles, corresponding to $1\sigma$ uncertainties for a Gaussian distribution.}
\end{deluxetable} %

\subsection{Reflex Response Results}\label{sec:vreflex_results}

\begin{figure*}
    \centering
    \includegraphics[height=6cm]{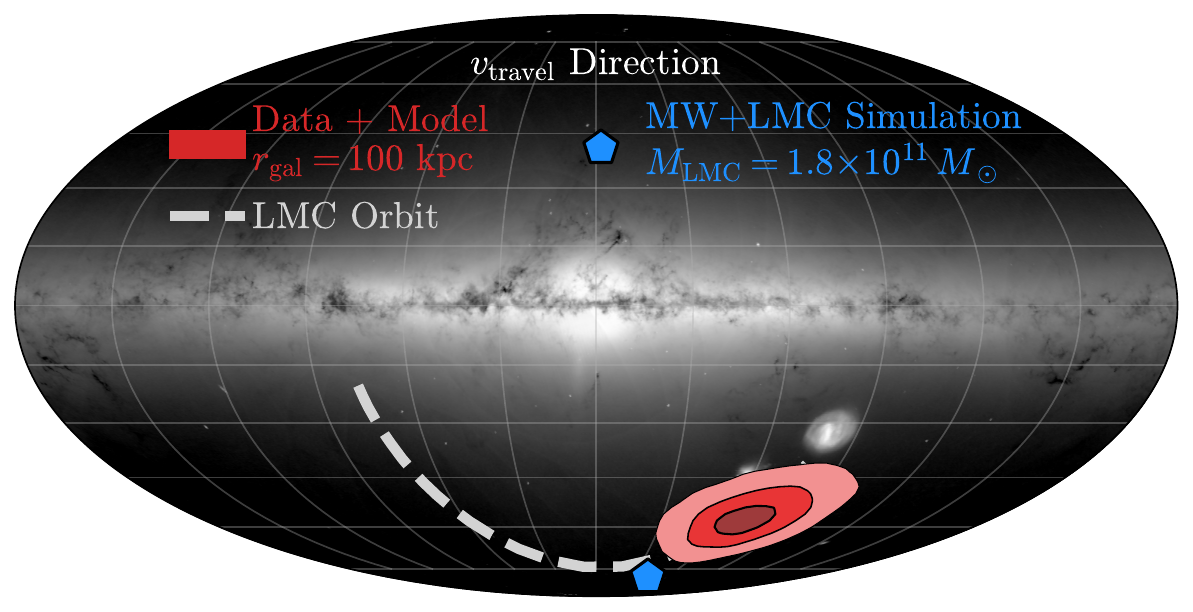}
    \includegraphics[height=7.5cm]{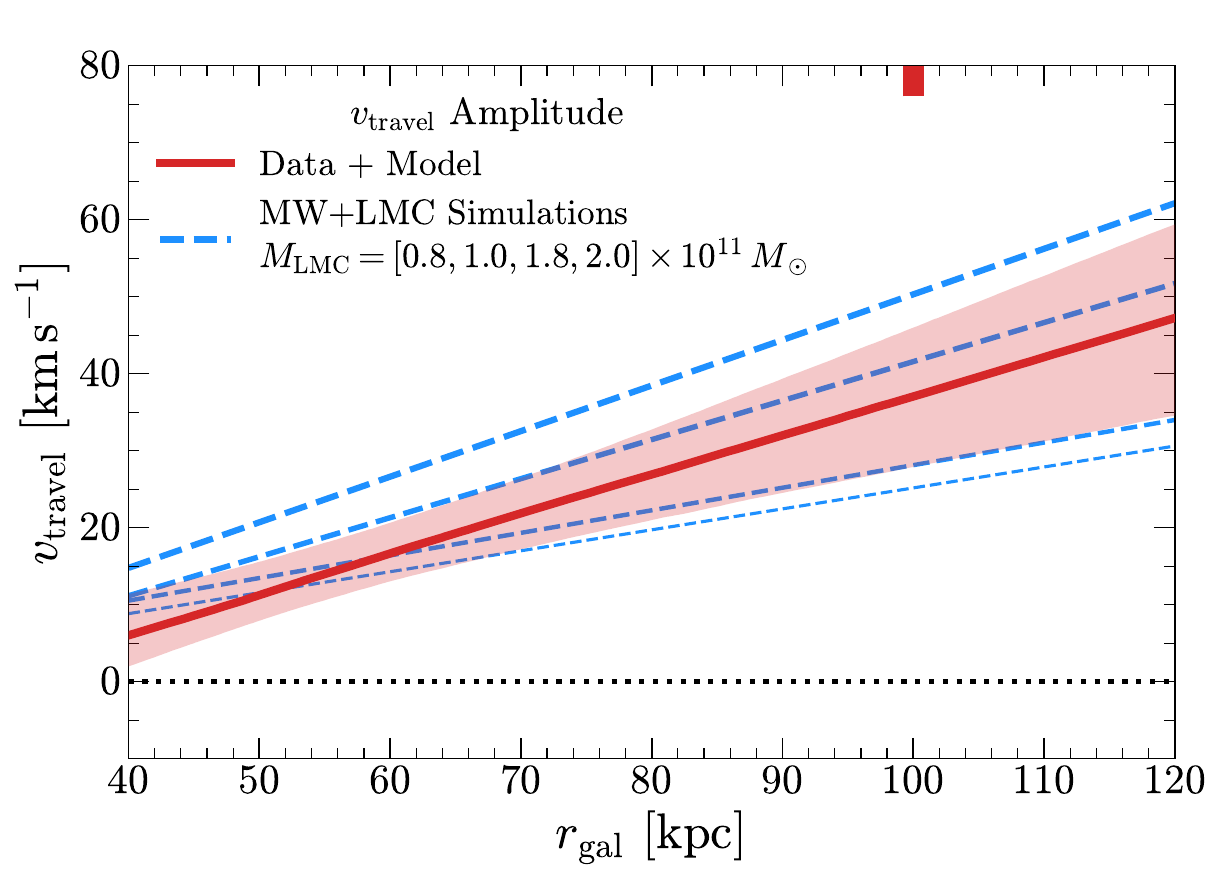}
    \caption{The travel velocity of the Milky Way relative to the outer halo, as measured by our $v_\mathrm{travel}$ continuity model as a function of Galactocentric distance $r_\mathrm{gal}$. 
    The MW+LMC simulations from \citetalias{Garavito-Camargo2019} are shown in blue, with linewidths in the bottom panel corresponding to LMC halo masses from $[0.8, 1.0, 1.8, 2.5] \times 10^{11}\,M_\odot$.
    \textbf{Top:} The apex direction of the apparent reflex motion in Galactic coordinates, evaluated from our best-fit continuity model at a reference distance of $r_\mathrm{gal} = 100$~kpc.
    The LMC's past orbit is overlaid in grey, and the background image shows the density of stars in \textit{Gaia} DR3. 
    Contours are placed at 40th, 70th, and 90th quantiles, corresponding to Gaussian-equivalent levels of 0.5$\sigma$, 1$\sigma$, and 1.5$\sigma$.
    The blue marker shows the apex direction inferred from the $1.8 \times 10^{11}\,M_\odot$ simulation of \citetalias{Garavito-Camargo2019}.
    \textbf{Bottom:} Amplitude of the travel velocity as a function of Galactocentric distance. The mean of posterior draws from our continuity model is shown, along with the $1\sigma$ credible interval.
    The reference distance used in the top panel is indicated with a tick mark along the top axis of the bottom panel.} 
    \label{fig:vreflex_results}
\end{figure*}

Figure~\ref{fig:vreflex_results} summarizes the key results from the continuity model measurement of \vtravel{}. 
The top panel illustrates the posterior distribution of the \vtravel{} direction in Galactic coordinates, evaluated at the reference distance of $r_\mathrm{gal} = 100$~kpc. 
As a reminder, the continuity model evaluated at this reference distance utilizes information from stars spanning $r_\mathrm{gal} = 40-120$~kpc.
Our measured travel velocity relative to the Galactic outskirts points directly at a past location along the LMC's orbit, close to the present-day location of the LMC. 

The bottom panel of Figure~\ref{fig:vreflex_results} shows the posterior distribution of the measured amplitude of \vtravel{} as a function of Galactocentric distance $r_\mathrm{gal}$, with the $1\sigma$ credible interval shaded. 
The \vtravel{} amplitude significantly grows as a function of distance, from $\approx 6$~\kms{} at $r_\mathrm{gal} = 40$~kpc to a maximum of $v_\mathrm{travel} = 47 \pm 12$ at $r_\mathrm{gal} = 120$~kpc. 
Along with the recent measurements of \citetalias{Yaaqib2024}, this is the first measurement of the growing reflex velocity as a function of distance.
Whereas \citetalias{Yaaqib2024} have a single bin containing all their stars beyond 50~kpc, our larger dataset and continuity model allow us to model the $v_\mathrm{travel}$ variation out to $120$~kpc. 

The MW+LMC simulations from \citetalias{Garavito-Camargo2019} are also shown in the bottom panel of Figure~\ref{fig:vreflex_results}. 
Since we only require the maximum-likelihood parameters (and not their uncertainties) from the simulations, we fit the simulations with \texttt{scipy.optimize} \citep{Virtanen2020} instead of \texttt{dynesty}. 
We have verified that measuring \vtravel{} directly from the simulation particles --- by averaging particle velocities in spherical shells, as done by \citetalias{Yaaqib2024} --- produces a similar slope, and the linear model is sufficient to capture the overall growth of \vtravel{}. 
However, fitting the simulation with the full model is important to disentangle the mean halo velocity terms from the $v_\mathrm{travel}$ amplitude. 
The bottom panel of Figure~\ref{fig:model_comparison} demonstrates that the amplitude of the travel velocity \vtravel{} is most consistent with simulations of a $\gtrsim 1.8 \times 10^{11}\,M_\odot$ LMC from \citetalias{Garavito-Camargo2019}, or an LMC that is $15\%$ the mass of the Milky Way. 

\begin{figure}
    \centering
    \includegraphics[width=\columnwidth]{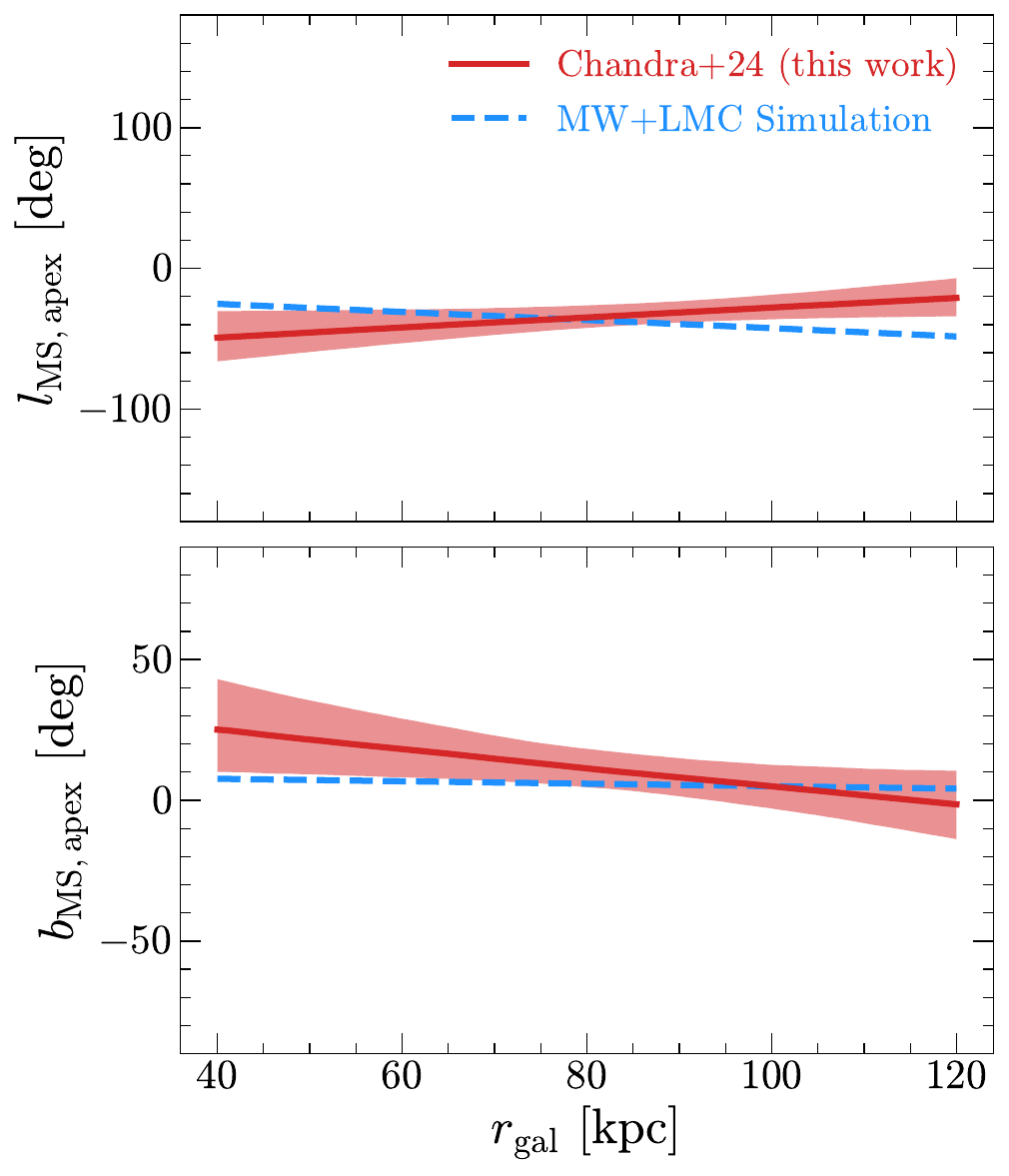}
    \caption{The apex direction of \vtravel{} measured in this work (red lines) as a function of Galactocentric radius $r_\mathrm{gal}$ with $1\sigma$ uncertainties shaded.
    The dashed blue line shows the same quantities from the $1.8 \times 10^{11}\,M_\odot$ \texttt{LMC3} simulation of \citetalias{Garavito-Camargo2019}.}
    \label{fig:apex_rgal}
\end{figure}

Figure~\ref{fig:apex_rgal} shows the apex direction of \vtravel{} as a function of Galactocentric radius. 
The $1.8 \times 10^{11}\,M_\odot$ LMC simulation from \citetalias{Garavito-Camargo2019} is shown for comparison. 
In both the data and simulations, there is little evolution in the apex direction as a function of distance.

\begin{figure}
    \centering
    \includegraphics[width=\columnwidth]{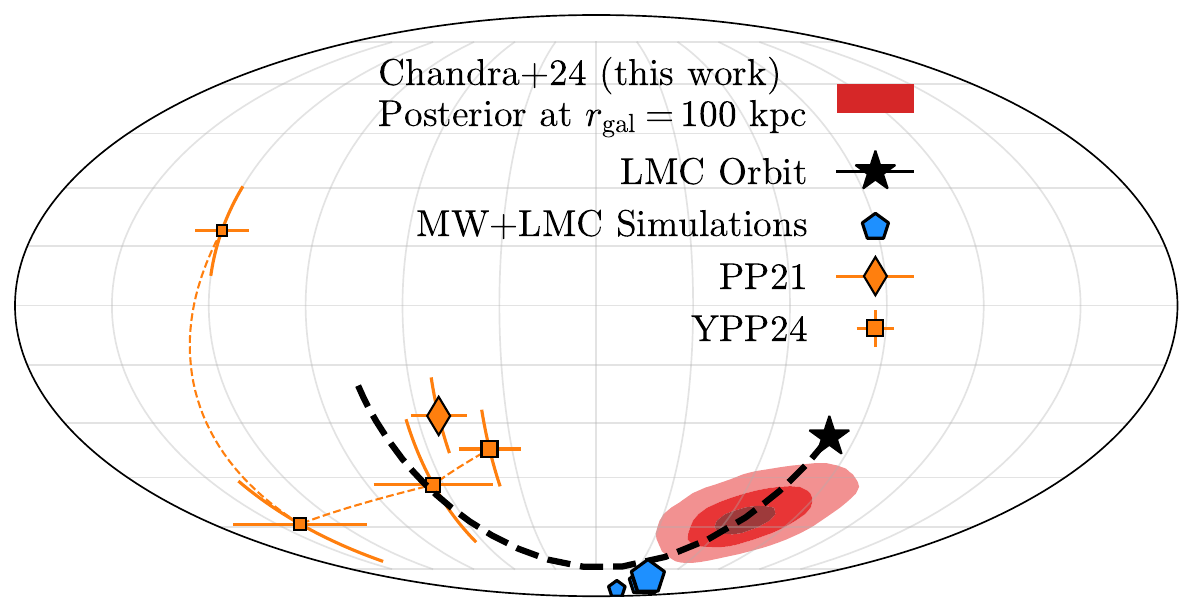}
    \includegraphics[width=\columnwidth]{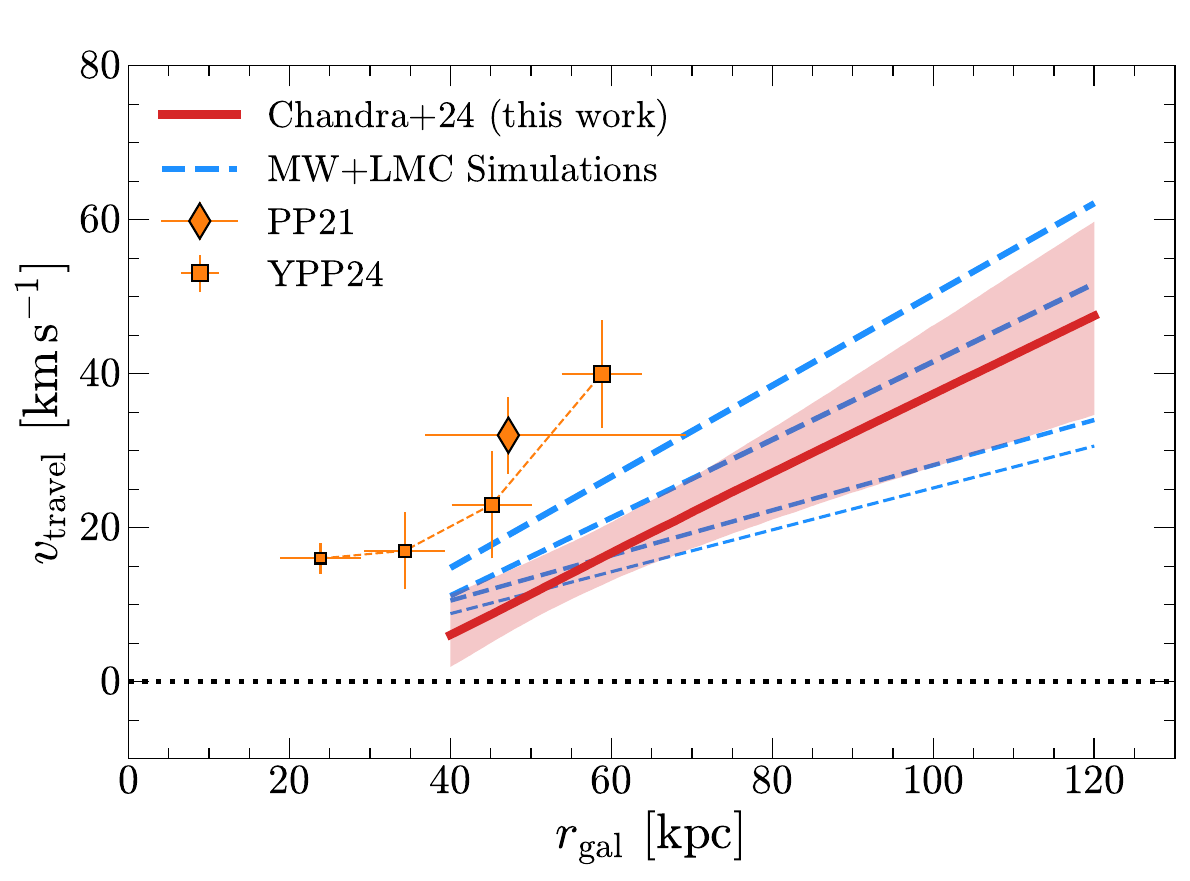}
    \caption{Contextualizing the results of our $v_\mathrm{travel}$ model with N-body simulations and past measurements. 
    The MW+LMC simulations from \citetalias{Garavito-Camargo2019} are shown in blue, with increasing marker sizes and linewidths  corresponding to LMC halo masses from $[0.8, 1.0, 1.8, 2.5] \times 10^{11}\,M_\odot$.
    Previous measurements from \citetalias{Petersen2021} and \citetalias{Yaaqib2024} are shown for comparison. 
    For the \citetalias{Yaaqib2024} measurements the marker size is proportional to distance bin they fitted, ranging from 25-60~kpc.
    }
    \label{fig:model_comparison}
\end{figure}

For the remainder of this section, we focus on the \vtravel{} continuity model evaluated at a reference distance of $r_\mathrm{gal} = 100$~kpc.
This distance provides a balance between being well-sampled by our dataset while having a confidently-detected reflex motion amplitude $\approx 40$~\kms{}. Figure~\ref{fig:model_comparison} contextualizes our measurement by comparing it to past measurements of the reflex velocity from \citetalias{Petersen2021} and \citetalias{Yaaqib2024}. 
Predictions from the self-consistent MW+LMC simulations of \citetalias{Garavito-Camargo2019} are also shown for the four LMC masses considered in that work. 

The top panel of Figure~\ref{fig:model_comparison} shows that the \citetalias{Garavito-Camargo2019} simulations predict a reflex velocity pointing towards a relatively recent position of the LMC's orbit, consistent with our measurement. 
On the other hand, the measurements of \citetalias{Petersen2021} and \citetalias{Yaaqib2024} point towards the opposing Galactic quadrant. 
The difference between those measurements and the \citetalias{Garavito-Camargo2019} simulations is acknowledged in those works (see also \citealt{Vasiliev2023}). 
The measurements in \citetalias{Yaaqib2024} get closer to the present-day location of the LMC at larger distances, and their appendices explore how different selection effects can non-trivially affect the inferred apex direction. 
Furthermore, the predicted \vtravel{} amplitude is quite small ($\lesssim 20$~\kms{}) for the distances covered by the majority of the \citetalias{Petersen2021} and \citetalias{Yaaqib2024} samples, making it more challenging to disentangle from other bulk motions in the halo than at larger distances.

\begin{figure}
    \centering
    \includegraphics[width=\columnwidth]{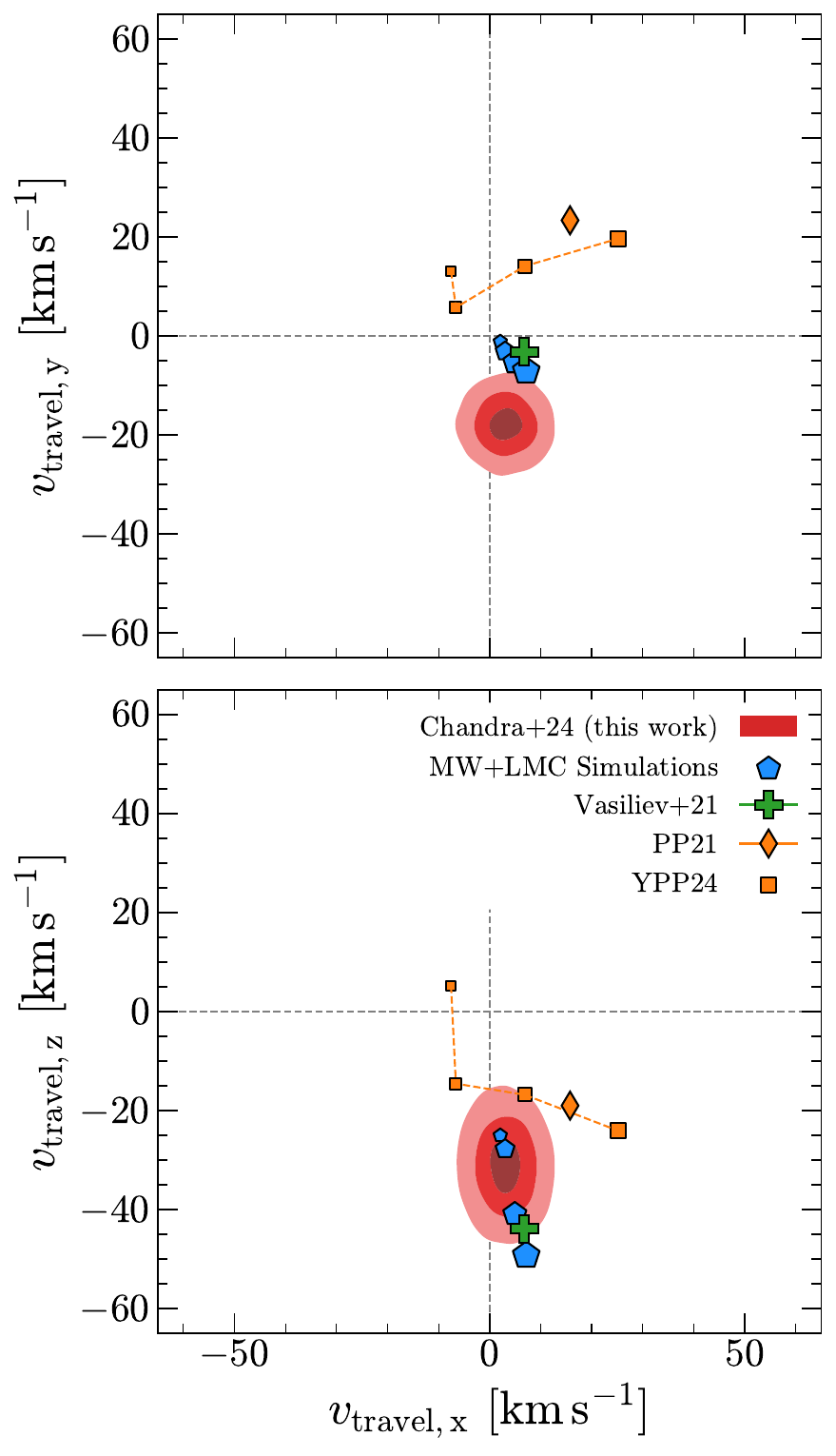}
    \caption{The mean velocity \vtravel{} of the inner MW relative to the Galactic outskirts, expressed in Galactocentric coordinates. The posterior distribution of our \vtravel{} continuity model is evaluated here at a reference distance of $100$~kpc.
    For comparison, we show the MW+LMC simulations of \citetalias{Garavito-Camargo2019} evaluated at the same distance, with increasing marker sizes corresponding to LMC halo masses from $[0.8, 1.0, 1.8, 2.5] \times 10^{11}\,M_\odot$.
    Previous measurements from \citet[][also evaluated at $r_\mathrm{gal} = 100$~kpc]{Vasiliev2021a}, \citetalias{Petersen2021}, and \citetalias{Yaaqib2024} are also shown.
    }
    \label{fig:vtravel_galcen}
\end{figure}

\begin{figure*}
    \centering
    \includegraphics[width=0.8\textwidth]{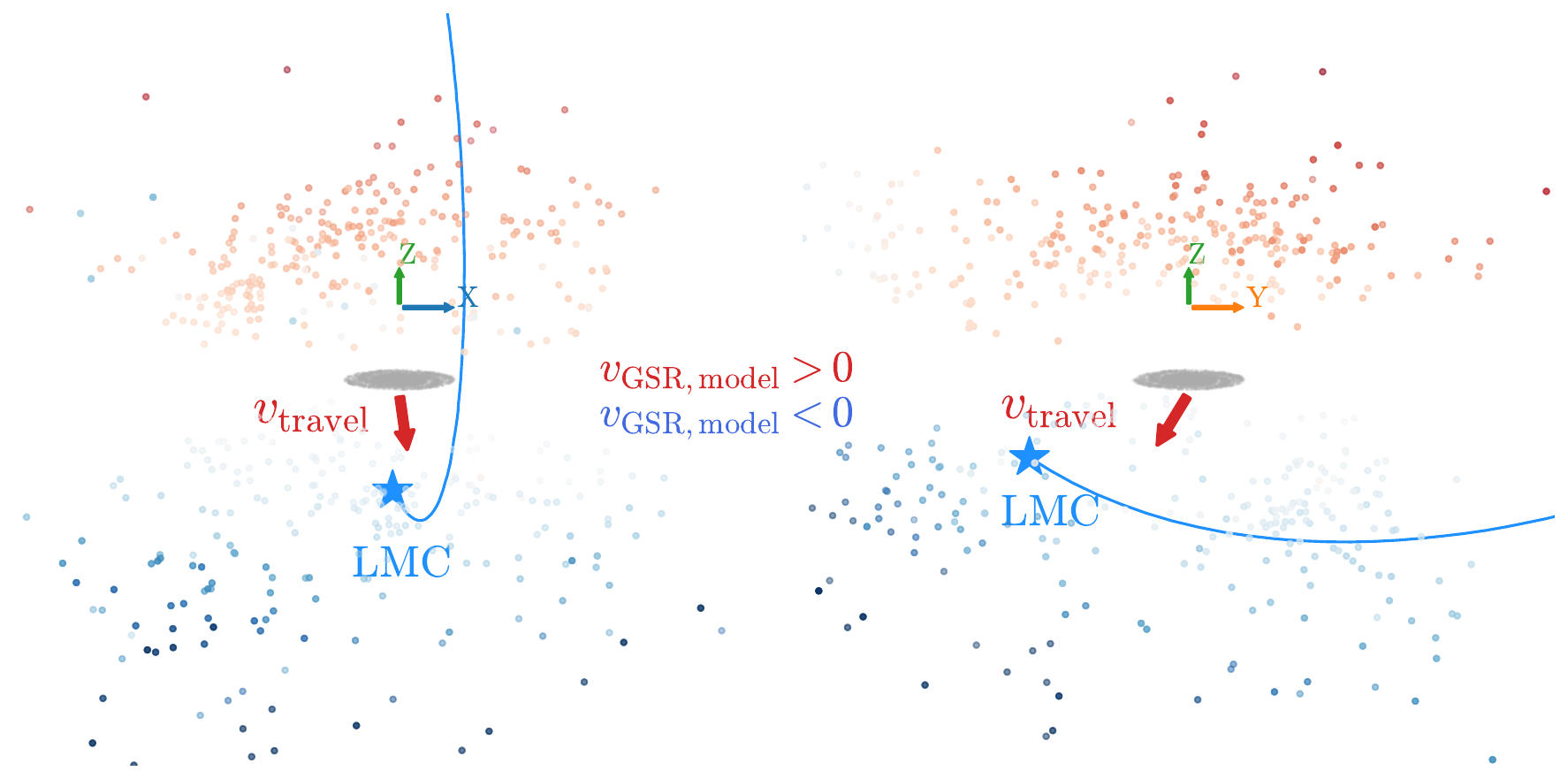}
    \caption{Visualization of our best-fit reflex velocity model in 3D Galactocentric coordinates. 
    A mock MW disk is shown, along with the measured 3D positions of stars in our dataset.
    Stars are colored by their corresponding $v_\mathrm{GSR}$ velocity evaluated from the best-fit $v_\mathrm{travel}$ continuity model. 
    The red arrow shows the measured $v_\mathrm{travel}$ vector of the inner galaxy relative to the halo at 100~kpc, arbitrarily scaled up to 500~Myr of motion.
    The past orbit of the LMC from \citetalias{Garavito-Camargo2019} is shown in blue. 
    See \href{https://bit.ly/mw-vtravel}{here} for an animated, rotating version of this figure.}
    \label{fig:vreflex_cartoon}
\end{figure*}

Figure~\ref{fig:vtravel_galcen} offers an alternative perspective of our measurements by showing the Galactocentric components of the $v_\mathrm{travel}$ velocity. 
As a reminder, \vtravel{} is the inferred velocity of the inner galaxy relative to the outer halo, which sources the apparent reflex velocity of those outer halo stars as measured by us. 
% Our best-fit \vtravel{} continuity model is evaluated at $r_\mathrm{gal} = 100$~kpc.
The corresponding \vtravel{} values for each MW+LMC simulation from \cite{Garavito-Camargo2019} are also evaluated at the reference distance of $r_\mathrm{gal} = 100$~kpc, along with past measurements of \vtravel{} from \cite{Vasiliev2021a}, \citetalias{Petersen2021}, and \citetalias{Yaaqib2024}. 
Note that \cite{Vasiliev2021a} measure \vtravel{} by fitting the orbit of the Sagittarius stream, whereas \citetalias{Petersen2021} and \citetalias{Yaaqib2024} directly fit the kinematics of halo stars. The \cite{Vasiliev2021a} measurement in Figure~\ref{fig:vtravel_galcen} was provided by E. Vasiliev (private communication).

Figure~\ref{fig:vtravel_galcen} demonstrates that although all measurements and simulations broadly agree regarding the $v_\mathrm{Z}$ velocity (unsurprisingly, since the LMC remains in the southern hemisphere throughout its recent orbit), there is a difference in the $v_\mathrm{Y}$ velocity. 
Specifically, we measure $v_\mathrm{Y} \lesssim 0$ in this work (as do \citealt{Vasiliev2021a}), whereas \citetalias{Petersen2021} and \citetalias{Yaaqib2024} measure $v_\mathrm{Y} > 0$. 
This difference is modest relative to the uncertainties in both the past and present measurements. 
In Appendix~\ref{sec:pp21fit}, we investigate this discrepancy in detail, finding that the measurements in \citetalias{Petersen2021} and \citetalias{Yaaqib2024} were likely influenced by the presence of Sagittarius Stream stars in their datasets (see also \citealt{Bystrom2024}). 
We conclude that the difference between our measurement and past measurements from \citetalias{Petersen2021} and \citetalias{Yaaqib2024} is primarily due to the different selection cuts used to remove Sagittarius stars from the datasets.

Figure~\ref{fig:vreflex_cartoon} provides a 3D illustration of the reflex motion measured in this work. 
A model of the MW disk is shown in the X-Z (left) and Y-Z (right) Galactocentric frame, with halo stars from our dataset shown at their true 3D locations, and the 3D orbit of the LMC reproduced from \citetalias{Garavito-Camargo2019}.
A red arrow denotes our measured \vtravel{} vector evaluated at $100$~kpc, arbitrarily scaled to reflect 500~Myr of motion. 
The stars in our dataset are plotted at their measured positions, and colored by their corresponding Galactocentric radial velocity in the best-fit \vtravel{} model. 
This illustration conveys the main results of this work: the inner MW exhibits a significant net motion towards a the recent orbit of the LMC --- relative to distant stars in the outer halo --- and the amplitude of this signal grows stronger as a function of distance. 

\subsection{The Global Velocities of the Outer Halo}\label{sec:bulkvel}

\begin{figure}
    \centering
    \includegraphics[width=\columnwidth]{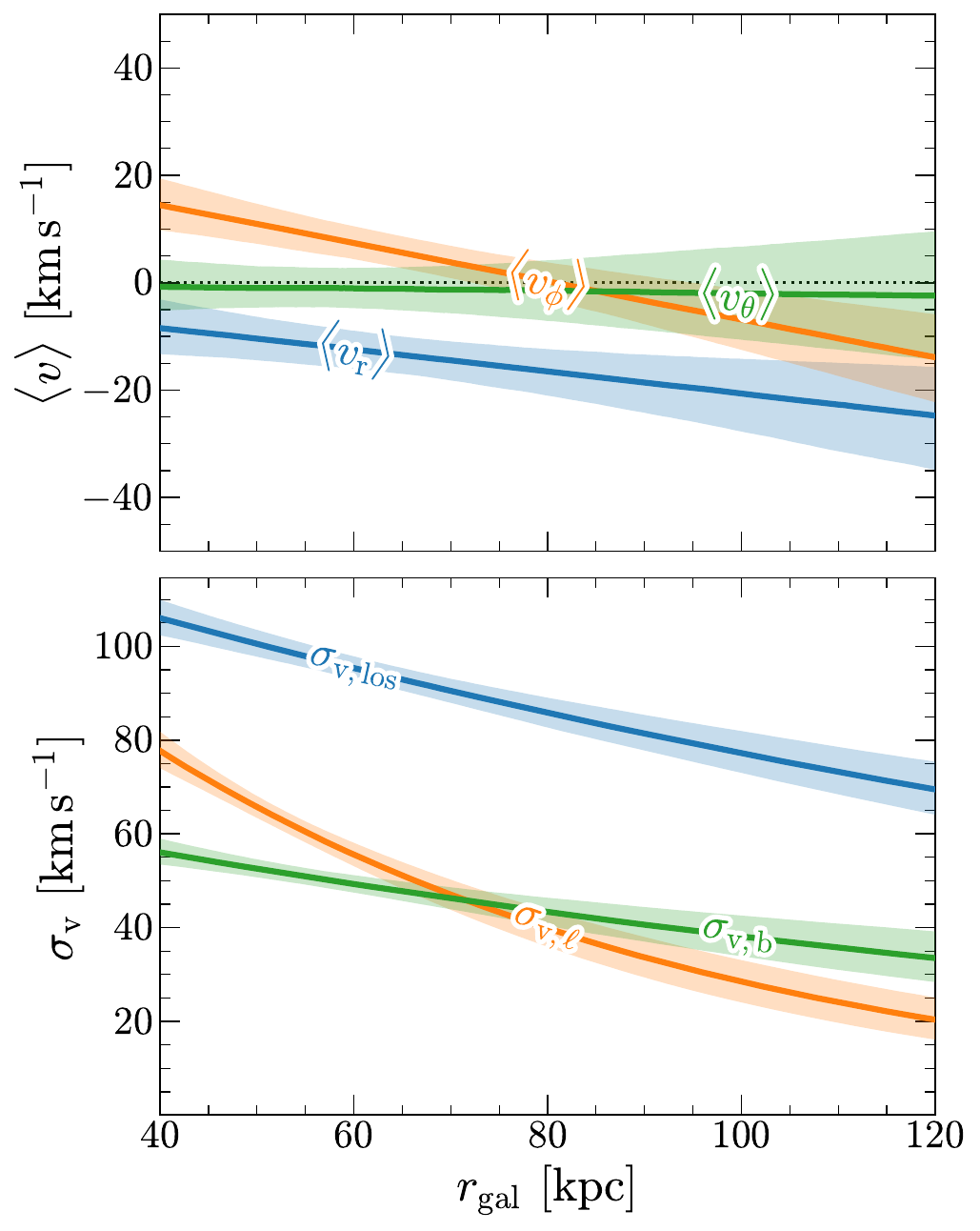}
    \caption{Spherical velocity components of the MW halo, fit to our all-sky dataset simultaneously with the reflex velocity model. 
    The top panel shows mean velocities in spherical Galactocentric coordinates, and the bottom panel shows intrinsic dispersions in Galactic velocities. 
    $1\sigma$ credible intervals are shaded. 
    }
    \label{fig:halo_vel}
\end{figure}

In addition to the reflex velocity of outer halo stars, our \vtravel{} model also measures the bulk motion (and intrinsic dispersion) of the outer halo as a function of distance. 
Figure~\ref{fig:halo_vel} shows the best-fit bulk halo velocity parameters from our model, the mean velocity components of the stellar halo (top panel) and their corresponding intrinsic dispersions (bottom panel). 
These measurements represent the sky-averaged motion of the MW halo, once the reflex response to the LMC has been accounted for.

The top panel of Figure~\ref{fig:halo_vel} shows significant trends in the average radial and azimuthal velocity of the outer halo. The halo exhibits a net `compression' signal ($\langle v_\mathrm{r} \rangle < 0$) in radial velocity at 40~kpc (e.g., \citetalias{Yaaqib2024}), growing to $\approx -20$~\kms{} at $\approx 120$~kpc. 
Simulations of the LMC infall predict a similar dynamical compression --- at the level of $-20$~\kms{} at these distances \citepalias{Yaaqib2024} --- suggesting that the LMC's gravitational pull may be responsible for this signature. 

Another interesting signature in the top panel of Figure~\ref{fig:halo_vel} is the non-zero azimuthal rotation measured by the $\langle v_\phi \rangle$ component. 
Whereas \citetalias{Petersen2021} and \citetalias{Yaaqib2024} measured a net negative $\langle v_\phi \rangle$ at all distances, our best-fit model has net positive $\langle v_\phi \rangle$ out to $\approx 80$~kpc, after which it switches to being negative.
Our relative trend matches \citetalias{Yaaqib2024}, with $\langle v_\phi \rangle$ becoming more negative as a function of distance. 
In our Galactocentric convention, this corresponds to a net retrograde halo that switches to prograde in the outskirts, although the uncertainties are large (see \citealt{Deason2017}). 

The bottom panel of Figure~\ref{fig:halo_vel} shows the radial trends of velocity dispersion fitted alongside our \vtravel{} model. 
Note that the continuity model assumed a linear trend of the log-dispersion as a function of $r_\mathrm{gal}$, producing the apparent curvature of the lines in the bottom panel of Figure~\ref{fig:halo_vel}. 
We emphasize that these are the intrinsic dispersions from our \vtravel{} model, and are therefore inherently de-convolved from the measurement uncertainties in radial velocities and proper motions. 
All components of the halo velocity dispersion fall as a function of distance, as has been seen in prior studies mapping the outer halo \citep[e.g.,][]{Bird2021}. 
It is apparent that the radial dispersion is significantly higher than the tangential dispersions across all Galactocentric radii. 
We have verified that directly calculating the velocity dispersion in radial bins --- and subtracting the median measurement error in each bin, in quadrature --- produces quantitatively similar profiles.
We do note that it is challenging to measure the intrinsic tangential dispersions at large distances, since they are sensitive to the adopted proper motion and distance measurement uncertainties.

\begin{figure}
    \centering
    \includegraphics[width=\columnwidth]{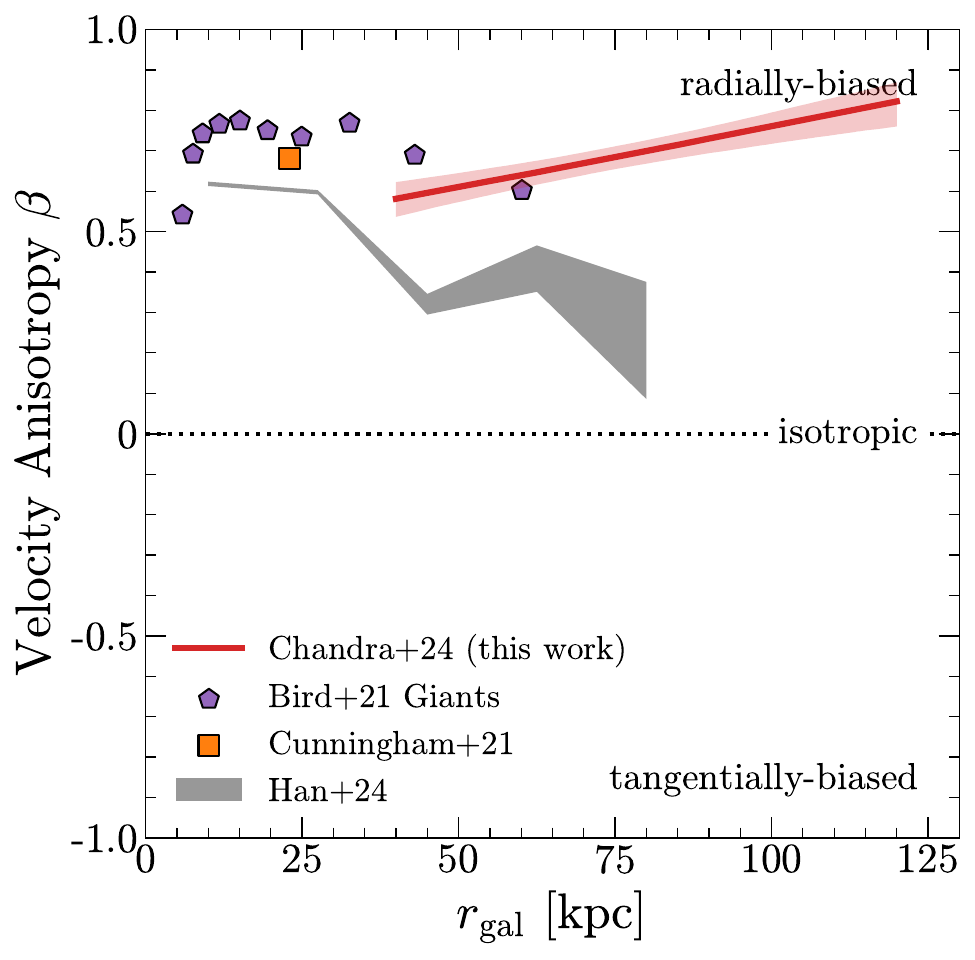}
    \caption{
    Velocity anisotropy parameter $\beta$ as a function of Galactocentric distance, as derived from the velocities shown in Figure~\ref{fig:halo_vel}. 
    Previous measurements are shown for comparison: \cite{Cunningham2019halo7d} from the HALO7D survey, \cite{Bird2021} using SDSS/LAMOST K giants, and \cite{Han2024} using the H3 Survey. 
    Our sample constrains $\beta$ twice as far as previous measurements, suggesting a radially anisotropic halo out to $\gtrsim 100$~kpc. 
    }
    \label{fig:beta}
\end{figure}

The radial bias (or lack thereof) of the stellar velocity ellipsoid is compactly summarized in the velocity anisotropy parameter \citep{Binney2008}: $$\beta \equiv 1 - \frac{\langle v_\phi^2 \rangle + \langle v_\theta^2 \rangle}{2 \langle v_r^2 \rangle}$$
This is a crucial parameter for dynamical mass estimates of the MW that use the Jeans formalism \citep[e.g.][]{Binney2008, Wang2020}.
We compute $\beta$ from $40-120$~kpc using the bulk halo velocities (and their intrinsic dispersions) shown in Figure~\ref{fig:halo_vel}.
For example, $\langle v_r^2 \rangle = \langle v_r \rangle^2 + \sigma_\mathrm{v,r}^2$.
Since our intrinsic dispersions are computed in the observed rather than Galactocentric frame, we make the simplifying assumption that $\sigma_\mathrm{v, r} \approx \sigma_\mathrm{v, los}$, 
$\sigma_\mathrm{v, \phi} \approx \sigma_\mathrm{v, \ell}$, and 
$\sigma_\mathrm{v, \theta} \approx \sigma_\mathrm{v, b}$ 
which is suitable for the distant stars in our sample. 
We have verified by fitting the simulations of \cite{Garavito-Camargo2019} that we can recover their input $\beta(r_\mathrm{gal})$ profile using our fitting methodology.
Our measured Galactocentric profile of $\beta$ is shown in Figure~\ref{fig:beta}, with some corresponding measurements from the literature displayed for comparison.

Simulations of hierarchical galaxy formation predict that at large distances, the stellar halo should have radially-biased orbits with $\beta > 0$ \citep[e.g.,][]{Diemand2005, Abadi2006, Sales2007, Debattista2008, Amorisco2017, Loebman2018}. 
We measure a significant radial bias in the motions of stars out to $r_\mathrm{gal} \approx 120$~kpc, with $\beta$ ranging from $\approx 0.6 - 0.8$.
These are the most distant all-sky measurements of the halo's velocity ellipsoid to date, supporting the idea that bulk of the MW halo was hierarchically assembled even in its most distant outskirts. 
Importantly, along with the reflex motion described earlier, these $\beta$ values should be incorporated in future studies that measure the dynamical mass of the MW using the Jeans equations \citep[see][for a recent review]{Wang2020}. 

\subsection{Local Effects of the LMC}\label{sec:wake}

Up to this point, we have mainly focused on the global response of the MW to the LMC, which is dominated by the reflex motion of the MW disk towards the LMC. 
As explored in detail by the simulations of \citetalias{Garavito-Camargo2019} and \citetalias{Garavito-Camargo2021}, the LMC is also predicted to impart significant \textit{local} perturbations along its past orbit. 
This `transient response' is a direct consequence of the LMC's gravitational influence and dynamical friction wake. 
Although \cite{Conroy2021} present evidence of a wake-like feature in the density of stars, the tell-tale signature of the transient response should be imprinted in the kinematic structure of distant halo stars (see \citetalias{Garavito-Camargo2019} for a detailed exploration).
These kinematic features should appear as higher-order (in angular scale) structures in velocity maps of the outer halo, above the dominant lower-order structure imprinted by the overall travel velocity (see Figure~\ref{fig:sim_vgsr_map}). 

\begin{figure}
    \centering
    \includegraphics[width=\columnwidth]{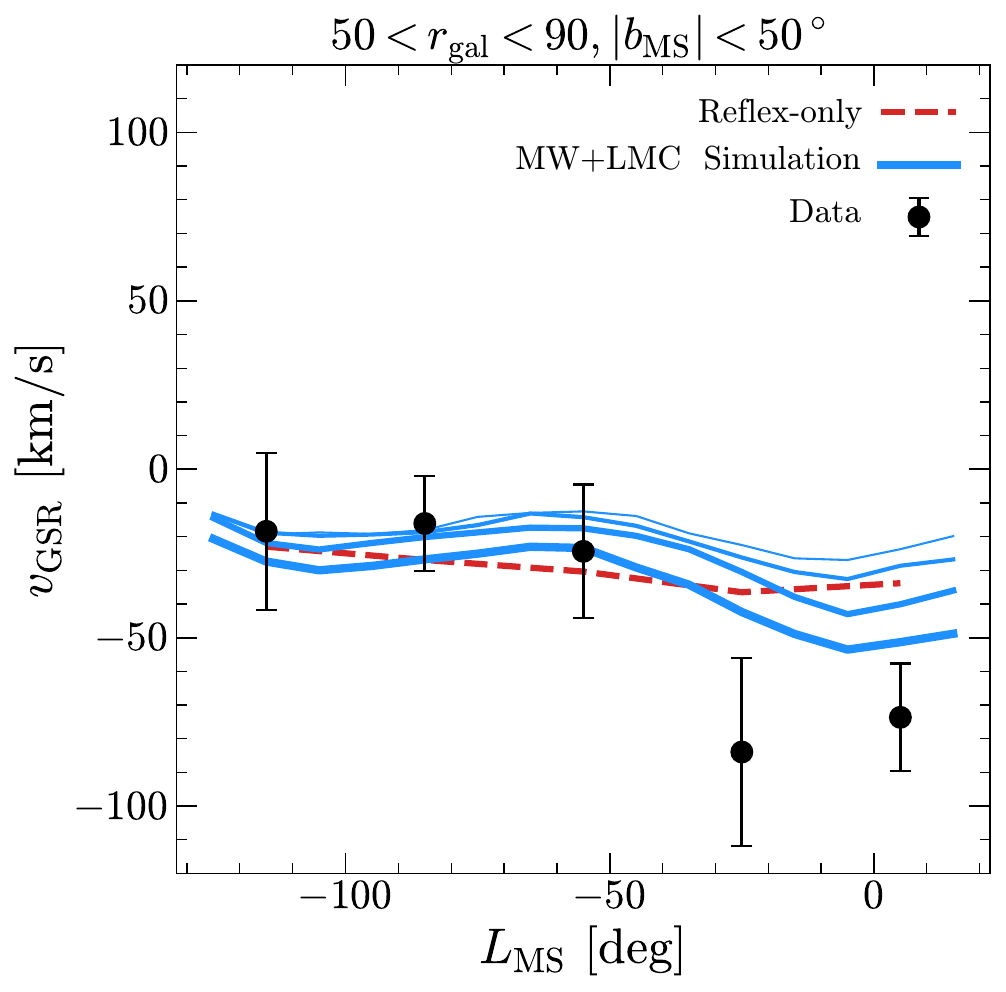}
    \includegraphics[width=\columnwidth]{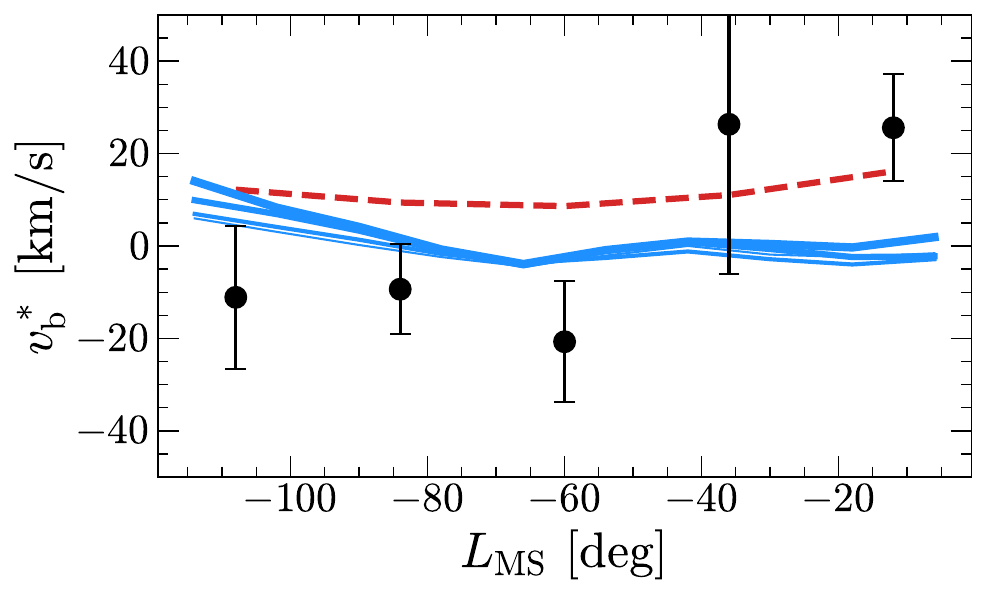}
    \includegraphics[width=\columnwidth]{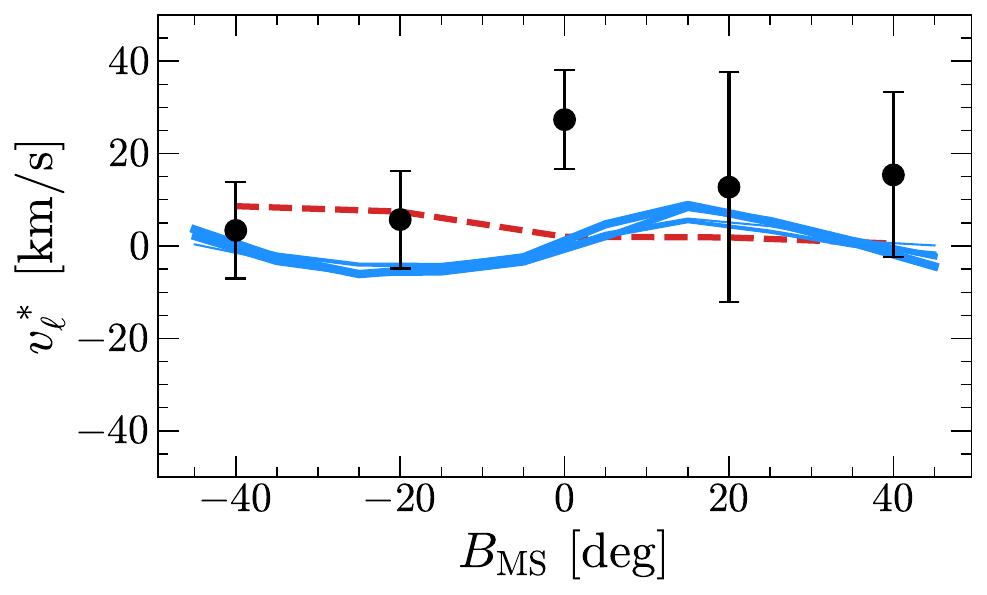}
    \caption{Kinematic signatures of the LMC-induced dynamical friction wake along the Galactocentric radial velocity (top panel), Galactic latitude (middle panel) and Galactic longitude (bottom panel) components. 
    The tangential velocities have been corrected for the solar reflex motion. 
    Binned mean measurements from our dataset are shown in black with $1\sigma$ uncertainties, and predictions from the N-body simulations of \citetalias{Garavito-Camargo2019} are shown in blue. 
    For the latter, thicker lines correspond to more massive LMC halo masses.
    Predictions from the pure-reflex $v_\mathrm{travel}$ model (with no local LMC perturbations) are shown in red. 
    }
    \label{fig:local_vgsr}
\end{figure}

To more closely examine the local velocity perturbations due to the LMC, we compare selected observed velocity trends to the simulations of \citetalias{Garavito-Camargo2019} in Figure~\ref{fig:local_vgsr}. 
The observed datapoints in Figure~\ref{fig:local_vgsr} are directly measured from our stellar dataset, i.e., they are independent of the \vtravel{} model described above. 
An alternative and valuable approach would be to examine the residuals of the data compared to the \vtravel{} model, but we elect to show directly observed quantities here to highlight the simplest observable signatures that future studies should search for.
The top panel shows the mean Galactocentric radial velocity along the $L_\mathrm{MS}$ coordinate, which is roughly aligned with the past orbit of the LMC \citep{Nidever2008}. 
The extent of this plot only contains stars in the southern Galactic hemisphere. 
The overall trend of decreasing $v_\mathrm{GSR}$ versus $L_\mathrm{MS}$ is a signature of the global reflex response (red line), but there are higher-order `wiggles' in the trend from N-body MW+LMC simulations (blue lines) that signal \textit{local} effects from the LMC. 
It is clear that even along the relatively precisely-measured radial velocity component, a much larger sample of stars will be required to disentangle the kinematic signature of the local wake from the global reflex motion. 

The middle and bottom panels of Figure~\ref{fig:local_vgsr} are similar to the top panel, but for the tangential velocity components. 
The bottom panel uses $B_\mathrm{MS}$ as the x-coordinate, since the predicted $v_\ell^\ast$ motions produce a `converging' S-shaped trend in this space. 
Even for luminous giants, the typical tangential velocity uncertainties are an order of magnitude larger than the radial velocity uncertainties. 
Consequently, our data are entirely unable to detect the local LMC signature in tangential velocities. 
Regardless, we include these panels to demonstrate the strongest tangential velocity trends that future works should search for.

\section{Discussion}\label{sec:discussion}

\subsection{Context and Limitations}

Before discussing the conclusions of our work, it is important to highlight the limitations of our dataset and methodology. 
As Figure~\ref{fig:sample_map} illustrates, our MagE survey has expanded spectroscopic coverage of the outer halo to the southern hemisphere, a major improvement over past samples. 
However, the bottom panel of Figure~\ref{fig:sample_map} also shows that the MagE survey is much more efficient at identifying stars beyond $\gtrsim 60$~kpc than corresponding northern surveys (H3 and SEGUE). 
Therefore, there is a strong need to observe more $\gtrsim 60$~kpc stars in the northern hemisphere, to achieve homogeneous sky coverage at these distances. 
Regardless, we have demonstrated (see Figure~\ref{fig:vgsr_rgal} and Appendix~\ref{sec:velmaps_distance}) that the deeper southern coverage should not negatively affect the main measurements in our paper. 

Although the dipole nature of the outer halo's Galactocentric radial velocity structure is apparent from the raw measurements (Figure~\ref{fig:sim_vgsr_map}), eventually we fit a kinematic model (motivated by \citetalias{Petersen2021}) to the 6D phase-space positions of our sample. 
Since the MW travel velocity \vtravel{} is elucidated in Galactocentric coordinates, any uncertainties in the Solar position and velocity would propagate into our measurements. 
As noted by \citetalias{Petersen2021}, increasing the Solar velocity around the MW results in a lower \vtravel{} measurement, and vice-versa. 

The reflex velocity measured in this work is most sensitive to the mass ratio between the MW and LMC, not the absolute mass of the LMC itself. 
Therefore, our key result is perhaps best expressed as evidence that the LMC is $\gtrsim 15\%$ the mass of the MW, since the \citetalias{Garavito-Camargo2019} simulations referred to in this work assume a MW virial mass $M_\mathrm{vir, MW} = 1.2 \times 10^{12}\,M_\odot$. 
Simulations with a proportionally more/less massive MW and LMC should broadly match our observed signature too. 
Furthermore, although our measurements suggest a massive LMC on first infall, \cite{Vasiliev2024} argue that the LMC may have had another (much more distant) pericenter passage in the past. 
Our observations cannot rule out this particular second-infall scenario, since the older pericenter passage has a negligible influence on the present-day reflex motion of the MW halo \citep[see][]{Vasiliev2023, Sheng2024, Vasiliev2024}. 

In our kinematic \vtravel{} model we retain the parameterization of the bulk halo velocities --- $\langle v_\mathrm{r} \rangle$, $\langle v_\mathrm{\phi}\rangle$, $\langle v_\mathrm{\theta} \rangle$ --- from \citetalias{Petersen2021} for a straightforward comparison to these previous measurements. 
However, we note that future work should ideally utilize a more physically-motivated model for the bulk motions. 
The $\langle v_\mathrm{\phi} \rangle$ and $\langle v_\mathrm{\theta} \rangle$ motions are challenging to interpret, and it might be preferable to model a single mean rotation component of the halo, with the amplitude and axis of rotation left as free parameters. 

The \vtravel{} measured here differs in the sign of $v_\mathrm{Y}$ compared to the past measurements of \citetalias{Petersen2021} and \citetalias{Yaaqib2024}.
In Appendix~\ref{sec:pp21fit}, we investigate this discrepancy in detail. 
Briefly, we find that the \citetalias{Petersen2021} and \citetalias{Yaaqib2024} measurements appear to be significantly influenced by the presence of Sagittarius Stream (Sgr) stars in their datasets (as first noticed by A. Bystr{\"o}m, private communication). 
These Sgr stars have an on-sky velocity distribution that closely mimics a dipole pointing in the quadrant opposing the LMC. 
In Appendix~\ref{sec:pp21fit}, we demonstrate that after removing this Sgr contamination, the \citetalias{Petersen2021} dataset does not have a detectable \vtravel{} signature (Figure~\ref{fig:pp21_clean}). 
Therefore, the most plausible explanation for the difference between our measurements and those from \citetalias{Petersen2021} and \citetalias{Yaaqib2024} is the presence of Sgr contamination in the earlier datasets. 

Another possible reason our measurement could differ from previous ones is the larger distance and sky coverage of our dataset.
In Appendix~\ref{sec:skycoverage} we test the effect of sky coverage by fitting limited subsamples of our data, finding that our SEGUE+H3+MagE dataset can recover $v_\mathrm{travel}$ even if only stars with $\delta > -20^\circ$ --- observable from the northern hemisphere --- are fitted, although the uncertainties are larger. 
However, if we only use SEGUE data, $v_\mathrm{travel}$ becomes undetected.
Therefore we concur with \citetalias{Yaaqib2024} that \vtravel{} can in principle be measured using only data observed from the northern hemisphere, but conclude that the increased sample size (and larger distance coverage) of the H3 and MagE datasets are vital contributors to our measurement.

Our agreement with simulations is by no means perfect, with the amplitude of our measured $v_\mathrm{Y}$ being larger than the simulations of \citetalias{Garavito-Camargo2019} and the Sgr measurements of \cite{Vasiliev2021a}. 
One caveat is that our inclusion of southern data could make our measurements more susceptible to \textit{local} halo perturbations of the LMC. 
Although the simulations of \citetalias{Garavito-Camargo2019} suggest that the all-sky velocity structure is quite smooth and dominated by the reflex motion itself, any deviations from the simulation (or unresolved substructure from the LMC halo itself) could bias our measurement. 
The LMC's local influence, including the presence of stripped stars from the Clouds, can be explicitly modeled in the future.

Finally, an implicit assumption in our measurements --- and the comparison to simulations --- is that there is a relatively smooth component of the MW halo, relative to which a reflex velocity can be measured. 
The \vtravel{} measurement uncertainties quoted in this work are purely statistical, assuming a particular kinematic model.
We have carefully removed known substructures including dwarf galaxies and the Sagittarius Stream from our dataset. 
However, additional unknown substructures may be present in the data. 
When fitting our \vtravel{} model, we iteratively clip stars that disagree with the model by $> 3\sigma$. 
These cuts should prevent small-scale undetected substructures from biasing our overall measurement. 
Regardless, given the emerging picture of the various constituents of the outer halo, it would be desirable for future simulations of the MW+LMC interaction to incorporate more realistic and structured models of the MW halo. 
In addition to including the Sagittarius Stream \citep[e.g.,][]{Vasiliev2021a}, simulations could utilize the latest observations of the Galactic halo's spatially-varying density profile \citep[e.g.,][]{Han2022b, Amarante2024} and velocity ellipsoid \citep[e.g.,][Figures~\ref{fig:halo_vel}-\ref{fig:beta} in this work]{Bird2021}.

\subsection{The Dynamical Disequilibrium of the Outer Halo}

The vast majority of dynamical analyses of our Galaxy --- for example, most measurements of the total mass of the MW --- assume that the stellar halo is in a state of equilibrium.
However, it remains an open question to what extent the MW has any `smooth' stellar halo component at these distances. 
Recent surveys have revealed that the inner halo is dominated by substructure from accreted dwarf galaxies, some well-mixed in phase-space, and some retaining coherent structure to this day \citep[e.g.,][]{Naidu2020, Malhan2022a}. 
Such a paradigm likely extends to the outer halo, where the fraction of surviving debris from smaller dwarf galaxies is higher \citep[e.g.,][]{Amorisco2017}. 

There is a growing picture that the density distribution of stars beyond $\approx 50$~kpc is highly anisotropic and structured (see Figure~11 in \citealt{Chandra2023b}, and also \citealt{Conroy2021} and \citealt{Amarante2024}). 
Apart from the dominant contribution of the Sagittarius stream, there exist prominent stellar overdensities including the Outer Virgo Overdensity, the Pisces Overdensity, and the elongated Pisces Plume. 
\cite{Chandra2023a} attribute the former two overdensities to the MW's most recent major merger, Gaia-Sausage Enceladus \citep{Helmi2017, Belokurov2018a}. 
The Pisces Plume appears to have a Magellanic origin, perhaps comprising both the dynamical friction wake of the infalling LMC, and the stripped Magellanic Stellar Stream \citep{Belokurov2019a, Conroy2021, Chandra2023b, Amarante2024}. 

Even if there exists a relatively smooth and well-mixed component of the stellar halo beyond $\approx 50$~kpc, it is a priori unlikely that the outer halo retains dynamical equilibrium, since the orbital timescales of these stars are longer ($\gtrsim 1$~Gyr) than the timescales over which the gravitational potential of the MW is changing due to the arrival of the Clouds \citep[e.g.,][]{Erkal2019a}. 
However, the precise amplitude and nature of this disequilibrium has been challenging to measure due to the paucity of all-sky stellar kinematic samples at these large distances. 

Prior to any detailed modeling, $\S$\ref{sec:velstruct} and Figures~\ref{fig:sim_vgsr_map}-\ref{fig:vtb_rgal} in this work clearly illustrate the dramatic departure from any notion of `dynamical equilibrium' beyond $\approx 50$~kpc. 
In $\S$\ref{sec:vreflex_model} we use a physical model to demonstrate that the kinematic disequilibrium of the Galactic outskirts can be explained --- to first order --- by the gravitational influence of the LMC. 
The kinematic influence of the LMC on our sample of stars is notably strong, suggesting an LMC dark matter halo on the massive side of recent estimates, $\gtrsim 15\%$ the mass of the Milky Way. 

Neglecting the LMC-induced disequilibrium of the halo can inflate dynamical mass estimates of the MW itself by up to 50\% \citep{Erkal2019a, CorreaMagnus2022, Kravtsov2024}. 
Even mass measurements from the so-called `timing argument' can be inflated by $10\%$ if the travel velocity is ignored \citep{Chamberlain2023}. 
Although we have made strides here towards observationally measuring this disequilibrium, the physical interpretation depends not only on the mass of the MW-LMC system, but also the past orbit of the LMC. 
Although the SMC is often ignored in simulations due to its low mass ($\approx 10\%$ of the LMC), it can substantially alter the implied past orbit of the LMC \citep{Vasiliev2023}. 
There is therefore a growing need for a comprehensive dynamical analysis that includes all four protagonists in the outer halo --- the MW, LMC, SMC, and Sagittarius --- a `tango for four' in the same spirit as \cite{Vasiliev2021a}. 
Our survey of the outer halo has contributed two strong observational constraints for this problem: the reflex motion measured in this work, and the Magellanic Stellar Stream \citep{Chandra2023b}.

\subsection{The Elusive Dark Matter Wake of the LMC}

The dynamical friction wake of the LMC is predicted to leave imprints in both the density and kinematics of halo stars along the past orbit of the LMC \citep[\citetalias{Garavito-Camargo2019};][]{Belokurov2019a, Conroy2021, Cavieres2024}. 
Although this wake has been detected in the density of stars --- forming the so-called `Pisces Plume' overdensity --- a confident detection of the wake's kinematic signature remains elusive. 
In Figure~\ref{fig:local_vgsr} we search for the LMC's local kinematic perturbation in radial and tangential velocities, over and above the bulk reflex motion. 
However, even in the relatively well-measured radial velocity direction, the detection is weak to non-existent given current observational errors. 

The middle and bottom panels of Figure~\ref{fig:local_vgsr} show predicted and observed tangential velocity signatures along the past orbit of the LMC.
% The predicted signals are discussed in detail by \citetalias{Garavito-Camargo2019}. 
These signatures are more straightforward to disentangle from the reflex-only model (compared to radial velocities), as indicated by the separation between the blue and red curves in Figure~\ref{fig:local_vgsr}.
Despite the size and fidelity of our giant sample, the observational errors are currently far too large to detect these local variations in the tangential velocity --- the typical uncertainty in each bin is 30~\kms{}, of the same order as the predicted peak-to-peak signal. 

Based on our results, detecting the tangential velocity signature of the LMC's dark matter wake at the 2$\sigma$ level will require $\gtrsim 3\times$ smaller velocity uncertainties, or a $\approx 10\times$ larger sample size ($\approx 8000$ stars beyond 40~kpc). 
Updated proper motions from the next \textit{Gaia} data release should also shrink proper motion uncertainties by almost a factor of two.
A mixture of the dataset presented here and ongoing surveys like SDSS-V \citep{Kollmeier2017} and DESI \citep{Cooper2023} should eventually be sensitive to the velocity signature of the LMC's dark matter wake. 

\section{Conclusions}\label{sec:conclusion}

We have assembled the largest all-sky dataset of $851$ luminous red giant stars from $40-160$~kpc with full 6D kinematics, including $\approx 300$ new stars in the southern hemisphere from our own tailored spectroscopic survey. 
In this work, we utilize this dataset to measure the motion of the inner galaxy relative to the outer halo, and find that it is induced by a massive and recently-infalling LMC. 
Our main conclusions are summarized below. 

\begin{itemize}

    \item We construct an all-sky dataset of $851$ red giant stars beyond $40$~kpc using spectroscopy from H3, SEGUE, and our own MagE outer halo survey. 
    All spectra are homogeneously analyzed with the same stellar parameter pipeline, resulting in the largest all-sky catalog of distant halo stars with 6D kinematic information (Figures~\ref{fig:dval}-\ref{fig:sample_map}). 

    \item The radial velocity distribution of distant halo stars has significant spatial structure (Figure~\ref{fig:sim_vgsr_map}). We measure an all-sky dipole radial velocity signature with an apex aligned towards the present-day location of the LMC.
    The velocity dipole grows as a function of distance, a hallmark signature of the expected `reflex response' of the MW halo to a massive LMC (Figure~\ref{fig:vgsr_rgal}). 

    \item We measure --- for the first time --- the growth of the tangential velocity signature of the reflex motion as a function of distance (Figure~\ref{fig:vtb_rgal}). 
    The tangential velocity along the Galactic latitude $\langle v_\mathrm{t,b}^\ast \rangle$ rises to $\approx 50$~\kms{} beyond 100~kpc, consistent with the response to a massive $\gtrsim 10^{11}\,M_\odot$ LMC. 
    
    \item We fit the 6D positions and velocities of 821 stars from $40 < r_\mathrm{gal} < 120$ to infer the amplitude and direction of the inner MW's motion relative to distant halo stars. 
    The apex direction of this travel velocity points towards the past orbit of the LMC, close to its present-day location (Figure~\ref{fig:vreflex_results}, top). 
    The amplitude of the travel velocity grows from $\sim 0$ at $r_\mathrm{gal} = 40$ up to $50 \pm 20$~\kms{} at $r_\mathrm{gal} = 120$~kpc (Figure~\ref{fig:vreflex_results}, bottom).
    A 3-D schematic of the measurement is shown in Figure~\ref{fig:vreflex_cartoon} (animated version \href{https://bit.ly/mw-vtravel}{here}).
    
    \item The \vtravel{} direction and amplitude measured in this work align with predictions from N-body simulations of the MW halo's response to the LMC. 
    The measured amplitude of the reflex velocity is most consistent with an LMC that is $\approx 15\%$ the mass of the Milky Way, based on the simulations of \citetalias{Garavito-Camargo2019} (Figure~\ref{fig:vreflex_results}).
 
    \item Along with \vtravel{} itself, the bulk velocity and velocity dispersion components of the halo are simultaneously measured (Figure~\ref{fig:halo_vel}). 
    We detect a strong radial `compression' signal ($\langle v_r \rangle \lesssim 20$~\kms{}), as well as some net azimuthal rotation in the outer halo.
    
    \item The intrinsic velocity dispersion of the MW halo declines as a function of distance in along all three spherical coordinates. 
    After accounting for the LMC-induced reflex motion, outer halo stars exhibit significantly radially-biased motions, with a rising profile of the velocity anisotropy parameter $\beta \approx 0.6 - 0.8$ from to $40-120$~kpc (Figure~\ref{fig:beta}).

    \item We search for the \textit{local} influence of the LMC on the mean radial velocities of MW halo stars along its past orbit (Figure~\ref{fig:local_vgsr}).
    A much larger dataset will be required to definitively measure these effects, at least twice as large to detect the radial velocity signature, and $\approx 10\times$ larger for the tangential velocity signature. 
    Future spectroscopic surveys --- combined with \textit{Gaia} DR4 --- should be sensitive to these signatures. 
    
\end{itemize}

\clearpage

\begin{acknowledgments}

We thank the referee, Michael Petersen, for thorough feedback that significantly improved this work.
VC gratefully acknowledges a Peirce Fellowship from Harvard University. RPN acknowledges support for this work provided by NASA through the NASA Hubble Fellowship grant HST-HF2-51515.001-A awarded by the Space Telescope Science Institute, which is operated by the Association of Universities for Research in Astronomy, Incorporated, under NASA contract NAS5-26555. CL acknowledges funding from the European Research Council (ERC) under the European Union’s Horizon 2020 research and innovation programme (grant agreement No. 852839). It is a pleasure to thank 
Amanda Bystr{\"o}m,
Nitya Kallivayalil,
David Nidever,
Theo O'Neill,
Ekta Patel,
Michael Petersen,
Yanjun Sheng,
Eugene Vasiliev,
and Rashid Yaaqib
for insightful conversations and feedback. 
This work benefited from discussions at the 2024 ``Milky Clouds over Manhattan'' Workshop at the Flatiron Institute Center for Computational Astrophysics. 

We are indebted to the CfA and MIT telescope time allocation committees for enabling our long-term survey of the Galactic outskirts. 
We thank the staff at Las Campanas Observatory --- including Yuri Beletsky, Carla Fuentes, Jorge Araya, Hugo Rivera, Alberto Past\'en, Mauricio Mart\'inez, Roger Leiton, Mat\'ias D\'iaz, Carlos Contreras, and Gabriel Prieto --- for their invaluable assistance. 
The H3 Survey is funded in part by NSF grant NSF AST-2107253.
The computations in this paper were run on the FASRC Cannon cluster supported by the FAS Division of Science Research Computing Group at Harvard University. 

This work has made use of data from the European Space Agency (ESA) mission {\it Gaia} (\url{https://www.cosmos.esa.int/gaia}), processed by the {\it Gaia} Data Processing and Analysis Consortium (DPAC, \url{https://www.cosmos.esa.int/web/gaia/dpac/consortium}). Funding for the DPAC has been provided by national institutions, in particular the institutions participating in the {\it Gaia} Multilateral Agreement. 
This research has made extensive use of NASA's Astrophysics Data System Bibliographic Services.

\end{acknowledgments}

\software{\texttt{numpy} \citep{Harris2020}, 
\texttt{scipy} \citep{Virtanen2020}, 
\texttt{matplotlib} \citep{Hunter2007}, 
\texttt{astropy} \citep{AstropyCollaboration2013, AstropyCollaboration2018, AstropyCollaboration2022},
\texttt{gala} \citep{gala,adrian_price_whelan_2020_4159870},
\texttt{MINESweeper} \citep{Cargile2020},
\texttt{ReflexMotion} \citepalias{Petersen2021}
}

\facilities{Magellan:Baade (MagE), Gaia, Sloan, PS1, 2MASS, WISE}

\bibliography{library, bib}
\bibliographystyle{aasjournal}

\clearpage
\appendix 

\section{The PP21 Dataset}\label{sec:pp21fit}

A key conclusion from our work is that the apex direction of the MW's motion points towards a very recent location along the past orbit of the LMC (Figure~\ref{fig:vreflex_results}). 
Although this apex direction agrees well with the results from \cite{Bystrom2024}, it is somewhat different from the direction reported in \citetalias{Petersen2021} and \citetalias{Yaaqib2024}. 
In particular, the latter two works report an apex direction that points towards the opposing Galactic quadrant as the LMC (Figure~\ref{fig:model_comparison}). 
In this section, we examine the dataset used by \citetalias{Petersen2021} to investigate the source of this difference.
We use the K giant dataset from \citetalias{Petersen2021} --- although that work also fitted BHB stars and satellite galaxies, the K giants were by far the most constraining due to their low proper motion uncertainties.
We find that the most plausible reason for our differing measurements is the presence of Sagittarius Stream stars in the \citetalias{Petersen2021} and \citetalias{Yaaqib2024} datasets.

\begin{figure}
    \centering
    \includegraphics[width=\columnwidth]{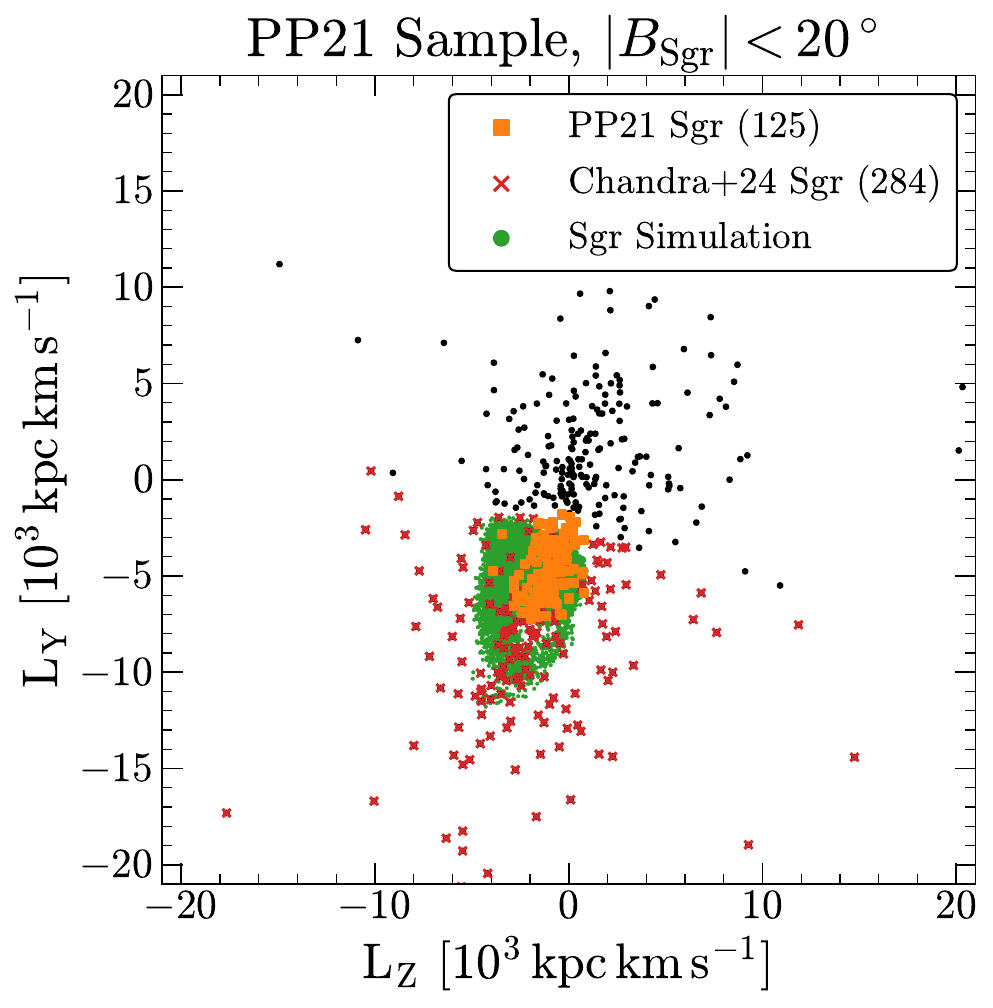}
    \caption{Comparing different angular momentum selections to identify and remove debris from the Sagittarius Stream. The entire sample from \citetalias{Petersen2021} is shown in black. Stars that they identify as Sgr members using a spherical angular momentum selection are shown in orange, and stars that our linear angular momentum cut would identify as Sgr debris are shown in red. 
    For reference, the Sgr simulation from \cite{Vasiliev2021a} is shown in green.}
    \label{fig:L_sgr}
\end{figure}

\citetalias{Petersen2021} and \citetalias{Yaaqib2024} utilized the angular momentum-based cuts from \cite{Penarrubia2021} to identify Sgr stars, selecting stars within 3000~\lunit{} of the Sgr progenitor's angular momentum: $[L_x=605, L_y=-4515, L_z=-1267]$~\lunit{}. 
Figure~\ref{fig:L_sgr} shows the K giant dataset from \cite{Petersen2021}, with stars satisfying their Sgr selection highlighted in orange. 
We only show stars within $20^\circ$ of the Sgr orbital plane, as defined by the \cite{Law2010} coordinate frame.
In red we show stars from their dataset that would fail our broader angular momentum selection drawn from \cite{Johnson2020}, which removes all stars with $L_\mathrm{Y} < -2.5 - 0.3 \times L_\mathrm{Z}$, where the angular momentum is in units of $10^3$\,\lunit{}. 
In the background, green points show Sgr stars from the simulations of \cite{Vasiliev2021a}, which were carefully fit to observations of the stream.

\begin{figure}
    \centering
    \includegraphics[width=\columnwidth]{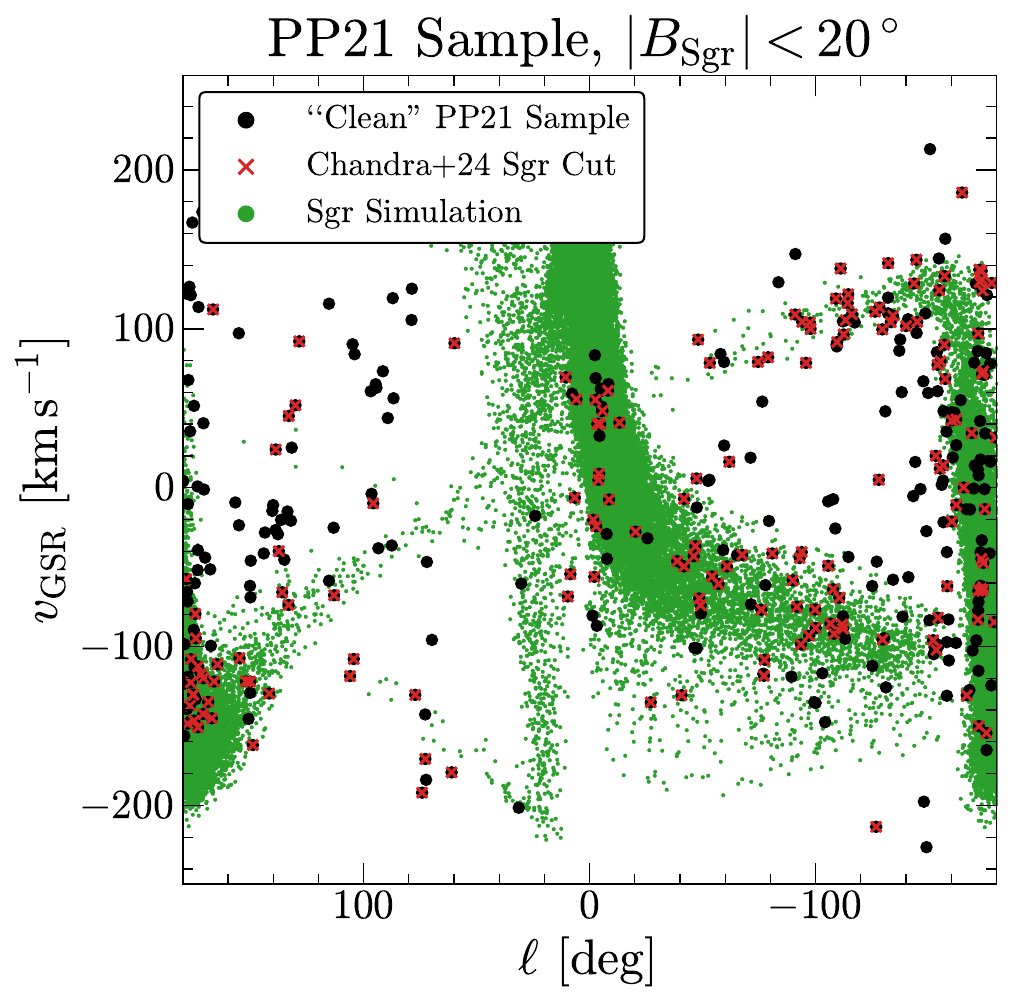}
    \includegraphics[width=\columnwidth]{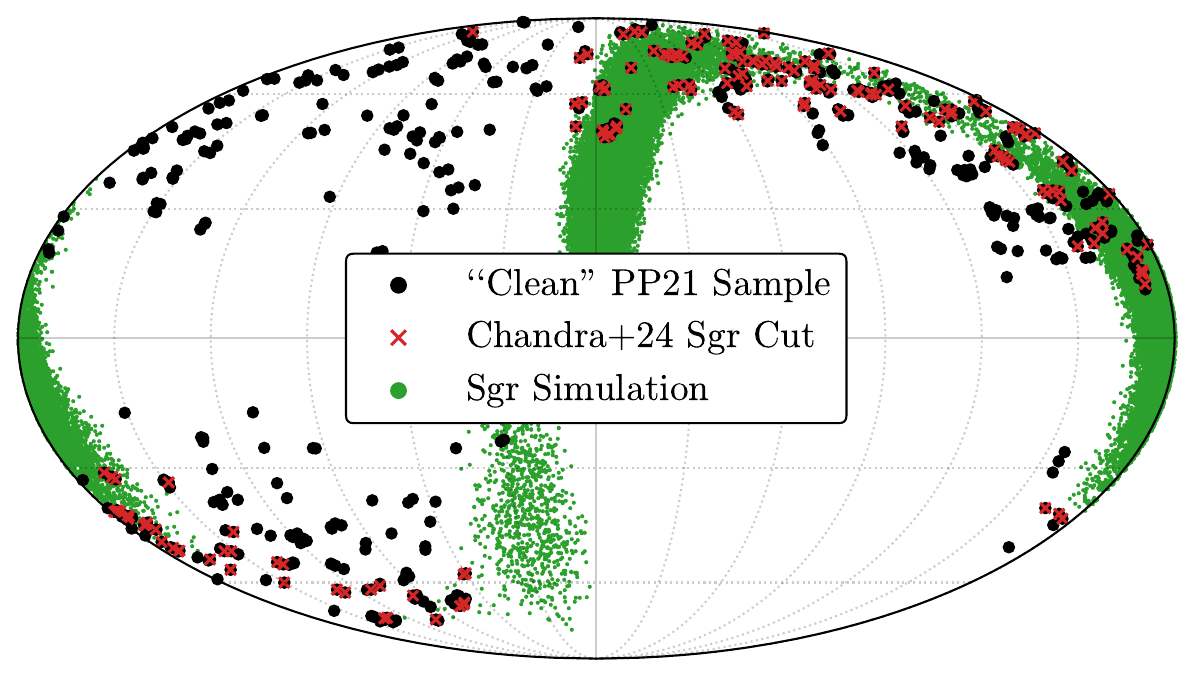}
    \includegraphics[width=\columnwidth]{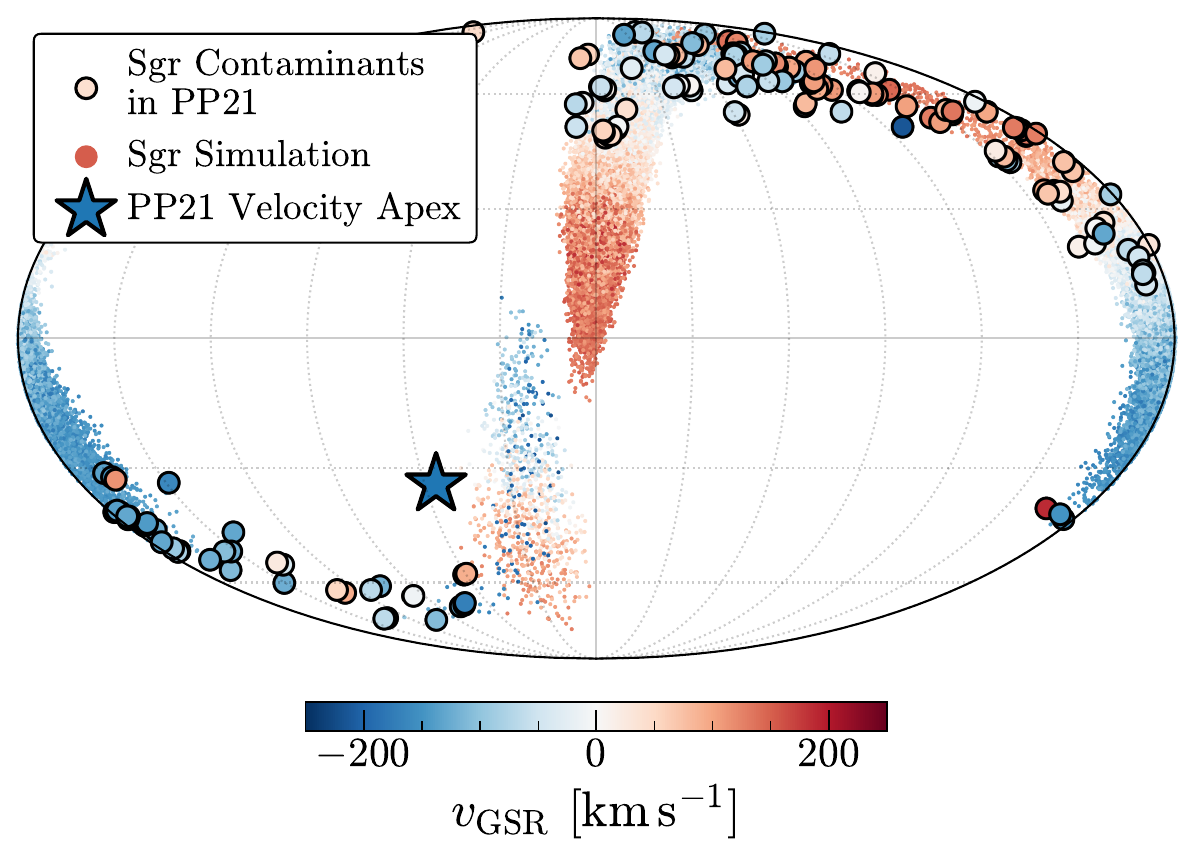}
    \caption{Illustrating the presence and impact of Sgr debris in the sample from \citetalias{Petersen2021}. 
    Top: Galactocentric radial velocity as a function of Galactic longitude. 
    Middle: On-sky distribution in Galactic coordinates. 
    We show the clean sample from \citetalias{Petersen2021} as black circles, highlighting with red crosses stars that our cuts would identify as Sgr debris.
    For reference, the Sgr simulation from \cite{Vasiliev2021a} is shown (green). 
    Bottom: Stars from \citetalias{Petersen2021} that we would classify as Sgr debris, colored by Galactocentric radial velocity. The \cite{Vasiliev2021a} simulation of Sgr stars is also shown with the same color scale. 
    } 
    \label{fig:pp21_clean}
\end{figure}

Figure~\ref{fig:L_sgr} shows that there is a marked overdensity of Sgr-like stars left in the \citetalias{Petersen2021} dataset after their Sgr cuts are applied. 
Note that the observed stars span a wider swath of angular momentum space than the \cite{Vasiliev2021a} simulations due to observational uncertainties, chiefly the distance uncertainty.
To further illustrate this point, we show the clean \citetalias{Petersen2021} dataset in Figure~\ref{fig:pp21_clean} after their Sgr cut has been applied. 
The top panel shows radial velocities as a function of Galactic longitude, and the middle panel shows an all-sky map in Galactic coordinates. 
In both panels, it is clear that a significant fraction of the K giant dataset fit by \citetalias{Petersen2021} is coincident with Sgr debris. 
In particular, of the 548 stars in the clean \citetalias{Petersen2021} dataset, our angular momentum cuts would have removed a further 165 stars as Sgr contaminants, 30\% of the dataset fit by \citetalias{Petersen2021}.

The bottom panel of Figure~\ref{fig:pp21_clean} illustrates how the 30\% Sgr contamination could bias the reflex motion apex direction inferred by \citetalias{Petersen2021} (and by extension, \citetalias{Yaaqib2024}). 
We show the 165 stars in their final sample that fall into the Sgr selection used in this work, colored by Galactocentric radial velocity. 
In the background are Sgr particles from the simulations of \cite{Vasiliev2021a}. 
The Sgr debris have a strong signature in radial velocities, which broadly appears like a dipole on the sky: moderately positive in the northern hemisphere, and strongly negative in the southern hemisphere. 
Furthermore, this dipole is oriented along the Sgr stream, with an apex direction (region of maximum negative Galactocentric radial velocity) in the Galactic quadrant opposing the LMC. 
This is the same quadrant as the apex direction inferred by \citetalias{Petersen2021} (indicated in Figure~\ref{fig:pp21_clean}, see also Figure~\ref{fig:model_comparison}). 
Clearly, the presence of these Sgr stars could significantly influence the apex direction inferred by \citetalias{Petersen2021}.

\begin{figure}
    \centering    
    \includegraphics[width=\columnwidth]{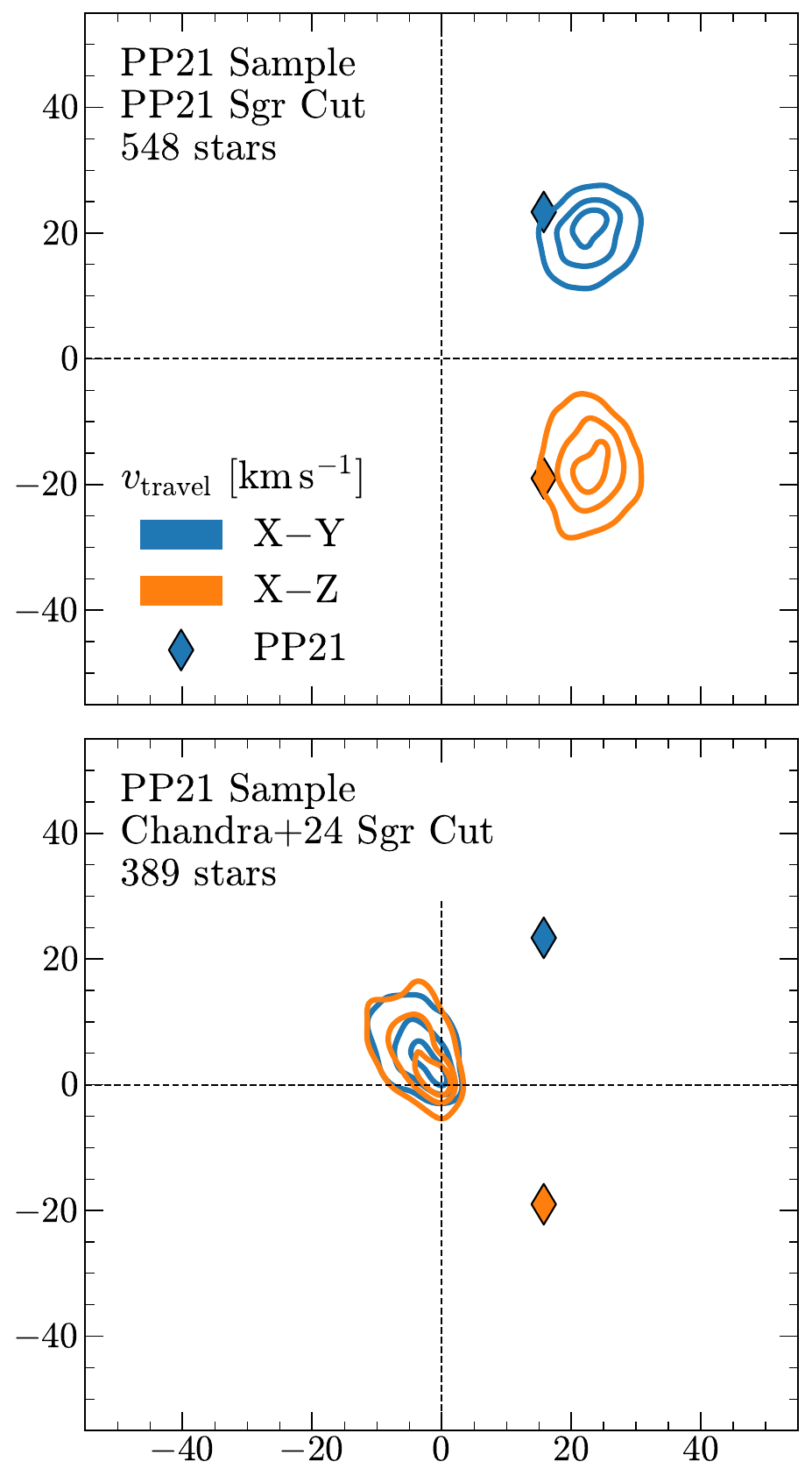}
    \caption{Posterior distribution of \vtravel{} in Galactocentric coordinates inferred by fitting the \citetalias{Petersen2021} dataset, before (top panel) and after (bottom panel) applying our angular momentum cuts to remove the $\approx 30\%$ residual Sgr contamination in that dataset. 
    The contours show our measurements, and the diamonds show the reported measurements in \citetalias{Petersen2021}. 
    Blue contours/markers correspond to $v_\mathrm{travel, x}$ on the x-axis and $v_\mathrm{travel, y}$ on the y-axis, and orange ones correspond the same in the X$-$Z plane. 
    }
    \label{fig:fitpp}
\end{figure}

To further demonstrate the influence of Sgr contamination, we fit the stars in the \citetalias{Petersen2021} dataset with our \vtravel{} model, and also fit their dataset after applying our more broad angular momentum selection (Figure~\ref{fig:L_sgr}). 
Following \citetalias{Petersen2021}, we use the distance-independent \vtravel{} model here, fitting all stars in the dataset with a single set of \vtravel{} parameters (see Appendix~\ref{sec:vreflex_all} for the analogous measurement on our own dataset). 
The resulting posterior distributions of \vtravel{} are shown in Figure~\ref{fig:fitpp}. 
The top panel shows fits to the \citetalias{Petersen2021} dataset after applying their Sgr cuts, and the bottom panel shows the same after our Sgr cuts are applied to their dataset. 
Once our Sgr cuts are applied, the apparent \vtravel{} signature in the \citetalias{Petersen2021} dataset goes away, with \vtravel{} being within $1\sigma$ of zero in all components. 
The apparent \vtravel{} measured by \citetalias{Petersen2021} (and by extension \citetalias{Yaaqib2024}) therefore appears to be primarily driven by the kinematics of Sgr contaminants in those datasets, although some residual \vtravel{} signal may remain.

The top panel of Figure~\ref{fig:fitpp} also shows that any differences with \citetalias{Petersen2021} are not due to a mis-specification of our \vtravel{} model, since we recover their quoted parameters within $1\sigma$ --- the small disagreement in $v_\mathrm{travel, x}$ is likely due to subtle differences in the sample selection or sigma-clipping methodology. 
We conclude that our modeling framework is consistent with that of \citetalias{Petersen2021} and \citetalias{Yaaqib2024}, and the differences in our results are primarily due to Sgr contamination in the earlier datasets (see also \citealt{Bystrom2024}, who arrive at the same conclusion).

\section{The Importance of All-Sky Coverage}\label{sec:skycoverage}

In Appendix~\ref{sec:pp21fit} we argue that the primary reason our results differ from past measurements made by \citetalias{Petersen2021} and \citetalias{Yaaqib2024} is our different strategies to remove contamination from the Sagittarius Stream, with apparent Sgr contamination visible in the earlier datasets (see Figure~\ref{fig:pp21_clean}). 
However, our dataset differs from past data in two other ways: the inclusion of the H3 Survey adds a significant number of outer halo stars in the northern hemisphere, and the inclusion of our MagE survey fills in a missing southern Galactic quadrant (see Figure~\ref{fig:sample_map}). 
In this section, we test the effect of sky coverage on our \vtravel{} measurement.

We repeat the \vtravel{} measurement described in $\S$\ref{sec:vreflex_model} (incorporating the linear continuity model from $40-120$~kpc) on two subsets of our data. 
The first subset emulates the selection function of surveys observed from the northern hemisphere (e.g., H3 and SEGUE): $\delta > -20^\circ$. 
The resulting subset is predominantly composed of H3 and SEGUE stars, although some MagE stars at equatorial latitudes are also present. 
The second subset emulates the SEGUE-only dataset used by \citetalias{Petersen2021}, albeit with our updated \minesweeper{} stellar parameters and broad cuts to remove Sagittarius contamination.

\begin{figure}
    \centering
    \includegraphics[width=\columnwidth]{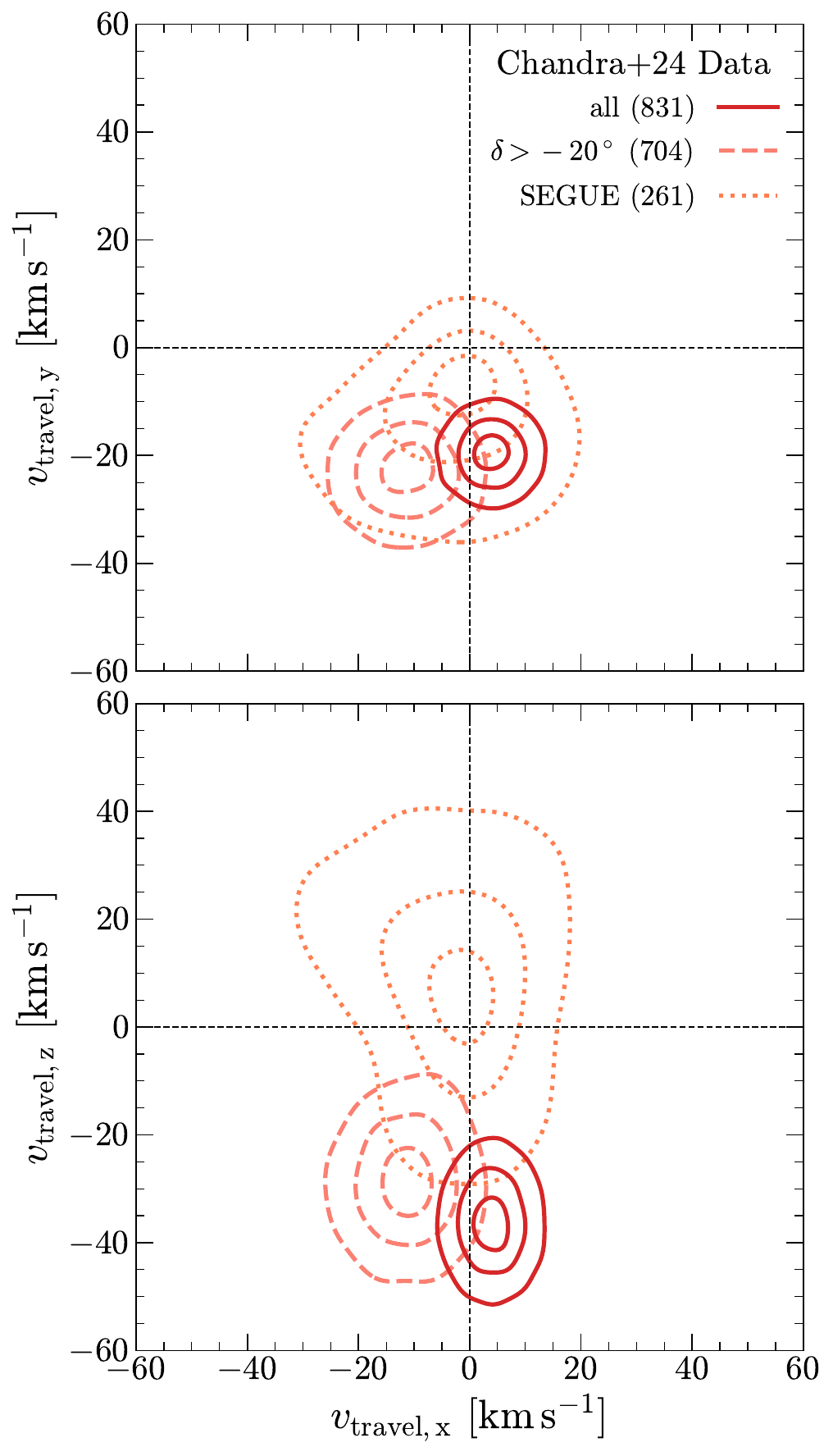}
    \caption{Posterior distribution of \vtravel{} in Galactocentric coordinates, inferred by fitting limited subsamples of our dataset. We show the measurement from using our full dataset (solid contour, identical to Figure~\ref{fig:vtravel_galcen}), from using only stars observable from the northern hemisphere (dashed), and from using only data from the SEGUE survey (dotted).}
    \label{fig:skycoverage}
\end{figure}

The resulting measurements of \vtravel{} in Galactocentic coordinates are shown in Figure~\ref{fig:skycoverage}, evaluated at $r_\mathrm{gal} = 100$~kpc analogous to Figure~\ref{fig:vtravel_galcen}. 
Using north-only data --- including our SEGUE, H3 and MagE data --- produces results somewhat consistent with our full dataset, albeit with larger uncertainties. 
Furthermore, the $v_\mathrm{travel,x}$ component is mildly biased relative to the full fit, and the $v_\mathrm{travel,y}$ signature is slightly weaker.

On the other hand, using SEGUE-only data --- after appropriately removing Sagittarius contamination using our angular momentum cuts (see Figure~\ref{fig:L_sgr}) --- results in a non-detection of the \vtravel{} signature. 
Note that this is analogous to the measurement shown in Figure~\ref{fig:fitpp} in that it only uses SEGUE data, but with some differences. 
We use our own \minesweeper{} stellar parameters rather than the \cite{Xue2015} parameters used by \citetalias{Petersen2021}.
Furthermore, in Figure~\ref{fig:fitpp} we fit a single \vtravel{} at all distances for consistency with \citetalias{Petersen2021}, whereas in Figure~\ref{fig:skycoverage} we use the full linear continuity model. 
There are less stars in our SEGUE dataset than in \citetalias{Petersen2021} since we apply more stringent SNR cuts (SNR~$>10$~per pixel) to ensure reliable stellar parameters. 
We have verified that there are no major systematic differences between our radial velocities and distances compared to the parameters from \cite{Xue2015}. 
Despite these differences, the conclusion from both these tests is similar: after Sagittarius contamination is removed, there does not appear to be a strong \vtravel{} signature in the SEGUE-only dataset.

In summary, the tests shown in Figure~\ref{fig:skycoverage} demonstrate that although it is feasible to detect \vtravel{} --- albeit with mildly biased orientations --- using only data observed from the northern hemisphere (as argued by \citetalias{Yaaqib2024}, and recently shown by \citealt{Bystrom2024} using DESI data), our SEGUE dataset alone appears insufficient to reliably measure \vtravel{}. 
The inclusion of H3 and MagE data substantially improves the north-only measurement, and the best measurement (unsurprisingly) comes from our all-sky dataset that incorporates all three surveys. 
These tests also demonstrate that the reflex response is a sensitive and challenging measurement that can be affected by a variety of factors including sky coverage, distance coverage, and contamination by substructure.

\section{Fitting Data-Matched MW+LMC Simulation}\label{sec:datamatch}

When fitting any model to data, it is important to consider the influence of the data's selection function. 
Our dataset has the advantage of sampling the entire sky, an improvement over past measurements. 
Furthermore, our MagE dataset systematically spans larger distances than past halo datasets.
However, as the top panel in  Figure~\ref{fig:sample_map} illustrates, this produces an on-sky asymmetry in the typical distance towards stars in our dataset. 
Note that for our \vtravel{} measurement, we restrict the data to $40-120$~kpc, which somewhat mitigates this asymmetry. 
However, it is important to verify that this selection effect does not affect the conclusions of this work. 

Throughout this work, we have shown for comparison \vtravel{} measurements made by fitting the simulations of \citetalias{Garavito-Camargo2019} (e.g., see Figure~\ref{fig:vreflex_results}). 
For this Appendix, we sub-sample the simulation particles to match the spatial distribution and size of our final dataset. 
Specifically, for each of the $821$ stars in our final observed sample, we select the nearest particle (in 3D Galactocentric coordinates) from the \citetalias{Garavito-Camargo2019} \texttt{LMC3} simulation, which has a $1.8 \times 10^{11}\,M_\odot$ LMC. 
This produces a mock dataset that exactly matches the spatial selection function of our observed dataset.
Since distances are expected to be our dominant source of measurement uncertainty, we assume 10\% distance uncertainties on this mock dataset.

\begin{figure}
    \centering
    \includegraphics[width=\columnwidth]{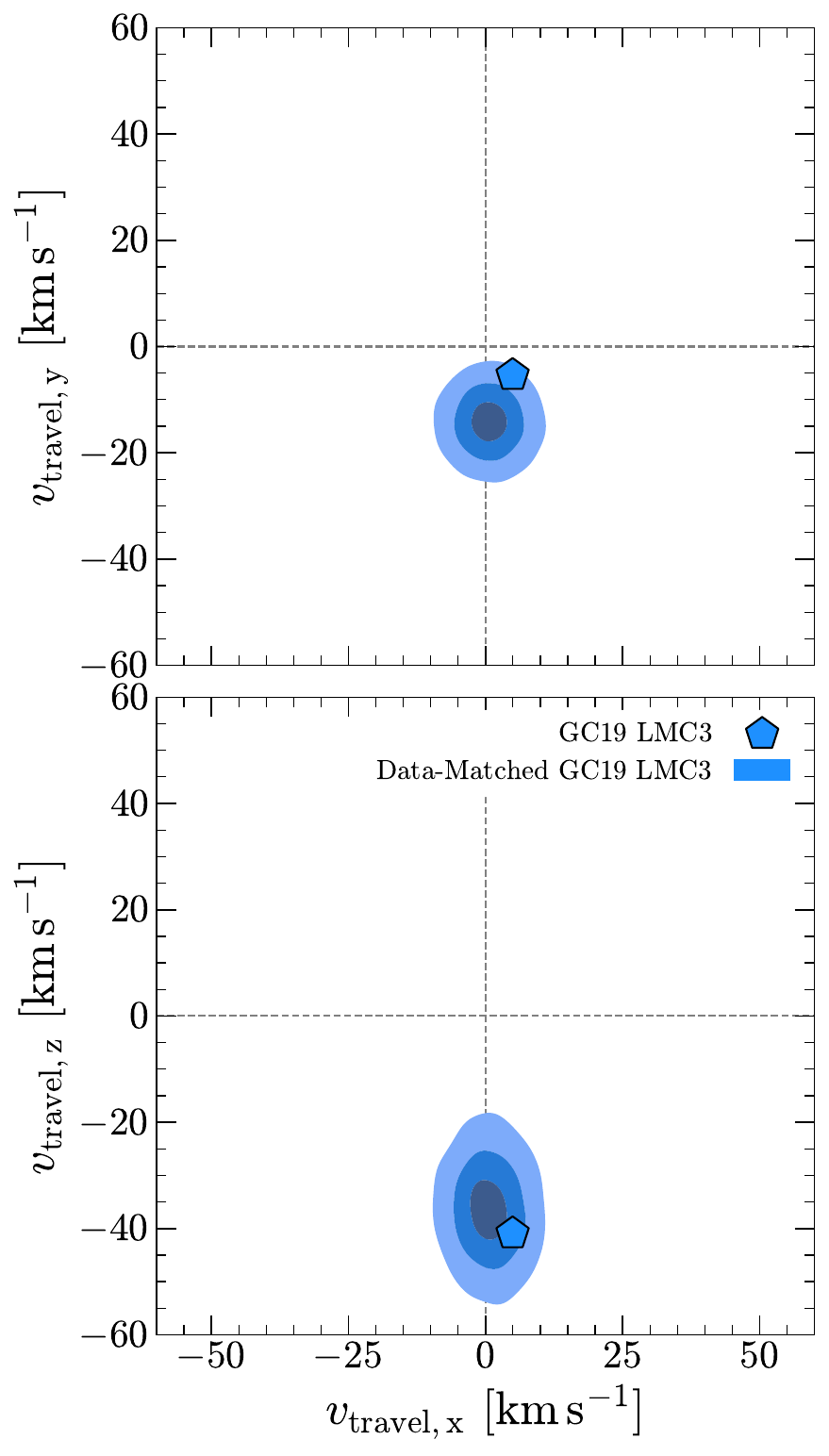}
    \caption{Posterior distribution of \vtravel{} in Galactocentric coordinates from our fit to the data-matched \citetalias{Garavito-Camargo2019} \texttt{LMC3} simulation.
    \vtravel{} is evaluated at the reference distance of $r_\mathrm{gal} = 100$~kpc, as in Figure~\ref{fig:vtravel_galcen}. 
    The `true' \vtravel{} inferred without any selection function applied is shown with a pentagon. 
    After the imposition our selection function and sample size, \vtravel{} is recovered to within 1.5$\sigma$ (the outermost contour).
    }
    \label{fig:simfit_dmatch}
\end{figure}

We fit the data-matched simulation using the \vtravel{} model described in $\S$\ref{sec:vreflex_model}, and show the resulting posterior contours in Figure~\ref{fig:simfit_dmatch}.
The \vtravel{} inferred from fitting the simulation without any imposed selection function is also shown.
We also verify that the distance-dependence of \vtravel{} is well-captured, following the intrinsic behavior of the simulation well within $1\sigma$ uncertainties. 
Finally, we verify that our result is not affected by our broad angular momentum cut to remove Sgr contamination (based on \citealt{Johnson2020a}). 
Although \cite{Johnson2020a} applied this cut to stars across the sky, we only apply it to stars within $25^\circ$ of the Sgr orbital plane, and removes $\approx 15\%$ of field stars from the simulated dataset. 
However, removing these stars does not meaningfully alter the best-fit $v_\mathrm{travel}$ parameters.
These tests support our claim that the size and spatial selection function of our dataset --- and our strategy to remove Sgr contamination --- does not limit or bias its ability to measure \vtravel{}. 

For ease of comparison to future works, we include in Table~\ref{tab:fit_summary_lmc3} our best-fit parameters to the \texttt{LMC3} simulation from \citet{Garavito-Camargo2019}. 
To determine these parameters, we randomly sub-sampled $10,000$ particles from the simulation, with no selection function or uncertainties applied. 
These parameters (and their analogs for the other three simulations from \citetalias{Garavito-Camargo2019}) are used as the reference comparisons throughout this work. 

\begin{deluxetable}{ccc}
\label{tab:fit_summary_lmc3}
\tablecaption{Summary of the best-fit parameters from applying our \vtravel{} continuity model to a random sub-sample of $10,000$~stars from the \texttt{LMC3} simulation of \citetalias{Garavito-Camargo2019}.}
\tablehead{\colhead{Parameter} & \colhead{$r_\mathrm{gal} = 40$~kpc} & \colhead{$r_\mathrm{gal} = 120$~kpc}}
\tablewidth{\columnwidth}
\startdata
$v_\mathrm{travel}$ [km$\,$s$^{-1}$] & $11$ & $51$ \\
$\ell_\mathrm{MS, apex}$ [deg] & $-25$ & $-48$ \\
$b_\mathrm{MS, apex}$ [deg] & $8$ & $4$ \\
$\langle v_\mathrm{r} \rangle$ [km$\,$s$^{-1}$] & $-5$ & $-20$ \\
$\langle v_\mathrm{\phi} \rangle$ [km$\,$s$^{-1}$] & $-1$ & $-3$ \\
$\langle v_\mathrm{\theta} \rangle$ [km$\,$s$^{-1}$] & $8$ & $9$ \\
${\sigma_\mathrm{v,r}}$ [km$\,$s$^{-1}$] & $115$ & $87$ \\
${\sigma_\mathrm{v,\ell}}$ [km$\,$s$^{-1}$] & $90$ & $62$ \\
${\sigma_\mathrm{v,b}}$ [km$\,$s$^{-1}$] & $85$ & $66$ \\
\enddata
\end{deluxetable} %

\section{All-Sky Velocity Maps vs Distance}\label{sec:velmaps_distance}

\begin{figure*}[!ht]
    \centering
    \includegraphics[width=\columnwidth]{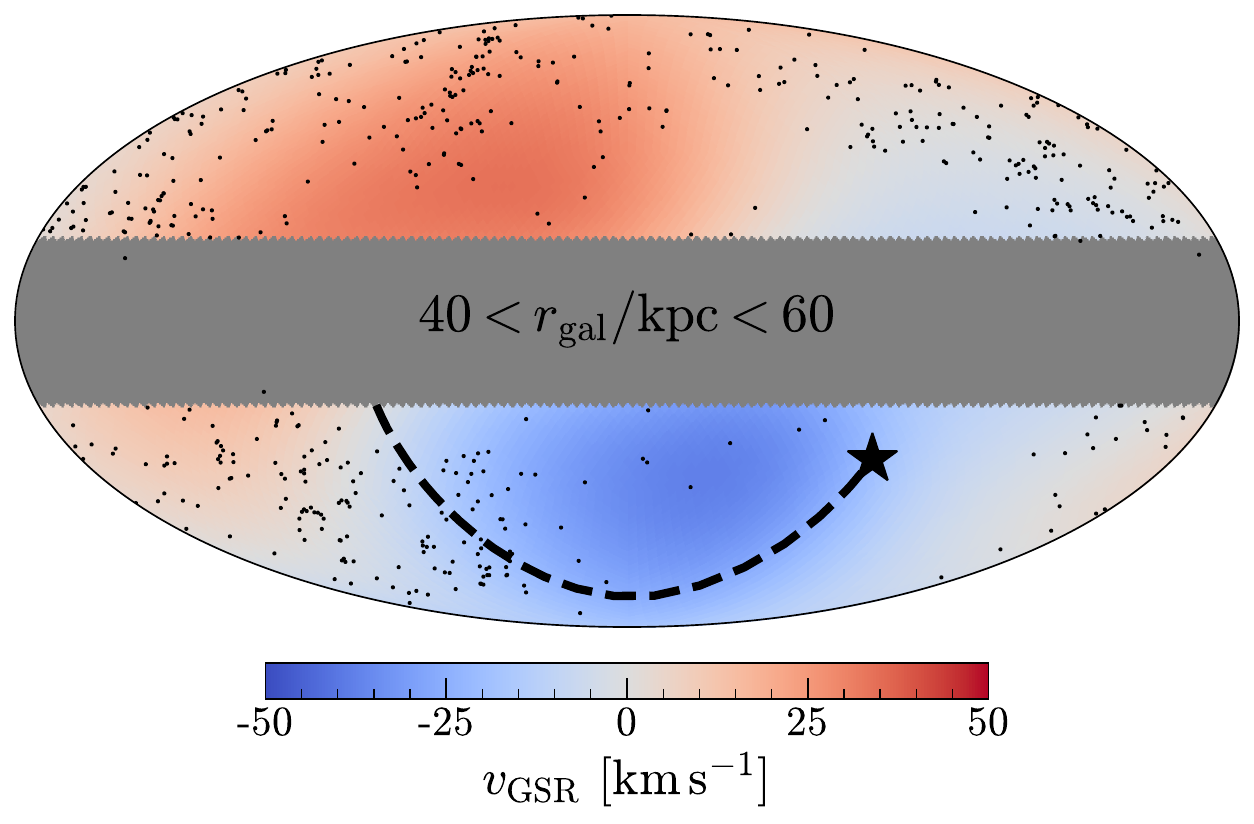}
    \includegraphics[width=\columnwidth]{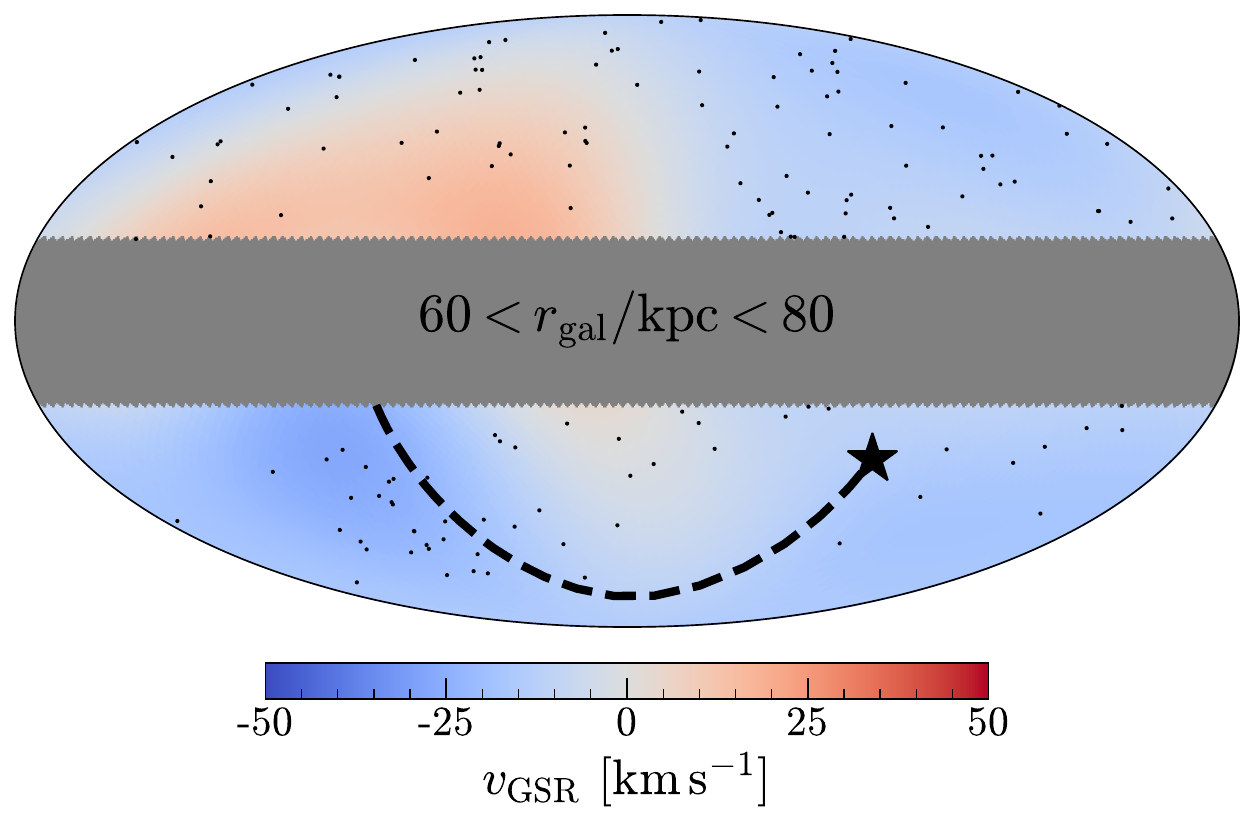}
    \includegraphics[width=\columnwidth]{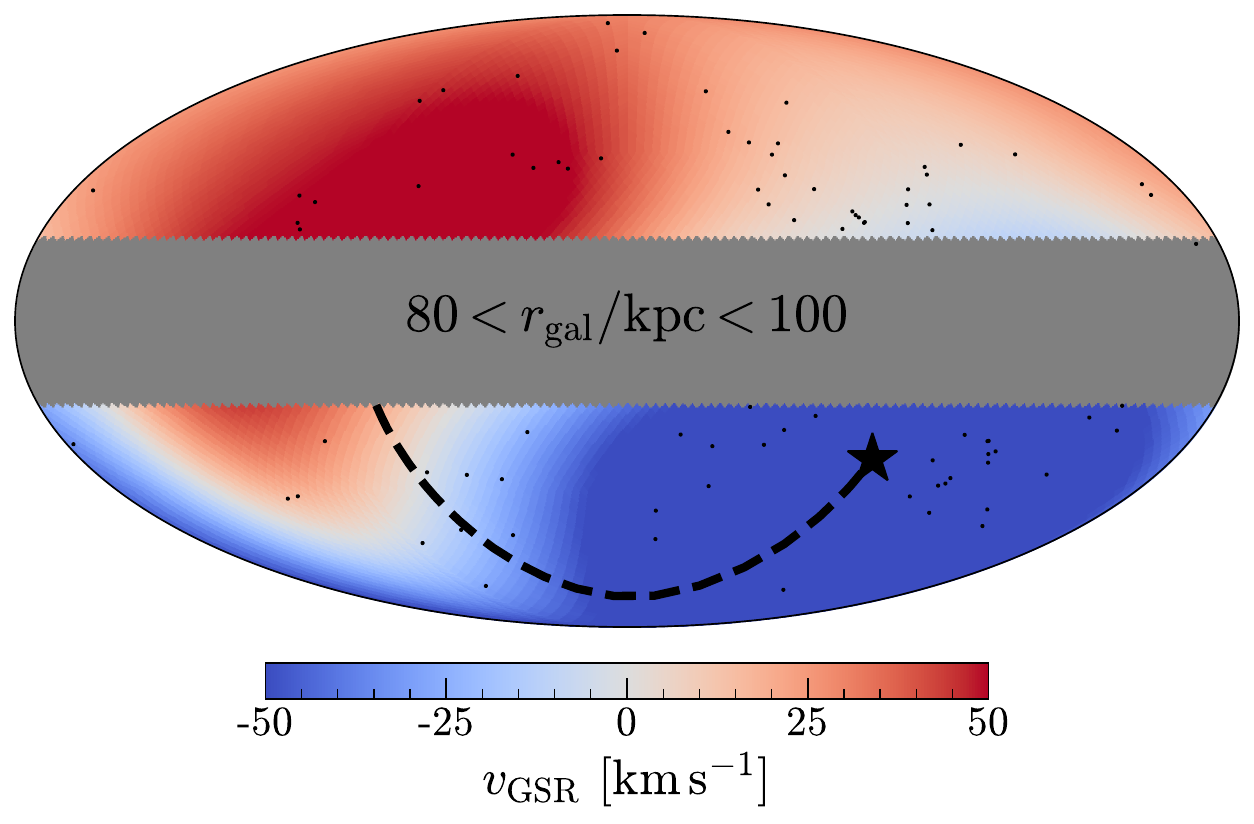}
    \includegraphics[width=\columnwidth]{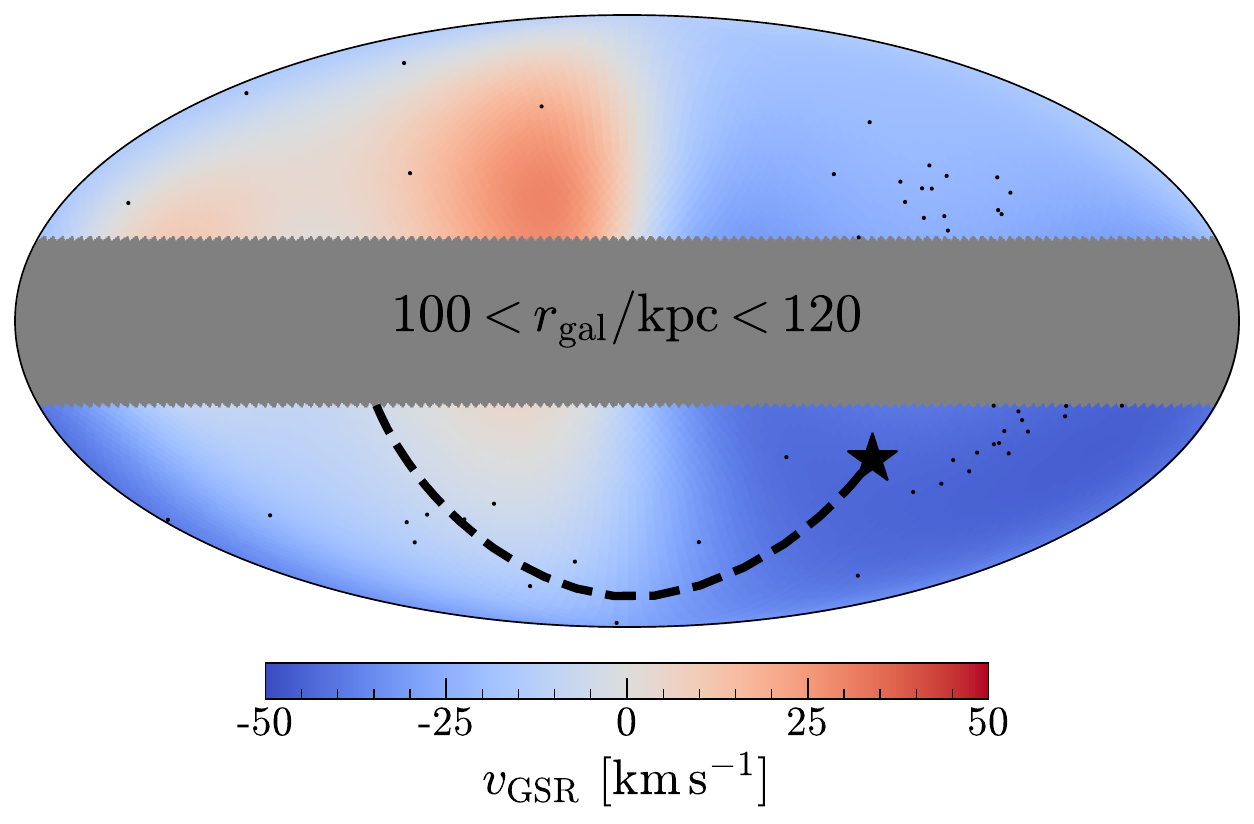}
    \caption{All-sky maps of the mean Galactocentric radial velocity $v_\mathrm{GSR}$ using our dataset, smoothed on $45^\circ$ scales. 
    These figures are computed identically to Figure~\ref{fig:sim_vgsr_map}, but are split into four disjoint $r_\mathrm{gal}$ bins.
    On-sky positions of stars in each distance bin are shown as black circles. 
    }
    \label{fig:vgsr_rbins}
\end{figure*}

As Figure~\ref{fig:sample_map} illustrates, our MagE survey in the southern hemisphere systematically pushes to further distances than the northern H3 and SEGUE samples used here. 
In the context of the all-sky velocity maps of Figure~
\ref{fig:sim_vgsr_map}, one possible concern is that the deeper southern coverage could bias these maps and produce artificial structures. 
To investigate this further, in Figure~\ref{fig:vgsr_rbins} we reproduce the mean velocity map of Figure~\ref{fig:sim_vgsr_map} in four disjoint $r_\mathrm{gal}$ bins. 
The stars in each bin are shown as black circles, demonstrating the underlying spatial distribution used to compute each velocity map. 

Although the maps shown in Figure~\ref{fig:vgsr_rbins} are naturally noisier than Figure~\ref{fig:sim_vgsr_map} --- which uses our entire sample --- the features are broadly consistent. 
% The $80-100$~kpc bin in particular has relatively uniform on-sky coverage, yet reproduces the basic structure seen in Figure~\ref{fig:sim_vgsr_map}. 
% Even if the MagE data are systematically more distant than the northern data in this bin, the difference in the reflex velocity expected (from simulations) over this bin's distance range is $\lesssim 10$~\kms{} (see Figure~\ref{fig:vreflex_results}). 
Figure~\ref{fig:vgsr_rbins} therefore provides a reassuring check that the overall velocity structure seen in Figure~\ref{fig:sim_vgsr_map} is not dominated by the inhomogeneous distance ranges spanned by the surveys combined here, but rather represents the underlying velocity structure of the halo. 
It is naturally desirable to follow up the results presented here with more spatially homogeneous datasets in the future. 
As Figure~\ref{fig:sample_map} and Figure~\ref{fig:vgsr_rbins} illustrate, there is now a strong need for more distant halo tracers in the \textit{northern} hemisphere to match the advances made by our MagE survey in the south.

\section{Distance-Independent Reflex Model}\label{sec:vreflex_all}

In $\S$\ref{sec:vreflex_model} we built on the work of \citetalias{Petersen2021} to develop a distance-dependent continuity model to measure the increasing reflex velocity as a function of $r_\mathrm{gal}$. 
Although this model is highly flexible, it also doubles the number of free parameters, and consequently the uncertainties on the measured parameters are larger. 

To ensure that we have robustly measured the basic direction and amplitude of the MW's LMC-induced travel velocity \vtravel{}, we fit all stars in our sample from $40-120$~kpc with a distance-independent \vtravel{} model. 
This model is directly analogous to that used by \citetalias{Petersen2021} and \citetalias{Yaaqib2024}, in that it uses a single value for each model parameter, instead of a linearly varying distance-dependent value (see $\S$\ref{sec:vreflex_model}).

\begin{figure}
    \centering
    \includegraphics[width=\columnwidth]{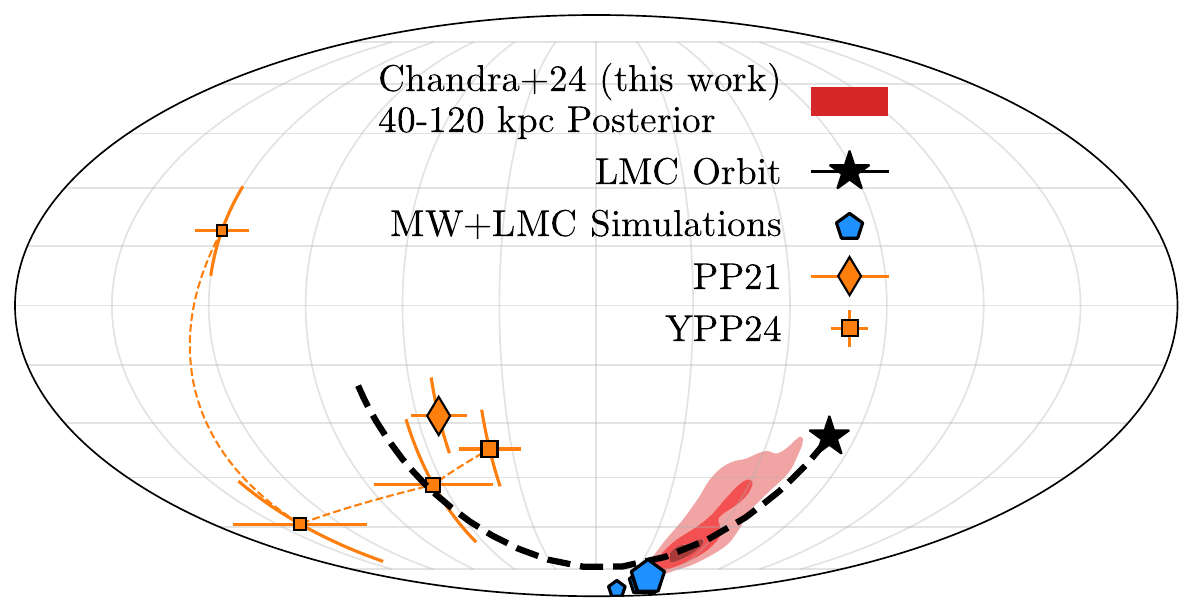}
    \includegraphics[width=\columnwidth]{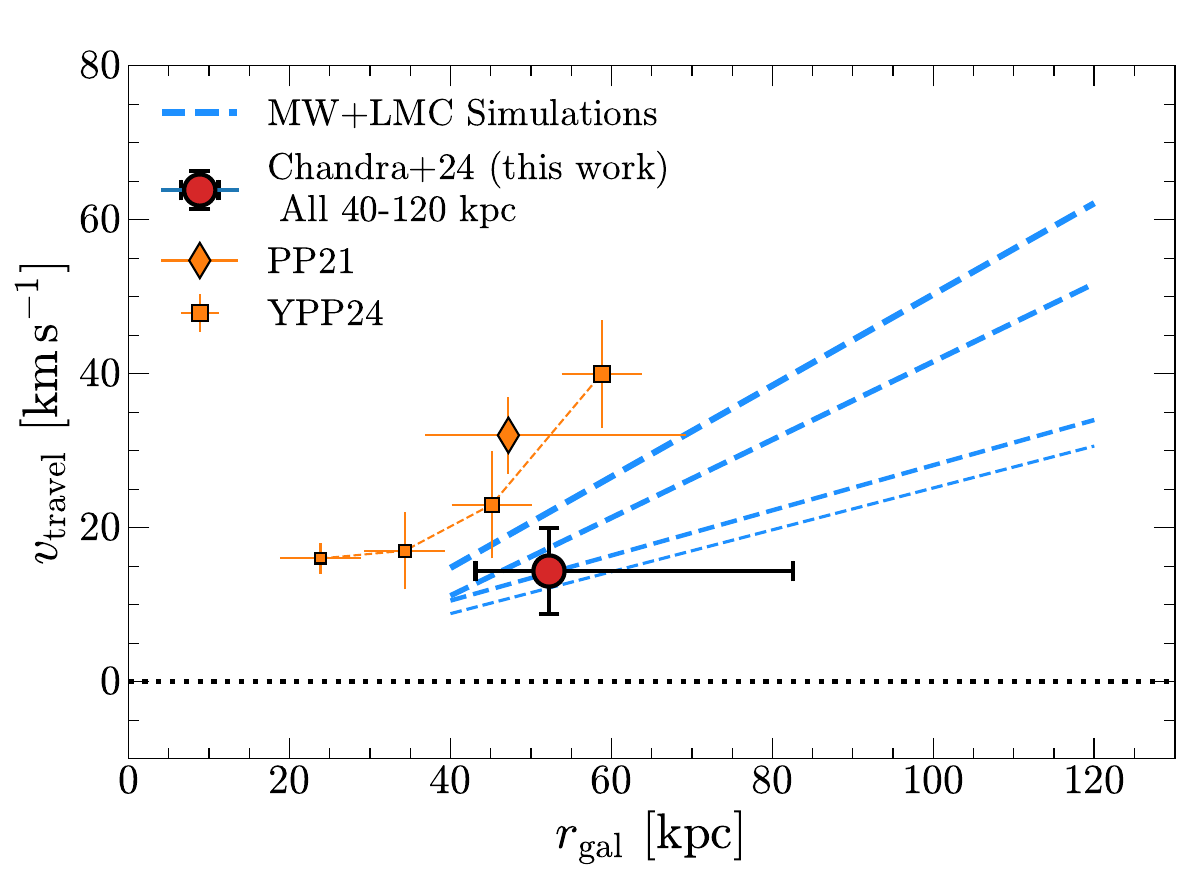}
    \caption{Results from the distance-independent \vtravel{} fit to all data from $40-120$~kpc. 
    The other elements are described in the caption for Figure~\ref{fig:vreflex_results}.
    In the bottom panel, the horizontal error bar denotes the 16th and 84th quantile of our dataset's $r_\mathrm{gal}$ distribution. 
    This distance-independent fit provides consistent results with the full distance-dependent model described in $\S$\ref{sec:vreflex_model}, most importantly regarding the apex direction of the reflex motion. 
    }
    \label{fig:vreflex_all}
\end{figure}

\begin{deluxetable}{cc}
\label{tab:fit_summary_allstars}
\tablecaption{Summary of the best-fit parameters from our \vtravel{} model that uses all stars from $40-120$~kpc.}
\tablehead{\colhead{\hspace{1cm}Parameter}\hspace{1cm} & \colhead{\hspace{1cm}Value}\hspace{1cm}}
\tablewidth{\columnwidth}
\startdata
$v_\mathrm{travel}$ [km$\,$s$^{-1}$] & $17^{+4}_{-4}$ \\
$\ell_\mathrm{apex}$ [deg] & $-86^{+35}_{-49}$ \\
$b_\mathrm{apex}$ [deg] & $-73^{+7}_{-10}$ \\
$\langle v_\mathrm{r} \rangle$ [km$\,$s$^{-1}$] & $-16^{+3}_{-3}$ \\
$\langle v_\mathrm{\phi} \rangle$ [km$\,$s$^{-1}$] & $6^{+2}_{-2}$ \\
$\langle v_\mathrm{\theta} \rangle$ [km$\,$s$^{-1}$] & $1^{+4}_{-4}$ \\
${\sigma_\mathrm{v,r}}$ [km$\,$s$^{-1}$] & $95^{+2}_{-2}$ \\
${\sigma_\mathrm{v,\ell}}$ [km$\,$s$^{-1}$] & $58^{+2}_{-2}$ \\
${\sigma_\mathrm{v,b}}$ [km$\,$s$^{-1}$] & $50^{+1}_{-1}$ \\
\enddata
\tablecomments{The marginalized posterior distribution of each parameter is summarized using the 16th, 50th, and 84th quantiles, corresponding to $1\sigma$ uncertainties for a Gaussian distribution.}
\end{deluxetable} %

The results of the distance-independent fit to all our data is shown in Figure~\ref{fig:vreflex_all}. 
As expected, the results are broadly consistent in direction and amplitude with the full continuity model fit shown in Figure~\ref{fig:vreflex_results}.
It is also clear that ignoring the distance-dependence of the reflex motion makes it very challenging to constrain the MW-LMC mass ratio with these measurements.
Most stars in the sample are at the lower end of the distance range, but the underlying reflex motion is strongest for the most distant stars. 
This further motivates our use of a distance-dependent continuity model in $\S$\ref{sec:vreflex_model}.

\section{Posterior Distribution of Distance-Dependent Reflex Model Parameters}\label{sec:corner}

\begin{figure*}[!h]
    \centering
    \includegraphics[width=0.9\textwidth]{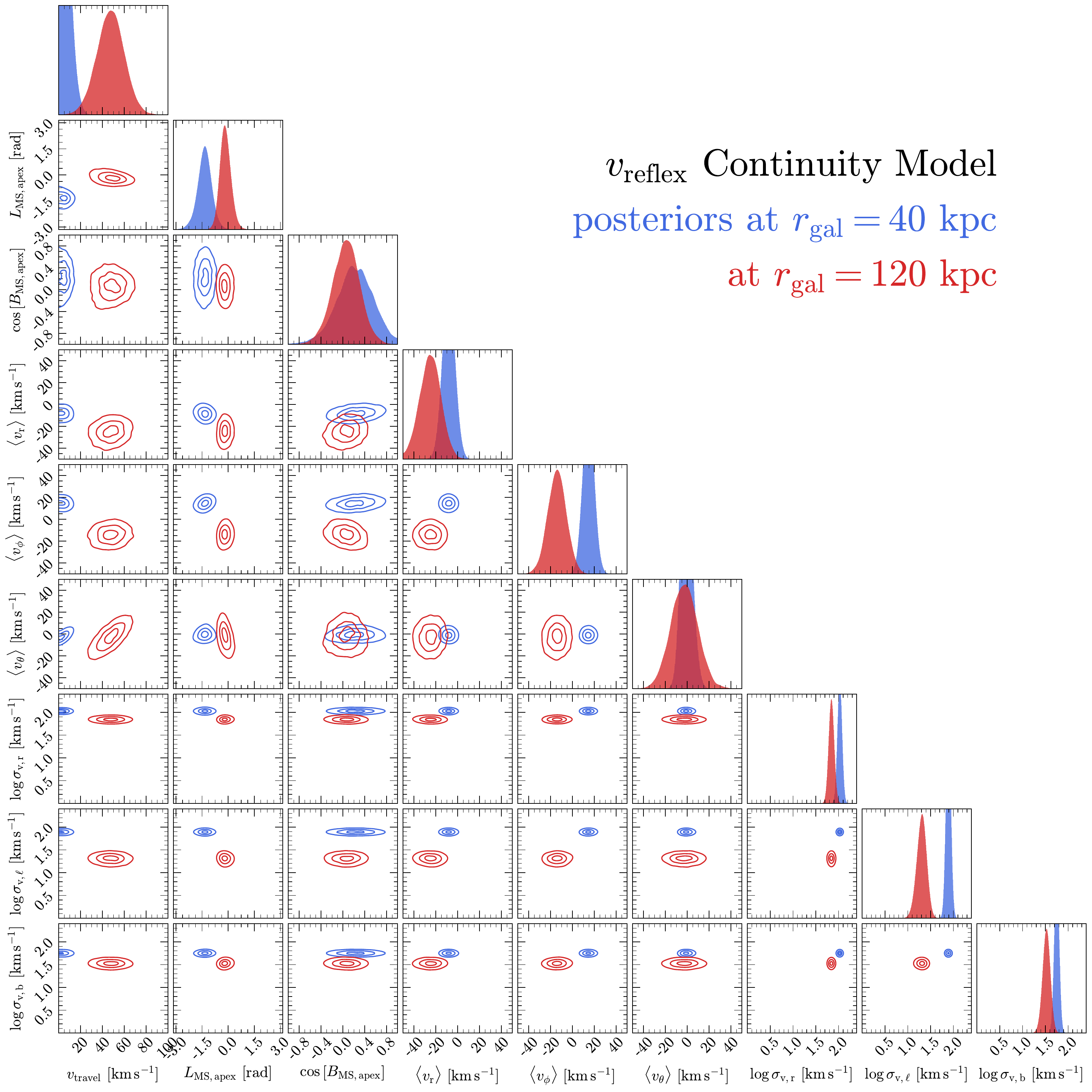}
    \caption{Posterior distribution of 18 model parameters from our distance-dependent $v_\mathrm{travel}$ continuity model (see $\S$\ref{sec:vreflex_model}). 
    Each parameter has a free value at $r_\mathrm{gal} = 40$~kpc (blue contours) and at $r_\mathrm{gal} = 120$~kpc (red contours), with the parameter assumed to vary linearly between those two values as a function of $r_\mathrm{gal}$.
    Contours correspond to 0.5$\sigma$, 1$\sigma$, and 1.5$\sigma$ Gaussian equivalents. 
    }
    \label{fig:cont_corner}
\end{figure*}

Figure~\ref{fig:cont_corner} shows the posterior distributions of all 18 parameters from our distance-dependent $v_\mathrm{travel}$ continuity model, described in $\S$\ref{sec:vreflex_model}.
These posterior distributions are quantitatively summarized in Table~\ref{tab:fit_summary}.

\begin{figure*}[!h]
    \centering
    \includegraphics[width=0.6\textwidth]{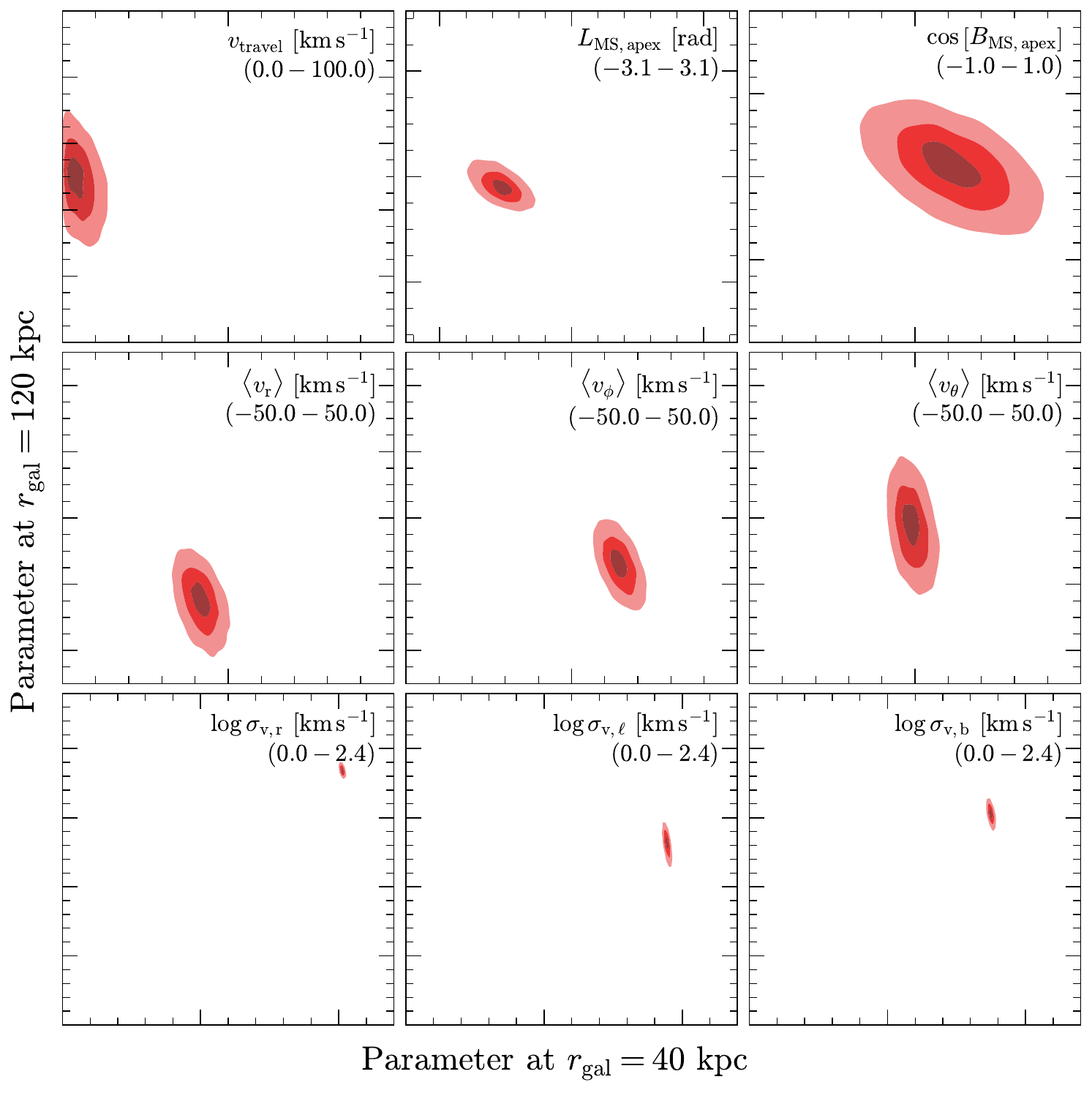}
    \caption{Posterior distributions for all parameters in the \vtravel{} continuity model, comparing the covariance between each parameter at $r_\mathrm{gal} = 40$~kpc and $r_\mathrm{gal} = 120$~kpc. 
    The axes are scaled to span the prior range of each parameter, and the axis limits are indicated in each panel.}
    \label{fig:nearfar_covar}
\end{figure*}

Figure~\ref{fig:nearfar_covar} shows the information that is missing from Figure~\ref{fig:cont_corner}, namely the covariance of each parameter at $r_\mathrm{gal} = 40$~kpc versus its counterpart at $r_\mathrm{gal} = 120$~kpc. 
There is modest correlation in all panels, which is to be expected for the same physical parameter at different distances. 
However, all parameters are well-constrained relative to their prior distributions at both distances, confirming that the linear continuity model is well-measured over the distance range considered.

\end{document}